\documentclass{ws-ijgmmp}
\usepackage[super,compress]{cite}
\usepackage{graphicx}
\usepackage{float}  
\usepackage{bm}
\usepackage{slashed}
\usepackage[T1]{fontenc}
\usepackage{hyperref}
\hypersetup{
    colorlinks=true, 
    linktoc=all,     
    linkcolor=blue,  
}
\hypersetup{linktocpage}
\begin{document}

%
\catchline{}{}{}{}{}
%

\title{Clifford algebra Cl(0,6) approach to beyond the standard model and naturalness problems}

\author{Wei Lu}


\maketitle

\begin{history}
\end{history}

\begin{abstract}
Is there more to Dirac's gamma matrices than meets the eye? It turns out that gamma zero can be factorized into a product of three operators. This revelation facilitates the expansion of Dirac's space-time algebra to Clifford algebra Cl(0,6). The resultant rich geometric structure can be leveraged to establish a combined framework of the standard model and gravity, wherein a gravi-weak interaction between the extended vierbein field and the extended weak gauge field is allowed. In conjunction with the composite Higgs model, we examine the vierbein field as a Cooper-pair-like fermion-antifermion condensation. Quantum gravity is realized indirectly via the quantized standard model spinor fields which underlie the composite space-time metric. We propose that the fundamental energy scales of the universe including the Planck scale are emergent and resulted from quantum condensations, thus possibly addressing the cosmological constant problem through an unconventional multi-scale renormalization procedure for multiplications of divergent Feynman integrals. The Clifford algebra approach also permits a weaker form  of charge conjugation without particle-antiparticle interchange, leading to a Majorana-type mass that conserves lepton number. Additionally, with reshuffling the traditional quark-lepton pairing pattern of three generations of fermions, we explore a three-Higgs-doublet model with Higgs VEVs 246 GeV, 42 GeV and 2.5 GeV which could explain the mass hierarchies of fermions. 
\end{abstract}


\tableofcontents
\markboth{Wei Lu}
{Clifford algebra  Cl(0,6) approach to BSM and naturalness problems}

\section{Introduction}
\label{sec:intro}
The mathematical imaginary number $i$ is ubiquitous in physics theories. In the case of quantum mechanics,  the imaginary number makes its appearance in the commutation relation of position operator $\hat X$ and momentum operator $\hat P$
\begin{align}
\label{eq:QM}
&[\hat X,  \hat P]= i\hbar,
\end{align}
where $\hbar$ is the Planck constant. Consequently, the {\it quantum} wave function is complex-valued. 

On the other hand, the imaginary number also shows up in the gauge transformation of a {\it classical} field
\begin{align}
\label{eq:phase}
&\psi \rightarrow e^{i\theta}\psi ,
\end{align}
which is essential in determining the electric charge property of $\psi$. 

We customarily treat the imaginary number $i$ in both examples as the same. It may come as a surprise that the imaginary number in the second case is different from the first one. The  imaginary number in gauge transformation~\eqref{eq:phase} is actually a unit pseudoscalar in disguise
\begin{align}
\label{eq:I4}
I &=  \gamma_0\gamma_{1}\gamma_{2}\gamma_{3},
\end{align}
where $\gamma_{a}$ are no other than the celebrated gamma operators discovered by Paul Dirac in 1928. The Dirac gamma operators satisfy the Clifford algebra $Cl(1,3)$ anticommutation relations
\begin{align}
\label{eq:gamma}
&\{ \gamma_{a}, \gamma_{b} \} = \gamma_{a}\gamma_{b}+\gamma_{b}\gamma_{a} = 2\eta_{ab},
\end{align}
where $\eta_{ab} = diag(1, -1, -1, -1)$.  

In view that $I^2 = -1$, pseudoscalar $I$ can be regarded as a surrogate for imaginary number $i$. As we will learn later in this paper, replacing imaginary number $i$ with pseudoscalar $I$ in gauge transformation~\eqref{eq:phase} such as $\psi \rightarrow \psi e^{I\theta}$ leads to a novel definition of charge conjugation without particle-antiparticle interchange. And for that matter, the original imaginary number $i$ shall be banished from the definition of classical fields such as classical spinor and Higgs fields. Instead, imaginary number $i$ should be reserved for its proper domain which is field quantization. 

Historically, the Dirac operators $\gamma_{a}$ are represented as gamma matrices. Due to the dichotomy between fermion states as columns and operators as matrices in the conventional formalism of quantum field theory (QFT), the aforementioned association of pseudoscalar with imaginary number would run into inconsistencies. This identification can only be achieved in an unconventional way by forgoing the traditional matrix and column representation and enlisting the aid of the Clifford algebra approach~\cite{HEST1,HEST2,PAV,Loun,DORA,Ablamowicz,BOUD,WB,Vaz,HEST3,Hestenes,CHIS1,CHIS2,Gull,TRAY,HEST4,Nesti2009,Pavsic,Pavsic2013,Pavsic2022,Castro,Gu,Trindade,Borstnik,Lasenby,Hitzer,WL1,WL2,WL3,WL4,WL5,WL6}, whereby both the algebraic spinor states and Dirac's gamma operators can be expressed in the same algebraic space. Note that we have already been using the Clifford algebraic non-matrix format, such as in Eq.~\eqref{eq:gamma}. 

\begin{subequations}\label{eq:gauge}
Clifford algebra, also known as geometric algebra or space-time algebra for the specific case of $Cl(1,3)$, is a potent mathematical tool that finds extensive applications in the physics arena. Remarkably, there is one more application of Clifford algebra $Cl(1,3)$ unbeknownst to Paul Dirac in 1928 which is the Clifford algebra $Cl(1,3)$ formalism of Lorentz gauge gravity. We know that gravity can be formulated as a Lorentz gauge theory~\cite{LORE,KIBB,SCIA,Hehl}  in terms of vierbein (a.k.a. tetrad, co-frame, or soldering form) and spin connection (a.k.a. Lorentz connection). According to the Clifford algebra $Cl(1,3)$ formalism of Lorentz gauge gravity~\cite{CF1, CF2,Rand}, the vierbein $\hat{e}_{\mu}$ and spin connection Lorentz gauge field $\hat{\omega}_{\mu}$  take values in the Clifford algebraic space
\begin{align}
\label{eq:gravityfields}
&\hat{e}_{\mu} = e_{\mu}^{a}\gamma_{a}, \\
&\hat{\omega}_{\mu} = \frac{1}{4}\omega_{\mu}^{ab}\gamma_{a}\gamma_{b},
\end{align}
\end{subequations}
where $ e_{\mu}^{a}$ and $\omega_{\mu}^{ab}$ are real-valued, $\omega_{\mu}^{ab} = -\omega_{\mu}^{ba}$,  and $a,b,\mu= 0, 1, 2, 3$. Throughout this paper the summation convention for repeated indices is adopted. The four distinct $\{\gamma_a\}$ and six distinct $\{\gamma_{a}\gamma_{b}; a<b\}$ are called vectors and bi-vectors of Clifford algebra. We denote vierbein as $\hat{e}_{\mu}$ and spin connection as $\hat{\omega}_{\mu}$ rather than $e_{\mu}$ and $\omega_{\mu}$ to accentuate the fact that they are Clifford-valued. 

The Lorentz gauge approach to gravity is also known as Einstein-Cartan gravity. The spin connection gauge field $\hat{\omega}_{\mu}$, associated with the local Lorentz group $SO(1, 3)$ (or $Spin(1, 3)$ when fermions are involved), plays the role of the gauge fields in Yang-Mills theory. In the gauge theory of gravity, the space-time metric $g_{\mu\nu}$ is defined as a composite field
\begin{equation}
\label{eq:metricG}
g_{\mu\nu} =\left\langle \hat{e}_{\mu}\hat{e}_{\nu}\right\rangle = e^{a}_{\mu}e^{b}_{\nu}\eta_{ab},
\end{equation}
where $\left\langle \ldots\right\rangle$ stands for the Clifford-scalar part of the enclosed expression. Thus vierbein $\hat{e}_{\mu}$ can be deemed as the ``square root'' of metric. 

Given that the gravity-related fields $\hat{e}_{\mu}$  and $\hat{\omega}_{\mu}$ are vector-valued and bi-vector-valued respectively in the Clifford algebraic space of $Cl(1,3)$, it's tempting to wonder whether the other interactions in nature such as the electroweak and strong gauge fields can take values in the Clifford algebraic space as well. The answer is a resounding yes, provided that one has to go beyond the confines of the familiar Clifford algebra $Cl(1,3)$. 

Learning from the above experience that we arrived at $Cl(1,3)$ via splitting the ``imaginary number'' $I$ into four  $\gamma_{a}$ operators in \eqref{eq:I4}, we may go one step further by decomposing Dirac's gamma zero operator $\gamma_0$ into its underlying components~\cite{WL1} 
\begin{align}
\gamma_0\ &=  \Gamma_{1}\Gamma_{2}\Gamma_{3},
\end{align}
where the additional trio of gamma operators \{$\Gamma_{1}$, $\Gamma_{2}$, $\Gamma_{3}$\} satisfy the anticommutation relations
\begin{align}
\label{eq:Gamma}
&\{ \Gamma_{i}, \Gamma_{j} \} = \Gamma_{i}\Gamma_{j}+\Gamma_{j}\Gamma_{i} = -2\delta_{ij},
\end{align}
and anticommute with Dirac's original trio \{$\gamma_{1}$, $\gamma_{2}$, $\gamma_{3}$\}
\begin{align}
\label{eq:mix}
&\{ \gamma_{i}, \Gamma_{j} \} =  \gamma_{i}\Gamma_{j}+\Gamma_{j}\gamma_{i}= 0.
\end{align}
Collectively, these six elements 
\begin{align}
\label{eq:gamma2}
\Gamma_{1}, \Gamma_{2}, \Gamma_{3}, \gamma_{1}, \gamma_{2}, \gamma_{3},
\end{align}
constitute the orthonormal vector basis of the real Clifford algebra $Cl(0,6)$, which is sometimes labeled as $Cl_{0,6}$ or $Cl_{0,6}(R)$ in the literature. Note that while the Clifford algebra has been extended, in our model we assume that the underlying space-time manifold remains 4-dimensional. 

Thanks to the recognition of $\gamma_0$ as a composite tri-vector, we are able to extend Dirac's Clifford algebra from $Cl(1,3)$ to $Cl(0,6)$, with $Cl(1,3)$ being a sub-algebra of $Cl(0,6)$. With it, we can define an algebraic spinor as a linear combination of all $2^{6}=64$ basis elements of $Cl(0,6)$. Considering that there are 16 Weyl fermions with $16\times2=32$ complex components (i.e. $64$ real components) within each of the three fermion families including the right-handed neutrino, an algebraic spinor of the {\it real} $Cl(0,6)$ with 64 {\it real} degrees of freedom is a perfect match for representing one generation of fermions. 

The geometrical wealth of Clifford algebra $Cl(0,6)$ can be exploited to establish a unified theory of gravity and the standard model~\cite{WL1,WL3,WL4} based on two central tenets: gauge symmetry and quantum condensation. The tenet of gauge symmetry stipulates  that both Lorentz and internal symmetries should be treated on an equal footing as gauge symmetries, while the   tenet of quantum condensation dictates that the fundamental energy scales of the universe such as the Planck scale are emergent and resulted from multi-spinor quantum condensation which is an ordered quantum phase induced by the dynamical symmetry breaking mechanism. 

According to the central tenet of gauge symmetry,  the spin connection Lorentz gauge field governing gravity and the Yang-Mills gauge fields governing the standard model interactions are associated with beyond the standard model (BSM) symmetries 
\begin{align}
 & Spin(1,3)_{L} \times Spin(1,3)_{R} \times Spin(1,3)_{WL} \times Spin(1,1)_{WR} \times U(1)_{WR} \nonumber \\
 &\times SU(3)_{C}\times U(1)_{B-L}, \label{eq:symmetry}
\end{align} 
where $Spin(1,3)_{L}$ and $Spin(1,3)_{R}$ are the left- and right-handed local Lorentz gauge groups, and the extended left-handed weak gauge group $Spin(1,3)_{WL}$ encompasses the standard model left-handed weak gauge group $SU(2)_{WL}$.  

The BSM symmetries \eqref{eq:symmetry} are accompanied by an extended vierbein valued in a 16-dimensional multivector subspace of $Cl(0,6)$ as opposed to the traditional vierbein \eqref{eq:gauge} valued in the 4-dimensional vector space of $Cl(1,3)$. The extended vierbein {\it transforms as a vector} under both Lorentz and the extended weak  $Spin(1,3)_{WL}$ gauge transformations. Therefore, a new kind of gravi-weak interplay is permitted between the extended vierbein and the extended weak gauge field. This gravi-weak interplay is otherwise impossible  since the traditional vierbein is {\it invariant} under the weak  $SU(2)_{WL}$ gauge transformation. Note that the gravi-weak interplay is not in conflict with the Coleman-Mandula theorem~\cite{Coleman}, since there is no nontrivial mixing between the Lorentz groups ($Spin(1,3)_{L}$ and $Spin(1,3)_{R}$) and the internal gauge groups including the extended weak group $Spin(1,3)_{WL}$. 

According to the second central tenet of quantum condensation, we investigate the vierbein field $\hat{e}_{\mu}$  as a composite entity emerging from a Cooper-pair-like fermion-antifermion quantum condensation via the dynamical symmetry breaking mechanism. Hence the standard model fermion fields are the origin of space-time metric. Given that vierbein can be viewed as the ``square root'' of metric, the standard model fermions can be considered as the ``quarter root'' of metric, which speaks to the fact that fermions have spin $1/2$ while gravitons have spin $2$.  Consequently, quantum gravity is realized indirectly via the quantized standard model spinor fields which underlie the composite space-time metric. Note that due to the chirality of the vierbein-related condensations and the chirality of spin connection fields associated with  left- and right-handed local Lorentz gauge groups $Spin(1,3)_{L}$ and $Spin(1,3)_{R}$, there are left- and right-handed gravitational interactions. 

We propose that all the regular Lagrangian terms, be it the fermion kinetic term or gravity and cosmological constant terms, are of quantum condensation origin. Therefore, both the Planck scale and the cosmological constant scale are emergent and resulted from quantum condensations. We advocate a multi-scale  renormalization procedure for quantum condensation, which is a paradigm shift from the conventional QFT renormalization approach (or the Wilsonian renormalization group approach) characterized by a single renormalization scale.  The cosmological constant problem can thus be evaded if we exercise extreme caution in the renormalization procedure that entails { \it multiplications} of divergent Feynman integrals. 

It's worth mentioning that Clifford algebra $Cl(0,6)$ is capable of accommodating enveloping groups larger than the BSM groups \eqref{eq:symmetry}. For example, the spin group $Spin(4,4)$  and the real symplectic group $Sp(8, R)$ are embedded in the $Cl(0,6)$ geometric structure. The Lorentz gauge group $Spin(1,3)$ and the extended weak group $Spin(1,3)_{WL}$ are two commuting subgroups of $Spin(4,4)$. The Pati-Salam's $SU(4)$~\cite{SU4}, which is a subgroup of the real symplectic group $Sp(8, R)$,  is isomorphic to the six-dimensional rotation group $Spin(6)$. We settle for a parsimonious set of subgroups~\eqref{eq:symmetry} due to chirality considerations and lack of experimental evidence (such as proton decay) supporting any larger unification groups such as $SU(4)$, whereas there are various clues suggestive of the BSM symmetries~\eqref{eq:symmetry}. 

According to the conventional wisdom, the usual candidates for gauge transformations are bi-vector-related rotations. Thus the symmetry group of $Cl(0,6)$ seems to be restricted to $Spin(6)$/$SO(6)$ generated by all the 15 bi-vectors of $Cl(0,6)$. Then how can $Cl(0,6)$ accommodate the BSM symmetries discussed above? This is because of two advantages of the Clifford algebra approach that are not enjoyed by the conventional approach. First of all, the permissible group generators of the Clifford algebra approach could involve both bi-vector and non-bi-vector Clifford elements, generalizing the traditional notion of bi-vector rotations. For example, the Lorentz boosts and the left-handed weak-boosts are generated by $Cl(0,6)$ 4-vectors rather than bi-vectors. 

Secondly, a Clifford algebraic spinor $\psi$ allows double-sided gauge transformations 
\begin{equation}
\psi\quad\rightarrow\quad V\psi U,
\end{equation}
where transformations $V$ and $U$ are {\it independent} of each other. And by definition,  $V$ and $U$ commute with each other.  Specifically, the gravi-weak symmetries in~\eqref{eq:symmetry} (the Lorentz gravity $Spin(1,3)_{L} \times Spin(1,3)_{R}$ and the extended weak symmetries $Spin(1,3)_{WL} \times Spin(1,1)_{WR} \times U(1)_{WR}$)  belong to the left-sided gauge transformation $V$, while the color and B-L symmetries in~\eqref{eq:symmetry} ($SU(3)_{C}\times U(1)_{B-L}$) belong to the  right-sided gauge transformation $U$. These double-sided gauge transformations are otherwise impossible in the conventional column fermion formalism. Note that  the left-handed and right-handed fermions transform independently under the chiral left-sided gauge transformations $Spin(1,3)_{L} \times Spin(1,3)_{WL}$ and $Spin(1,3)_{R}\times Spin(1,1)_{WR} \times U(1)_{WR}$ respectively, whereas the left-handed and right-handed fermions transform in unison under the right-sided gauge transformations  $SU(3)_{C}\times U(1)_{B-L}$.   

All in all, the above two features of the Clifford algebra approach are the underlying mechanism which enables us to explore symmetries that go beyond the conventional approach. 

In this paper, we present a detailed account of how BSM symmetries~\eqref{eq:symmetry} are broken by a cascade of symmetry breaking processes, triggered by the nonzero vacuum expectation values (VEV) acquired by the Clifford-valued vierbeins and Higgs fields. The first stage of symmetry breaking starts with the vierbeins acquiring nonzero flat space-time VEVs, or equivalently with the metric $g_{\mu\nu}$ acquiring the Minkowski VEV $\eta_{\mu\nu}$. As a result, the pseudo-weak gauge symmetry (the coset $Spin(1,3)_{WL}/SU(2)_{WL}$ and $Spin(1,1)_{WR}$) , local Lorentz gauge symmetry and diffeomorphism symmetry are lost. The residual gauge symmetries are $SU(2)_{WL} \times U(1)_{WR} \times SU(3)_{C} \times U(1)_{B-L}$, plus a remnant {\it global} Lorentz symmetry which is the  synchronization of the global portion of Lorentz gauge and diffeomorphism symmetries. 

The next step of symmetry breaking is triggered by the Majorana-Higgs field, which is a Higgs-like field in addition to the standard model Higgs field.  At this stage, the Majorana-Higgs field assumes a nonzero VEV and breaks the local gauge symmetries down to the standard model symmetries $SU(3)_{C}  \times SU(2)_{WL} \times U(1)_Y$. Consequently, the neutrino is endowed with a lepton number-conserving Majorana mass which is much heavier than the Dirac mass. A very small effective mass can thus be derived for the neutrino via the seesaw mechanism~\cite{NEUT}.

At the last stage of symmetry breaking, the standard model electroweak Higgs field acquires nonzero VEV and breaks the standard model symmetries down to $SU(3)_{C} \times U(1)_{EM}$, where $U(1)_{EM}$ is the electromagnetic gauge symmetry. In the context of spontaneous breaking of Peccei-Quinn-like symmetries, the fermion Dirac mass hierarchies can be explained by a three-Higgs-doublet model with the help from reorganizing the lepton-quark configurations of the three generations of standard model fermions. For instance, the first generation electron and neutrino are paired up with the third generation top and bottom quarks (rather than the first generation up and down quarks) to form one unique fermion generation in our model, which is very different from the traditional fermion family assignments.

One point we want to highlight is that  all the ingredients of our model, such as fermions, gauge fields, vierbeins, and Higgs fields, share the same Clifford algebraic space of $Cl(0,6)$. For instance, the electromagnetic gauge field $\hat{A}_\mu$ is pseudoscalar-valued 
\begin{align}
\label{eq:EM}
\hat{A}_{\mu} &=  q A_{\mu} I,
\end{align}
where $A_{\mu}$ is real-valued, and we include charge $q$ (such as $q=-1$ for electron) in the definition of $\hat{A}_\mu$. The pseudoscalar $I$ is the 6-vector of $Cl(0,6)$ which is the same $Cl(1,3)$ pseudoscalar $I$ in Eq.~\eqref{eq:I4} after expanding the tri-vector $\gamma_0$ into its constituents
\begin{align}
\label{eq:I}
I &=  \gamma_{0}\gamma_1\gamma_2\gamma_3 = \Gamma_{1}\Gamma_{2}\Gamma_{3}\gamma_1\gamma_2\gamma_3.
\end{align}

The electromagnetic gauge field $\hat{A}_\mu$ as shown above and the gravity-related fields $\hat{\omega}_{\mu}$/$\hat{e}_{\mu}$ in Eq.~\eqref{eq:gauge} take values in various Clifford algebraic subspaces with the same six gamma operators $\{ \Gamma_{1}, \Gamma_{2}, \Gamma_{3}, \gamma_{1}, \gamma_{2}, \gamma_{3}\}$ as the unifying building blocks. The Clifford algebraic gauge field of gravity and Yang-Mills fields are connected with each other through their interactions with the common fermion fields, vierbein fields and Higgs fields, all valued in Clifford algebraic space.  

Therefore, the model outlined in this paper is indeed a cohesive unified theory of the standard model and gravity. Note that the Clifford algebra approach differs from the conventional grand unified theories (GUTs)~\cite{GeorgiGlashow,Georgi,Baez,Boyle,Chester} and gravi-GUTs~\cite{Percacci,NestiPercacci,Chamseddine,KrasnovPercacci,Krasnov1,Krasnov2,MaiezzaNesti,Chkareuli,Konitopoulos} which demand that the gauge coupling constants should be unified. In the Clifford algebra approach, the allowable symmetries are the ones that preserve the invariance properties of spinor bilinears. Hence the symmetries of the model are in a sense derived rather than postulated. The permitted symmetries usually involve a direct product of different groups, suggesting that the individual gauge coupling constants are not necessarily related to each other. To a certain extent, the Clifford algebra approach as advocated here predicts the unpredictability of Weinberg angle $\theta_W$.  As such, the fusion delineated in this paper is more of a unification via spinors and less of a unification via symmetry groups. 

This paper is structured as follows: In Section~\ref{sec:Clifford}, we introduce the algebraic spinors of $Cl(0,6)$ and explore beyond the standard model gauge symmetries. In Section~\ref{sec:SSB}, we investigate spontaneous symmetry breaking due to the non-degenerate vacuum expectation values of various bosonic fields, and study the fermion mass hierarchies and the lepton number-conserving Majorana mass. In Section~\ref{sec:quantization}, we contemplate naturalness problems through the lens of quantum condensations and examine the extended gauge symmetries which enable gravi-weak interaction. In the last section we draw our conclusions. Throughout this paper, we adopt the natural units $c = \hbar = 1$. 

\section{Clifford Algebra Cl(0,6) and Gauge Symmetries}
\label{sec:Clifford}
In this section, we introduce algebraic spinors of the real Clifford algebra $Cl(0,6)$ and explore beyond the standard model gauge symmetries. We proceed with expressing the gauge fields and curvatures as multivectors of $Cl(0,6)$. We round out the section with a detailed formulation of the Lagrangian of the algebraic spinors, as well as the Lagrangian of the Yang-Mills and gravitational interactions in terms of Clifford algebra. 

Note that the {\it complexified} Clifford algebra $Cl(6,C)$~\cite{Casalbuoni,Zenczykowski,Furey1,Furey2,Furey3,Furey2023a,Furey2023b,Furey2023c,Stoica,Gillard,Gresnigt2020,Singh,Todorov,Gresnigt,Reynoso} and its isomorphic equivalences such as $Cl(7,0)$~\cite{TRAY} and the complex $8\times8$ matrix $M(8,C)$ algebra~\cite{Gording},  either as standalone algebra or in association with octonions or sedenions, have been investigated in connection with the color $SU(3)_c$ symmetry and the other standard model symmetries. Nevertheless, in our approach~\cite{WL1,WL3,WL4} we choose to stick with the {\it real} $Cl(0,6)$  which is based on two reasons: The first reason is that an algebraic spinor of the {\it real} $Cl(0,6)$ with 64 {\it real} basis elements is a perfect match with one generation of 16 Weyl fermions (including the right-handed neutrino) in the standard model with 32 complex degrees of freedom (i.e. $64$ real degrees of freedom).

The second reason for choosing the {\it real} $Cl(0,6)$ is that we regard the spinor and gauge fields as classical fields {\it prior to field quantization}. As explained in the introduction section, we consign the imaginary number $i$ to the realm of quantum theory. Therefore, the imaginary number $i$ should play no role in the definition of the classical spinor fields, gauge fields and their symmetry transformations, as demonstrated by the replacement of imaginary number $i$ with the pseudoscalar $I$  in the  electromagnetic $U(1)$ gauge transformation~\eqref{eq:phase}. In our model, we rigorously enforce the rule of avoiding the quantum $i$ in the definition of {\it classical} spinors, gauge fields, and the corresponding gauge-covariant derivatives. Note that we can nonetheless maintain a ``{\it pseudo-complex representation}'' with the pseudoscalar $I$ acting as the surrogate imaginary number.  As a side note, in our earlier papers~\cite{WL1,WL3,WL4} we have used notations for imaginary number (denoted as $\hat{i}$) and pseudoscalar (denoted as $i$)  which are different from notations in the current paper. 

Pioneered by Hestenes~\cite{HEST1}, one original objective of the Clifford algebra approach to physics is to abandon the imaginary number {\it altogether} and replace it with certain element of Clifford algebra. The endeavor of using an effective operator to proxy the imaginary number has been fairly successful in a wide variety of physics domains~\cite{Hestenes,Gull,VolovikZubkov,Pavsic2022}. However, it has been realized~\cite{WL4,WL5} that the initiative championed by Hestenes has its limitations: When it comes to field quantization and quantum loop integral calculations, the imaginary number is indispensable and can not be replaced by a Clifford element. For example, a self-energy loop diagram would yield an imaginary-valued quantum correction due to proper contour integral on the complex plane of Feynman propagators. Therefore,  in Section~\ref{sec:quantization} on field quantization, we will formally introduce the imaginary number $i$  in the Clifford functional integral formalism~\cite{WL5} when we study quantum phenomena. In the same Section~\ref{sec:quantization}, we will try to demystify the unexpected appearance of the imaginary number $i$ in the classical fermion Lagrangian as the quantum fingerprint left on the classical world. 

It is worth noting that the real 6-D Clifford algebra has 7 different signatures: Clifford algebras $Cl(0,6)$,  $Cl(3,3)$ and $Cl(4,2)$ are isomorphic to the real $8\times8$ matrix $M(8,R)$ algebra. On the other hand, Clifford algebras $Cl(6,0)$, $Cl(5,1)$, $Cl(2,4)$ and $Cl(1,5)$ are isomorphic to the quaternion $4\times4$ matrix $M(4,H)$ algebra. In the literature~\cite{Wilson}, the merits of various signatures of Clifford algebras have been discussed in connection with the standard model symmetries. 

Given the isomorphisms noted above, the $Cl(0,6)$-based model can be faithfully translated to the $Cl(3,3)$, $Cl(4,2)$ or $M(8,R)$ format, which would be basically the equivalent model. For example, in terms of the $Cl(0,6)$ orthonormal vector basis $\{ \Gamma_{1}, \Gamma_{2}, \Gamma_{3}, \gamma_{1}, \gamma_{2}, \gamma_{3}\}$, the orthonormal vector bases of $Cl(3,3)$ and $Cl(4,2)$ are $\{  \Gamma_{0}\Gamma_{1}, \Gamma_{0}\Gamma_{2}, \Gamma_{0}\Gamma_{3}, \gamma_{1}, \gamma_{2}, \gamma_{3}\}$ and $\{ \gamma_{1}\gamma_{0}, \gamma_{2}\gamma_{0}, \gamma_{3}\gamma_{0}, \Gamma_{0}, \Gamma_{1}, \Gamma_{2}\}$ respectively, where $\gamma_{0}=\Gamma_{1}\Gamma_{2}\Gamma_{3}$ and $\Gamma_{0}=\gamma_{1}\gamma_{2}\gamma_{3}$. As such, the time-like $\gamma_{0}$ can be defined as the 6-vector (pseudoscalar) of $Cl(3,3)$, whereas $\gamma_{0}$ can be defined as the 4-vector (product of the first 4 vectors) of $Cl(4,2)$.  The mapping between $Cl(0,6)$ and $M(8,R)$ can also be worked out, which will not be explored here in this paper since the physics essence of our model has already been elegantly captured by the purely Clifford algebraic formalism.  

\subsection{The algebraic spinor representation of fermions}
\label{subsec:Spinor}
As mentioned in the introduction section, one generation of the standard model fermions can be represented by the algebraic spinor of Clifford algebra $Cl(0, 6)$~\cite{WL1,WL3,WL4}. The goal of this subsection is to demonstrate how individual fermions, such as electrons, neutrinos, and quarks, are linked to the algebraic spinor without resorting to the traditional column representation of fermions. Note that this subsection concerns the classical spinor field, while quantization of spinor field will be  introduced  in Section~\ref{sec:quantization}. Following our principle stated earlier, in this subsection we strictly enforce the regime of not allowing the imaginary number $i$ in the definition of the classical algebraic spinor and the projections thereof. Therefore, we are ensured to stay within the real $Cl(0,6)$ algebraic space. 

For $Cl(0, 6)$, there are ${\binom{6 }{k}} $ independent $k $-vectors. To wit, there are one single scalar $1$ which is a $0$-vector, 6 vectors  (e.g. $\gamma_1$), 15 bi-vectors (e.g. $\gamma_1\gamma_2$), 20 tri-vectors (e.g. $\gamma_0 = \Gamma_1\Gamma_2\Gamma_3$), 15 4-vectors (e.g. $\gamma_1\gamma_2 I=-\Gamma_1\Gamma_2\Gamma_3\gamma_3$), 6 5-vectors (e.g. $\gamma_1 I=\Gamma_1\Gamma_2\Gamma_3\gamma_2\gamma_3$), and finally one single pseudoscalar $I=\Gamma_1\Gamma_2\Gamma_3\gamma_1\gamma_2\gamma_3$ which is a 6-vector. In total, there are $2^{6}=64$ independent basis elements given by the set of all $k $-vectors. The algebraic spinor $\psi$ is a multivector which can be expressed as a linear combination of all the $64$ basis elements
\begin{align}
\label{eq:psi}
\psi =&\psi_1 + \psi_2\Gamma_1 + \psi_3\Gamma_2 + \cdots + \psi_{62}\Gamma_2 I + \psi_{63}\Gamma_1 I + \psi_{64} I,
\end{align}
where the 64 linear combination coefficients $\{ \psi_n; n = 1, 2, \cdots, 64\}$ are {\it real} Grassmann numbers (with $\psi_n\psi_m+\psi_m\psi_n = 0$ and $\psi_n$ commuting with all 64 Clifford algebra Cl(0,6) basis elements)  satisfying the complex conjugation relation
\begin{align}
\label{eq:real}
\psi_n^* =&\psi_n.
\end{align}

As we noted in the introduction section, while the Clifford algebra has been extended to $Cl(0,6)$, the underlying space-time manifold remains 4-dimensional. Therefore, if we write out the coordinate $x$ explicitly for the algebraic spinor $\psi(x)$,  $x$ still represents the 4-dimensional space-time. Also note that the fermionic algebraic spinor as expressed in~\eqref{eq:psi} should not be confused with a bispinor, which is effectively bosonic (Grassmann-even) and can be also expanded in terms of the 64 elements of $Cl(0,6)$. While the bosonic bispinor components may transform as Lorentz scalar/pseudoscalar/vector/tensor, it should be emphasized that the fermionic Grassmann-odd algebraic spinor $\psi$ transforms as a spinor only, albeit $\psi$ comprises all the $Cl(0,6)$ basis elements. We will revisit the important subject of bosonic bispinors in later sections when we investigate scalar/tensor Higgs fields and (extended) vector vierbein fields as effective descriptions of bispinors. 

A few comments are in order at this point regarding the real Grassmann numbers $\psi_n$. First of all, due to the fermion nature of the algebraic spinor $\psi$, it's mandatory that $\psi_n$ should be Grassmann-odd. This requirement is not obvious when we write down the Dirac equation for $\psi$, where there is no multiplication between spinors.  In other words, defining $\psi_n$ as real number or as real Grassmann number does not make any difference as long as there is no multiplication between spinors. However, the Grassmann-odd characteristic of $\psi$ becomes essential when it comes to the Lagrangian involving multiplication between spinors. Since we are going to use the Lagrangian format to investigate multi-fermion condensations extensively in our model, we are compelled to employ the real Grassmann $\psi_n$. As we will learn later in this paper, the Majorana mass Lagrangian term is allowed only if the algebraic spinor is Grassmann-odd, or otherwise the Majorana mass Lagrangian term is identically zero. 

Secondly, it's customary to adopt {\it complex} Grassmann numbers in the conventional QFT where the reality condition~\eqref{eq:real} does not hold. The way to make contact with the conventional complex Grassmann numbers is to reorganize $\psi_n$ in pairs, such as
\begin{align}
\label{eq:complex}
\psi_1 + \psi_{64} I,
\end{align}
where pseudoscalar $I$ is acting as a proxy for the imaginary number $i$. Therefore, the combination $\psi_1 + \psi_{64} I$ is tantamount to the conventional complex Grassmann number. Another example is the $\psi_2\Gamma_1   + \psi_{63}\Gamma_1 I$ pair, which can be rewritten as $\Gamma_1(\psi_2    + \psi_{63} I)$. Hence $\psi_2  + \psi_{63} I$ is the proxy  ``complex'' Grassmann number. Note that pseudoscalar $I$ has to appear on the right side of $\Gamma_1$. We will come back to this point a bit later. 

The algebraic spinor $\psi$ of the real $Cl(0,6)$ with $2^{6}=64$ real components corresponds to the union of all 16 Weyl fermions in one fermion generation of the standard model (plus right-handed neutrino) endowed with $16\times2=32$ complex components (i.e. $64$ real components). For most part of this paper, our discussion is restricted to one generation/family of fermions. The Clifford algebraic structure of the fermions as well as the gauge symmetries are the same for the three generations. In other words, the three families are three replicas.  That said, in Section~\ref{subsec:Hierarchy} on the fermion mass hierarchy, we will propose three electroweak symmetry-breaking Higgs fields that couple to the three generations of fermions in different patterns. 

How do we connect one generation of fermions, such as electrons, neutrinos, and quarks with $\psi$? First of all, let's distinguish between the left-handed and right-handed fermions in the setting of Clifford algebra. We propose that fermions with left (right) chirality correspond to Clifford-odd (even) portion of the algebraic spinor~\cite{WL1}
\begin{align}
\psi_{L} &= \frac{1}{2}(\psi+ I\psi I), \label{eq:chiralL}\\
\psi_{R} &= \frac{1}{2}(\psi - I\psi I). \label{eq:chiralR}
\end{align}
That is to say, the left-handed $\psi_{L}$ is composed of 6 vectors,  20 tri-vectors, and 6 5-vectors (32 degrees of freedom in total)
\begin{align}
\psi_{L} &= \psi_2\Gamma_1  + \cdots  + \psi_{63}\Gamma_1 I
\end{align}
whilst the right-handed $\psi_{R}$ is composed of one scalar, 15 bi-vectors, 15 4-vectors, and one pseudoscalar (32 degrees of freedom in total)
\begin{align}
\psi_{R} &= \psi_1 + \cdots  + \psi_{64} I,
\end{align}
The projection of Clifford-odd (even) portion of $\psi$ leverages the property that the pseudoscalar $I$ anticommutes with Clifford-odd elements, and commutes with Clifford-even
elements
\begin{align*}
I\psi_{L} &= -\psi_{L}I, \\
I\psi_{R} &= \psi_{R}I.
\end{align*}
The operation of $-I\psi I$ as in Eq.~\eqref{eq:chiralR} can be mapped to the traditional $\gamma^5\psi$ operation, where $-I$ as in $-I\psi$ plays the role of the conventional pseudoscalar (more discussion will be provided on the minus sign of $-I$ in Section~\ref{subsec:Lorentz}), while $I$ as in $\psi I$ plays the role of the conventional imaginary number $i$.  Note that the positioning of $I$ relative to the algebraic spinor $\psi$ does matter. This speaks to the fact that the electromagnetic field, as a pseudoscalar in Eq.~\eqref{eq:EM}, should always be  applied to the right side of a spinor so that $I$ is equivalent to the imaginary number $i$ in the traditional setting. And for that matter, when we attempt to connect the real Grassmann numbers in our model with the traditional complex Grassmann numbers, we ought to make sure that the surrogate ``imaginary number'' $I$ should always appear on the right side of any Clifford element. 

We would like to emphasize that being Grassmann-odd/even and being Clifford-odd/even are two separate notions. Fermionic spinor fields are Grassmann-valued, while bosonic fields including gauge and Higgs fields are real-valued. On the other hand, being fermion or boson has no bearing on being Clifford-odd/even. For instance, according to the above spinor chirality definition, the left-handed spinor $\psi_{L}$ is Clifford-odd, while the right-handed spinor $\psi_{R}$ is Clifford-even. And we will learn later that the bosonic fields can also be either Clifford-odd or Clifford-even. For example, the bosonic Majorana-Higgs field and the bosonic vierbein field are both Clifford-odd, while the other bosonic fields such as the standard model Higgs field and all the gauge fields are Clifford-even.

\begin{subequations}\label{eq:color}
Now we are ready to identify $\psi$ with electrons, neutrinos, and quarks. Specifically, the projection operators for the three colors of red, green, and blue quarks are given by 
\begin{align}
P_{r} &= \frac{1}{4}(1+I\gamma_1\Gamma_1-I\gamma_2\Gamma_2-I\gamma_3\Gamma_3), \label{IDEM3}\\
P_{g} &= \frac{1}{4}(1-I\gamma_1\Gamma_1+I\gamma_2\Gamma_2-I\gamma_3\Gamma_3), \label{IDEM4}\\
P_{b} &= \frac{1}{4}(1-I\gamma_1\Gamma_1-I\gamma_2\Gamma_2+I\gamma_3\Gamma_3), \label{IDEM5}
\end{align}
while the lepton projection operator is defined as
\begin{align}
P_{l} &= \frac{1}{4}(1+I\gamma_1\Gamma_1+I\gamma_2\Gamma_2+I\gamma_3\Gamma_3). \label{IDEM2}
\end{align} 
\end{subequations}
In the context of $SU(4)$, the lepton projection operator $P_{l}$ can be regarded as the projection to the fourth color~\cite{SU4}. Leveraging the fact that $I$ can be factorized into the 6 $Cl(0,6)$ vectors~\eqref{eq:I}, it can be readily verified that the four color projections $P_{l}$, $P_{r}$, $P_{g}$, $P_{b}$ are orthogonal to each other and satisfy
\begin{align}
&P_{l} + P_{r}+P_{g}+P_{b}= P_{l} + P_{q} = 1, \label{IDEM7}
\end{align} 
where $P_{q} = P_{r} + P_{g} + P_{b}$ is the quark projection operator. Note that the bi-vectors $\gamma_i\Gamma_i$ appearing in the color projectors suggest an interesting interplay between the trialities  of  \{$\gamma_{1}$, $\gamma_{2}$, $\gamma_{3}$\}/\{$\Gamma_{1}$, $\Gamma_{2}$, $\Gamma_{3}$\} and three colors of quarks. Figuratively speaking, the three colors of quarks go hand in hand with the three space dimensions. In Section~\ref{subsec:Gauge}, we will further explore the significance of these color projectors in terms of carving out the color group $SU(3)_{C}$ from Pati-Salam's $SU(4)$. It's worth mentioning that the color projectors can be equivalently expressed via products of idempotents $ \frac{1}{2}(1 \pm \gamma_0\gamma_i\Gamma_i)$~\cite{WL1}, which will not be detailed here. 

For the purpose of differentiating between weak isospin up-type and down-type fermions, we introduce another set of orthogonal projection operators 
\begin{align}
&P_{\pm} = \frac{1}{2}(1\pm I{\Gamma_1}{\Gamma_2}), \label{IDEM6}
\end{align}
which sum up to
\begin{align}
&P_+ + P_-= 1. \label{IDEM8}
\end{align} 

We identify projections of the algebraic spinor $\psi$
\begin{equation}
\psi = \psi_L + \psi_R = (P_+ + P_-)(\psi_L + \psi_R)(P_{l} + P_{r} + P_{g} + P_{b}),
\end{equation}
with left-handed neutrino, electron, and quarks
\begin{equation}
\label{eq:left}
\left\{
\begin{array}{rl}
\nu_L &= P_+ \psi_L P_{l}, \\
u_{L, r} &= P_+ \psi_L P_{r}, \quad u_{L, g} = P_+ \psi_L P_{g},  \quad u_{L, b} = P_+ \psi_L P_{b},  \\
e_L &= P_- \psi_L P_{l}, \\
d_{L, r} &= P_- \psi_L P_{r}, \quad d_{L, g} = P_- \psi_L P_{g},  \quad d_{L, b} = P_- \psi_L P_{b},
\end{array}\right.
\end{equation}
and right-handed neutrino, electron, and quarks
\begin{equation}
\label{eq:right}
\left\{
\begin{array}{rl}
\nu_R &= P_- \psi_R P_{l}, \\
u_{R, r} &= P_- \psi_R P_{r}, \;\;\; u_{R, g} = P_- \psi_R P_{g}, \;\;\; u_{R, b} = P_- \psi_R P_{b},  \\
e_R &= P_+ \psi_R P_{l}, \\
d_{R, r} &= P_+ \psi_R P_{r}, \;\;\; d_{R, g} = P_+ \psi_R P_{g}, \;\;\; d_{R, b} = P_+ \psi_R P_{b}.\end{array}
\right.
\end{equation}

Since the algebraic spinor $\psi$ of $Cl(0,6)$ has $2^{6}=64$ real components, each of the above 16 distinctive projections possesses 4 real degrees of freedom, matching with the 16 standard model Weyl fermions with each having $2$ complex components (i.e. $4$ real degrees of freedom). 

Note that the definition of isospin $I_{3}$ for a given standard model fermion $\psi_f$ is given by
\begin{align}
&\psi_f (I_{3} I) = \frac{1}{2}\Gamma_1\Gamma_2\psi_f. \label{Iso}
\end{align}
With the help from the property that 
\begin{align}
&\Gamma_1\Gamma_2P_{\pm}\psi_f = {\mp} I P_{\pm}\psi_f, \label{Iso2}
\end{align}
it can thus be verified that the isospin values of the fermions in Eq.~\eqref{eq:left} and Eq.~\eqref{eq:right} are consistent with those of the standard model. When the $I$ in Eq.~\eqref{Iso2} appearing on the left side of $\psi_f$ is moved to the right side of $\psi_f$ as in Eq.~\eqref{Iso}, its sign is changed for the Clifford-odd left-handed fermions. This is the underlying reason why there is a flip of sign in $P_{\pm}$ between the left- and right-handed fermions when $P_{\pm}$ is assigned to the isospin up-type and down-type fermions respectively in Eq.~\eqref{eq:left} and Eq.~\eqref{eq:right}.

It's also worth mentioning that attempts have been made to associate species of fermions with minimal left ideals of the Clifford algebraic spinor where projection operators are restricted to acting on the right side of an algebraic spinor. Obviously, our fermion assignment scheme above departs from the minimal left ideal approach, given that the projection operators used to  identify the standard model fermions are applied on both the right side and the left side of the algebraic spinor in Eq.~\eqref{eq:left} and Eq.~\eqref{eq:right}. 

In summary, we have identified individual fermions with the projections of the $Cl(0,6)$ algebraic spinor without any reference to the column representation. And for that matter, there will not be a single expression of matrix or column in our model. In other words, every expression discussed in our model is purely algebra-based. Nevertheless, the mappings between the $Cl(0, 6)$ formalism and the conventional matrix/column representation can be worked out~\cite{WL1}, which will not be detailed in this paper. 

\subsection{Beyond the standard model symmetries}
\label{subsec:Gauge}
As explained in the introduction section, we strive to walk a careful line between being too ambitious and being too conservative when it comes to choosing the suitable gauge symmetries for our model. The aim of this subsection is to give an account of our thought process in selecting the BSM symmetry groups employed in this paper. The in-depth investigation of the  Clifford-valued gauge fields corresponding to these symmetries will be presented in Section \ref{subsec:Lagrangian}, while the details of the spontaneous symmetry breaking process  will be examined   in Section~\ref{sec:SSB}.

The conventional way of model building is to postulate the symmetry group upfront, and then proceed to find the fermion representation. In the case of Clifford algebra approach, it's the other way around. We choose the Clifford algebra $Cl(0,6)$ as the first step which is tantamount to staking out the fermion space. The allowable symmetry groups are thus tightly constrained by the spinor space. This is a desirable feature of the Clifford algebra approach, since the symmetry groups are in a sense derived, rather than postulated. 

So how do we determine the allowable symmetries? It hinges on the invariance property of spinor bilinear
\begin{align}
\left\langle \bar{\psi} \psi\right\rangle_{1, I}, \label{inner}
\end{align}
where $\left\langle \ldots\right\rangle_{1, I}$ stands for the Clifford-scalar and -pseudoscalar parts of the enclosed expression. It is the Clifford algebraic counterpart of the conventional Dirac inner product $\bar{\psi}\psi$ between the column spinors. For the rest of the paper, we will exclusively use $\left\langle \ldots\right\rangle$ which stands for the Clifford-scalar part of the enclosed expression, since we can get the pseudoscalar part via $\left\langle I \ldots \right\rangle$ if needed. The Dirac conjugate $\bar{\psi}$ in Eq.~\eqref{inner} is defined as
\begin{equation}
\bar{\psi} = \psi^{\dagger}\gamma_0,
\end{equation}
and the Hermitian conjugate satisfies
\begin{align}
\label{eq:hermitian}
(AB)^{\dagger} &= B^{\dagger}A^{\dagger}, 
\end{align}
for any $A$ and $B$ valued in the Clifford algebraic space, regardless of $A$ and $B$ being Grassmann-even or Grassmann-odd. With the six $Cl(0,6)$ basis vectors~\eqref{eq:gamma2} defined as anti-Hermitian ($\Gamma_i^{\dagger} = -\Gamma_i$ and $\gamma_i^{\dagger} = -\gamma_i$), the Hermitian conjugate of any Clifford element can thus be determined by recursively applying \eqref{eq:hermitian}. For example,
\begin{equation}
{\gamma}^{\dagger}_0 = (\Gamma_1\Gamma_2\Gamma_3)^{\dagger} = \Gamma_3^{\dagger}\Gamma_2^{\dagger}\Gamma_1^{\dagger} = -\Gamma_3\Gamma_2\Gamma_1 =  \gamma_0.
\end{equation}

Since the fermion Lagrangian comprises the Dirac inner product $\left\langle \bar{\psi} \psi\right\rangle$ or some variants thereof, the allowable symmetry transformations are the ones under which these sorts of Dirac inner products are invariant. Let's start with investigating the general gauge transformation
\begin{equation}
\psi\quad\rightarrow\quad V\psi U,
\end{equation}
where the Clifford-valued $V$ and $U$ are two {\it independent} gauge transformations. As an example, the following Dirac inner product transforms as
\begin{equation}
\left\langle \bar{\psi}I\psi I\right\rangle  \quad\rightarrow\quad \left\langle (U^{\dagger}\psi^{\dagger}V^{\dagger})\gamma_0 I (V\psi U) I\right\rangle = \left\langle \psi^{\dagger}(V^{\dagger}\gamma_0 I V)\psi (UIU^{\dagger})\right\rangle,
\end{equation}
which is invariant if
\begin{align}
&V^{\dagger}\gamma_0 I V = \gamma_0 I, \label{EQ1}\\
&UIU^{\dagger} = I, \label{EQ2}
\end{align}
where we have used the property $\left\langle AB \right\rangle= \left\langle BA \right\rangle$ provided that A and B are Grassmann-even. If we restrict our discussion to gauge transformations continuously connected to identity, the general solutions of the above equations are
\begin{align}
&V= e^{\theta^n(x) T_n}, \label{solution1}\\
&U= e^{\epsilon^n(x) K_n}, \label{solution2}
\end{align}
where $\{T_n; n=1, \cdots, 28\}$ are the generators of the spin group $Spin(4,4)$ (double cover of the rotation group $SO(4,4)$)
\begin{equation}
\gamma_a, \gamma_a\gamma_b, \Gamma_a\Gamma_b, I\Gamma_i, \gamma_i\gamma_j\Gamma_k, \label{LROT}
\end{equation}
and $\{K_n; n=1, \cdots, 36\}$ are the generators of the real symplectic group $Sp(8, R)$ (i.e. all 15 bi-vectors, all 20 tri-vectors, and the pseudoscalar)
\begin{equation}
\gamma_i\gamma_j, \Gamma_i\Gamma_j, \gamma_i\Gamma_k, \gamma_0, \Gamma_0, \gamma_i\gamma_j\Gamma_k, \Gamma_i\Gamma_j\gamma_k, I, \label{RROT}
\end{equation}
where $i, j, k= 1,2,3,  a, b = 0,1,2,3, i > j, a > b$. The tri-vector $\Gamma_0$ is defined as $\Gamma_0 = \gamma_1\gamma_2\gamma_3$, thus the basis $\{\Gamma_a; a = 0, \cdots, 3 \}$ parallels $\{\gamma_a; a = 0, \cdots, 3 \}$. The {\it real-valued} transformation parameters $\theta^n(x)$ and $\epsilon^n(x)$ are space-time dependent, since we are dealing with {\it local } gauge transformations. For the sake of brevity,  we will omit the $x$ label hereafter with the understanding that the space-time dependency is implied. 

Note that the $e^{\theta^n T_n}$ ($e^{\epsilon^n K_n}$) format of gauge transformation is known as the mathematicians' convention, whereas in the physicists' convention the gauge transformations take the form $e^{i\theta^n T_n}$ ($e^{i\epsilon^n K_n}$) with an extra $i$ in front of $\theta^n T_n$ ($\epsilon^n K_n$). As such, the physicists' Hermitian Lie group generators are translated into the mathematicians' anti-Hermitian generators, and vice versa. Following the overarching principle stated earlier, we should stay clear of introducing the quantum imaginary number $i$ in the definition of classical fields and their corresponding gauge transformations. Therefore, we choose the mathematicians' convention where there is no extra $i$ in the definition of gauge transformations. And for that matter, the identified gauge generators  $T_n$ and $K_n$ in Eqs.~\eqref{LROT} and \eqref{RROT} are also independent of imaginary number $i$ as demonstrated above. 

A few comments are in order. First of all, in the literature the usual candidates for gauge transformations are bi-vector-related rotations. The above $T_n$ and $K_n$ involve non-bi-vector Clifford elements such as $\gamma_0$ (tri-vector), $\gamma_i$ (i=1,2,3, vector) and the Lorentz boosts $\gamma_{0}\gamma_{i}$ (4-vector) in $T_n$, which go beyond the confines of bi-vectors. While the traditional rotation transforms a vector into another vector, the generalized ``rotation'' via the non-bi-vector-valued Clifford elements could potentially transform vectors into multivectors. 

Secondly, the two gauge transformations $V$ and $U$ are {\it independent} of each other. They are applied to the left side and right side of the algebraic spinor $\psi$, respectively.  Since by definition $V$ and $U$ commute with each other, the corresponding symmetry groups are the direct product $Spin(4,4) \times Sp(8, R)$ ($Spin(4,4)$ commutes with $Sp(8, R)$). The availability of the double-sided gauge transformations is one of the advantages of the Clifford algebra approach compared with the conventional column fermion formalism. This advantage has historically been leveraged in various Clifford algebraic models~\cite{TRAY,HEST4,Pavsic2013}. 

One interesting observation is that the 10 generators of de Sitter group $SO(1,4)$ (or the double cover $Spin(1,4)$ when spinors are involved) are contained in $T_n$~\eqref{LROT}, namely
\begin{equation}
\label{eq:desitter}
 \gamma_a, \gamma_a\gamma_b. 
\end{equation}
We know that there is another flavor of gauge gravity theory which is based on (anti-) de Sitter group~\cite{DESI,Wise,Westman}. It enjoys the desirable property that vierbein $\hat{e}_{\mu}$ and spin connection $\hat{\omega}_{\mu}$ in Eq.~\eqref{eq:gauge} jointly constitute the gauge fields of de Sitter group. As a comparison, in Lorentz gauge gravity theory, vierbein $\hat{e}_{\mu}$ is not a gauge field, albeit spin connection $\hat{\omega}_{\mu}$ is indeed the gauge field of Lorentz group. Vierbein is instead regarded as an add-on to Lorentz gauge gravity theory. This incentivized us to adopt de Sitter gauge theory of gravity in our first paper~\cite{WL1} on Clifford algebra $Cl(0,6)$.  However, there is a downside in this approach: Given that 4 $\gamma_a$ out of 10 de Sitter group generators in Eq.~\eqref{eq:desitter} are Clifford-odd, the associated gauge transformations mix left- and right-handed spinors which are Clifford-odd and -even, respectively. Consequently, the left- and right-handed spinors have to transform in sync, which disagrees with chirality of weak interaction. This is a major shortcoming of our first paper~\cite{WL1} on Clifford algebra $Cl(0,6)$. 

In our subsequent papers~\cite{WL3, WL4}, we improved our model and circumvented the above limitation by demanding that only the Clifford-even sub-algebras of $T_n$ and $K_n$ are permitted, which enables us to accommodate the chirality of weak interaction by virtue of decoupling the gauge transformations of the left- and right-handed spinors. The trade-off is that we have to settle for the Lorentz gauge gravity theory where vierbein $\hat{e}_{\mu}$ is an add-on transforming as a vector of the Lorentz symmetry. It is seemingly a disadvantage compared with the de Sitter gauge gravity theory. However, in Sections \ref{subsec:CC} and \ref{subsec:chiral} we will learn that it is a blessing in disguise, since the vierbein is never meant to be a fundamental gauge field. It's actually an emergent quantity arising from  quantum condensation. 

With the restriction to the Clifford-even sub-algebras of $T_n$ and $K_n$, we are left with the following symmetries~\cite{WL3}
\begin{align}
\label{eq:largest}
&\psi_L\quad\rightarrow\quad V_L\psi_L U_L, \\
\label{eq:largest2}
&\psi_R\quad\rightarrow\quad V_R\psi_R U_R,
\end{align}
where
\begin{align}
&V_L= e^{\theta_L^n T_{n}}, \quad U_L= e^{\epsilon_L^n K_{ n}} \label{solution3}, \\
&V_R= e^{\theta_R^n T_{n}}, \quad U_R= e^{\epsilon_R^n K_{n}} \label{solution4}.
\end{align}
The 6+6=12 Clifford-even $\{T_{n}; n=1, \cdots, 12\}$ comprise the generators of $Spin(1,3) \times Spin(1,3)_{Weak}$
\begin{align}
&\gamma_a\gamma_b; \quad \Gamma_a\Gamma_b,
\end{align}
where  $a, b = 0,1,2,3, a > b$, and the 15+1=16 Clifford-even $\{K_{ n}; n=1, \cdots, 16\}$ comprise the generators of $Spin(6)_{Pati-Salam} \times U_I(1)$
\begin{align}
\label{eq:PS}
\gamma_i\gamma_j, &\Gamma_i\Gamma_j, \gamma_i\Gamma_k; \quad I,
\end{align}
where $i, j, k= 1,2,3, i > j$. 
Note that the {\it real-valued} gauge transformation parameters such as $\theta_L^n$ ($\epsilon_L^n$) and $\theta_R^n$ ($\epsilon_R^n$) are independent of each other. Therefore the left- and right-handed spinors $\psi_L$ and $\psi_R$ transform independently. Considering that the same copies of the symmetry group generators $T_{n}$ and $K_{n}$ are employed for the left- and right-handed spinors, the model is left-right symmetric. 

Since the vierbein $\hat{e}_\mu$ transforms as a vector under the Lorentz gauge transformation
\begin{align}
\label{eq:vector}
\hat{e}_\mu & \quad\rightarrow\quad e^{\frac{1}{4}\theta^{ab}\gamma_a\gamma_b} \; \hat{e}_\mu  \; e^{-\frac{1}{4}\theta^{ab}\gamma_a\gamma_b},
\end{align}
a spinor bilinear term such as 
\begin{align}
\label{eq:kinetic}
\left\langle \bar{\psi} \hat{e}_\mu \psi \right\rangle,
\end{align}
can be verified to be gauge invariant under the above gauge transformations restricted to the Clifford-even sub-space (with the exception of the weak-boosts generated by \{$\Gamma_0\Gamma_i$\} which will be revisited in Section~\ref{sec:quantization}). This invariance is of paramount importance, given that the fermion-related Lagrangian is of similar form and should respect these symmetries. Note that $\hat{e}_\mu$ should always appear on the left side of spinor $\psi$ (or on the right side of $\bar{\psi}$), since Lorentz gauge transformation is applied on the left side of $\psi$.

The above gauge  symmetry groups are the largest ones permissible by a {\it chiral} algebraic spinor of $Cl(0,6)$. The spin group $Spin(6)_{Pati-Salam}$ is generated by all the 15 bi-vectors of $Cl(0,6)$.  It is isomorphic to Pati-Salem's $SU(4)$~\cite{SU4}, which encompasses $SU(3)_{C} \times U(1)_{B-L}$. The $Spin(1,3)_{Weak}$ group comprises the regular weak $SU(2)$ group generated by \{$\Gamma_i\Gamma_j$\}  as well as the weak-boosts generated by \{$\Gamma_0\Gamma_i$\}. These are the counterparts of the spacial rotations generated by \{$\gamma_i\gamma_j$\} and the Lorentz boosts generated by \{$\gamma_0\gamma_i$\}. It can be readily verified that the two sets of generators $\{\gamma_a\gamma_b\}$ and $\{\Gamma_a\Gamma_b\}$ (with  $a, b = 0,1,2,3, a > b$) commute with each other. Therefore, the left-sided gauge symmetries are the direct product between these two symmetries $Spin(1,3) \times Spin(1,3)_{Weak}$, which is the subgroup of the encompassing spin group $Spin(4,4)$ in~\eqref{LROT}. 

The BSM symmetry bonanza noted above is tantalizing.  From a practical point of view, do we have any inkling of symmetries beyond the local Lorentz and standard model gauge symmetries? The observation of neutrino oscillations\cite{FUK, AHM, EGU} provided an interesting clue. It implies that neutrinos have nonzero masses beyond the plain vanilla standard model. Curiously, the neutrino masses are much smaller than that of the other standard model fermions. The seesaw mechanism is hence proposed as an explanation~\cite{NEUT}, which invokes the right-handed neutrinos endowed with large Majorana masses. In light of these suggestive evidences, we whittle down to a minimum subset of groups which could accommodate a Higgs-like mechanism to generate the Majorana masses. 

Therefore, our choice of symmetry groups are~\cite{WL3}
 \begin{align}
 \label{eq:symmetry1}
 &Spin(1,3) \times SU(3)_{C} \times SU(2)_{WL} \times U(1)_{WR} \times U(1)_{B-L}.
\end{align}
Note that in Section~\ref{subsec:chiral} on emergent chiral vierbeins, we will expand the above groups to accommodate the extended vierbeins and the extended weak interactions. But for now, we will stay with the above unextended symmetry groups. 

The symmetry groups in Eq.~\eqref{eq:symmetry1} are the direct product of the spin connection's Lorentz $Spin(1,3)$ with six distinct generators (as in $\psi  \rightarrow  e^{\theta^n T_{n}} \psi$)
\begin{align}
\label{eq:gaugeA}
&\frac{1}{2}\gamma_a\gamma_b, 
\end{align}
where  $a, b = 0,1,2,3, a > b$, and the left-handed weak interaction's $SU(2)_{WL}$ with three generators (as in $\psi_L  \rightarrow e^{\theta_L^n T_{n}} \psi_L$)
\begin{align}
\label{eq:gaugeB}
&\frac{1}{2}\Gamma_2\Gamma_3, \quad \frac{1}{2}\Gamma_3\Gamma_1, \quad \frac{1}{2}\Gamma_1\Gamma_2,
\end{align}
and the right-handed weak interaction's $U(1)_{WR}$ with one generator (as in $\psi_R  \rightarrow e^{\theta_R^n T_{n}} \psi_R$)
\begin{align}
\label{eq:gaugeC}
&\frac{1}{2}\Gamma_1\Gamma_2, 
\end{align}
and the strong interaction's $SU(3)_{C}$ with eight generators (as in $\psi  \rightarrow \psi e^{\epsilon^n K_{n}}$)
\begin{equation}
\label{eq:SU3}
\begin{array}{rl}
&\frac{1}{4}(\gamma_1\Gamma_2 + \gamma_2\Gamma_1), 
\quad \frac{1}{4}(\Gamma_1\Gamma_2 + \gamma_1\gamma_2), 
\quad \frac{1}{4}(\Gamma_1\gamma_1 - \Gamma_2\gamma_2), \\
&\frac{1}{4}(\gamma_1\Gamma_3 + \gamma_3\Gamma_1),
\quad \frac{1}{4}(\Gamma_1\Gamma_3 + \gamma_1\gamma_3),  \\
&\frac{1}{4}(\gamma_2\Gamma_3 + \gamma_3\Gamma_2),
\quad \frac{1}{4}(\Gamma_2\Gamma_3 + \gamma_2\gamma_3),  \\
&\frac{1}{4\sqrt{3}}(\Gamma_1\gamma_1 + \Gamma_2\gamma_2 - 2\Gamma_3\gamma_3),
\end{array}
\end{equation}
and the BL interaction's $U(1)_{B-L}$ with one generator (as in $\psi  \rightarrow  \psi e^{\epsilon^n K_{n}}$)
\begin{align}
\label{eq:J}
&\frac{1}{2}J=\frac{1}{6}(\gamma_1\Gamma_1 + \gamma_2\Gamma_2 + \gamma_3\Gamma_3).
\end{align}
Note that some multipliers are applied to the generators to facilitate the gauge field definitions in Section~\ref{subsec:Lagrangian}. Due to the chirality of the weak interaction, the gauge transformation parameters for the left- and right-handed spinors ($\theta_L^n$ and $\theta_R^n$) are kept independent for $SU(2)_{WL}$ and $U(1)_{WR}$, whereas the other gauge transformation parameters of $Spin(1,3)$, $SU(3)_{C}$, and $U(1)_{B-L}$ are synchronized between the left- and right-handed spinors. 

The left-sided and right-sided gauge transformations are applied to the  left side and right side of the Clifford algebraic spinor, respectively.  The $Spin(1,3) \times SU(2)_{WL} \times U(1)_{WR}$ symmetries in~\eqref{eq:symmetry1} belong to the left-sided gauge transformations, while the $SU(3)_{C}\times U(1)_{B-L}$ symmetries in~\eqref{eq:symmetry1} belong to the  right-sided gauge transformations. By definition, the left-sided and right-sided gauge transformations commute with each other. Therefore, the total gauge symmetry in~\eqref{eq:symmetry1} is a direct product of $Spin(1,3) \times SU(2)_{WL} \times U(1)_{WR}$ and $SU(3)_{C}\times U(1)_{B-L}$. 

Let's look at some instances: The right-sided color $SU(3)_{C}$ gauge transformations~\eqref{eq:SU3} commute with the left-sided local Lorentz gauge transformations $Spin(1,3)$~\eqref{eq:gaugeA} and  left-sided weak gauge transformations $SU(2)_{WL}$~\eqref{eq:gaugeB} since they are applied to the different sides of the algebraic spinor, despite the fact that the generators in Eq.~\eqref{eq:SU3} and generators in Eq.~\eqref{eq:gaugeA}/\eqref{eq:gaugeB} seemingly do not commute with each other. In the same vein, since the color projection operators~\eqref{eq:color} are applied to the right side of the algebraic spinor,  they do not break the Lorentz symmetries which are applied to the left side of the algebraic spinor. 

There are three Clifford elements which play a pivotal role in the symmetry determination. The first element $\gamma_0$ is hard-wired into the definition of Dirac inner product $\left\langle \bar{\psi} \psi\right\rangle = \left\langle {\psi}^{\dagger}\gamma_0 \psi\right\rangle$. It facilitates pinning down the Lorentz group $Spin(1,3)$. The second element $\Gamma_1\Gamma_2$ is embedded in the definition of isospin~\eqref{Iso}, thus it picks out the isospin direction (we would like to highlight the fact that $\Gamma_1\Gamma_2$ commutes with the Lorentz generators $\gamma_a\gamma_b$, hence isospin projection does not break Lorentz symmetry). The third critical Clifford element is the BL interaction's $J$~\eqref{eq:J}. It is instrumental in separating out  $U(3) = SU(3)_{C} \times U(1)_{B-L}$ from the encompassing $Spin(6)_{Pati-Salam}$ which is isomorphic to $SU(4)$. Mathematically speaking, this is a specific case of a general procedure~\cite{EMBE,DORA} of separating out  $U(n)$ from $Spin(2n)$. 

When $J$ is applied to the four color projection operators~\eqref{eq:color}, it has the nice property that 
\begin{subequations}\label{eq:Jcharge}
\begin{align}
&P_lJ = -P_l I, \label{BL2} \\
&P_{r}J = \frac{1}{3}P_{r}I, \quad P_{g}J = \frac{1}{3}P_{g}I, \quad P_{b}J = \frac{1}{3}P_{b}I,\label{BL1} 
\end{align}
\end{subequations}
which means that $J$ is tantamount to
\begin{align}
\label{eq:BL}
&  J = (B-L)I, 
\end{align}
where $B$ and $L$ are baryon and lepton numbers, respectively. Therefore, $J$ indeed corresponds to the BL interaction. The definition of the four color projection operators~\eqref{eq:color} as well as the definition of the color algebra~\eqref{eq:SU3} are both predicated on how  $J$ is structured. The ansatz ensures that applying any generator in the color algebra to the lepton projector $P_l$ is identical to zero, hence leptons are invariant (singlets) under the color gauge transformations. 

If we start from our choice of symmetries~\eqref{eq:symmetry1} as a given, the symmetry breaking mechanism from symmetries~\eqref{eq:symmetry1} all the way down to $SU(3)_{C} \times U(1)_{EM}$ can be readily worked out. The follow is a high-level preview of the major symmetry breaking patterns. The details of these symmetry breaking processes in the Clifford algebraic setting will be elaborated in Section~\ref{sec:SSB}. 

The cascade of symmetry breaking processes begins with the vierbein acquiring a nonzero flat space-time VEV, which breaks the gauge symmetries down to 
\begin{align}
\label{eq:symmetry2}
 &SU(3)_{C} \times SU(2)_{WL} \times U(1)_{WR} \times U(1)_{B-L}.
\end{align}
As a result, the local Lorentz gauge symmetry is lost. The next step of symmetry breaking is triggered by the Majorana-Higgs field, which is a Higgs-like field in addition to the standard model Higgs field.  At this stage, the Majorana-Higgs field assumes a nonzero VEV and breaks the local gauge symmetries down to the standard model symmetries 
\begin{align}
\label{eq:symmetry3}
 &SU(3)_{C}  \times SU(2)_{WL} \times U(1)_Y,
\end{align}
where $U(1)_Y$ is the hypercharge gauge symmetry specified by the synchronized double-sided gauge transformation
\begin{align}
\label{eq:symmetryHyper}
&\psi_{L}\rightarrow \psi_{L} e^{\frac{1}{2}\epsilon_Y J}, \\
&\psi_{R}\rightarrow e^{\frac{1}{2}\epsilon_Y\Gamma_1\Gamma_2}\psi_{R} e^{\frac{1}{2}\epsilon_Y J},
\end{align}
where a shared rotation angle $\epsilon_Y$ synchronizes the double-sided hypercharge gauge transformation. At the third stage of symmetry breaking, the electroweak Higgs fields acquire nonzero VEVs and break the standard model symmetries down to 
\begin{align}
\label{eq:symmetry3}
&SU(3)_{C} \times U(1)_{EM},
\end{align}
where $U(1)_{EM}$ is the electromagnetic gauge symmetry characterized by the synchronized double-sided gauge transformation
\begin{align}
&\psi\rightarrow e^{\frac{1}{2}\epsilon_{EM}\Gamma_1\Gamma_2}\psi e^{\frac{1}{2}\epsilon_{EM} J}, \label{EM}
\end{align}
where a shared rotation angle $\epsilon_{EM}$ synchronizes the double-sided electromagnetic gauge transformation. Based on the definition of the pseudoscalar-valued electromagnetic gauge field $\hat{A}_\mu$ in Eq.~\eqref{eq:EM}, the electric charge $q$ of a given standard model fermion $\psi_f$ can thus be obtained via
\begin{align}
&e^{\frac{1}{2}\epsilon_{EM}\Gamma_1\Gamma_2}\psi_f e^{ \frac{1}{2}\epsilon_{EM}J} = \psi_f e^{q \epsilon_{EM}I}. \label{EM2}
\end{align}
Thanks to the properties of $\Gamma_1\Gamma_2$ in Eq.~\eqref{Iso2} and $J$ in Eq.~\eqref{eq:Jcharge}, the electric charges can be readily calculated as $q = 0, -1, \frac{2}{3}$, and $-\frac{1}{3}$ for neutrino, electron, up quarks, and down quarks according to the definitions in Eq.~\eqref{eq:left} and Eq.~\eqref{eq:right}, which are perfectly aligned with the standard model electric charge assignments. 

We shall underscore the fact that the algebraic spinors and all the gauge group generators are valued in the {\it real} Clifford space. Any reference of the imagine number $i$ in the conventional formalism with regard to gauge transformations (and gauge fields) and spinors can be replaced by the pseudoscalar $I$ acting on the right side of the spinor, as illustrated in the definition of electric charge above.

\subsection{Charge conjugation without particle-antiparticle interchange}
\label{subsec:charge}
The charge conjugation $C$ changes the sign of charges. In the conventional matrix formalism, the $C$ conjugation of a fermion $\psi$ in the Weyl basis is expressed as
\begin{align}
\label{eq:chargeC}
C: \quad \psi  \rightarrow \psi_c = -i\gamma_2 \psi^\star,
\end{align}
where $\psi^\star$ is the complex conjugate of $\psi$. Because of the complex conjugate operation, $C$ converts a particle into its corresponding antiparticle.

Can we decouple  charge conjugation from complex conjugate and thus evade particle-antiparticle interchange?  Such a decoupling is indeed possible in the Clifford algebra approach, thanks to the identification of the imaginary number $i$ with the pseudoscalar $I$ acting on the right side of a spinor. Considering the definition of electric charge $q$ in Eq.~\eqref{EM2}, we can define a weaker form of charge conjugation
\begin{align}
\label{eq:weakerC}
C': \quad \psi  \rightarrow \psi_{c'} =   (I\Gamma_2\Gamma_3)\psi \gamma_0. 
\end{align}
Note that $I\Gamma_2\Gamma_3$ and $\gamma_0$ in the above definition could be replaced with the general $I\Gamma_2\Gamma_3e^{\theta\Gamma_1\Gamma_2} $ and $\gamma_0e^{\epsilon I}$, where $\theta$ and $\epsilon$ are two arbitrary phase factors. The weaker form of charge conjugation satisfies the property $(\psi_{c'})_{c'} = \psi$. It does not involve complex conjugate, hence there is no particle-antiparticle switching. In Section~\ref{subsec:Majorana}, this property of $C'$ will be leveraged to construct a Majorana mass term that conserves lepton number.

According to Eq.~\eqref{EM2}, it can be easily checked that $\psi_{c'}$ transforms as
\begin{align}
\label{eq:chargeChange}
&\psi_{c'} \rightarrow \psi_{c'} e^{-q \epsilon_{EM}I}. 
\end{align}
under the electromagnetic gauge transformation. Therefore, the sign of electric charge is indeed changed. It's driven by the fact that $\gamma_0$ in the definition of $C'$ anticommutes with the unit pseudoscalar $I$
\begin{align}
\label{eq:anti}
&e^{q \epsilon_{EM}I}\gamma_0= \gamma_0 e^{-q \epsilon_{EM}I}.
\end{align}
This sort of mathematical maneuvering is otherwise impossible in the conventional formalism where the electromagnetic gauge transformation is associated with the imaginary number $i$ as in Eq.~\eqref{eq:phase}. Since $i$ commutes with any operator, the only way to change sign of $i$ is to invoke complex conjugate. Consequently, charge conjugation in the conventional formalism is inextricably linked to particle-antiparticle interchange.

It can be verified that $C'$ does not change isospin~\eqref{Iso} or color~\eqref{eq:color} of any standard model fermion. Since $I\Gamma_2\Gamma_3$ commutes with the Lorentz transformation generators $\gamma_a\gamma_b$, the Lorentz transformation properties of $\psi_{c'}$ remain the same as $\psi$. Note that $C'$ changes the chirality of $\psi$, since $C'$ involves the multiplication of the Clifford-odd tri-vector $\gamma_0$ and it turns a left-handed fermion into a right-handed one, and vice versa. 

The charge conjugation $C'$ of the gauge transformation parameters (and thus gauge fields) can be defined as
\begin{align}
\label{eq:weakerCgauge}
&C': \quad \theta^nT_n  \rightarrow  \theta^n_{c'}T_n= \theta^n (I\Gamma_2\Gamma_3) T_n (I\Gamma_2\Gamma_3), \\
&C': \quad \epsilon^nK_n  \rightarrow \epsilon^n_{c'}K_n=  \epsilon^n  \gamma_0 K_n\gamma_0,
\end{align}
with the understanding that the left side-type  $T_n$ should include the generators of  $Spin(1,3) \times SU(2)_{WL} \times U(1)_{WR}$ and the right side-type  $K_n$ should include the generators of  $SU(3)_{C} \times U(1)_{B-L}$. Some notable examples are that the electromagnetic gauge field $\hat{A}_\mu$ is $C'$-odd, while the gravity-related spin connection Lorentz gauge field $\hat{\omega}_{\mu}$ is $C'$-even. 



\subsection{Clifford-valued gauge fields and Lagrangian of the world}
\label{subsec:Lagrangian}
Having established the fermion representation and the symmetry structure, we are now well-positioned to investigate the Clifford-valued gauge fields corresponding to these symmetries,  as well as Lagrangians  and actions on the 4-dimensional space-time manifold. As we noted in the introduction section, while the Clifford algebra has been extended to $Cl(0,6)$, the underlying space-time manifold remains 4-dimensional. With a view toward writing down the diffeomorphism-invariant actions for general covariance, we are going to make extensive use of differential forms. We define gauge field as 1-forms and gauge forces as curvature 2-forms.  Diffeomorphism invariance/general covariance can be ensured if the action for the curved space-time is expressed as an integration of 4-forms on the 4-dimensional space-time manifold. 

The vierbein 1-form $\hat{e}$  and spin connection 1-form $\hat{\omega}$ of the Lorentz gauge theory of gravity are
\begin{align}
\label{eq:gravity2}
&\hat{e} = \hat{e}_{\mu}dx^{\mu}, \\
&\hat{\omega} = \hat{\omega}_{\mu}dx^{\mu}, 
\end{align}
where $\mu= 0, 1, 2, 3$, and the Clifford-valued $\hat{e}_{\mu}$ and $\hat{\omega}_{\mu}$ are given in Eq.~\eqref{eq:gauge}. The Clifford-valued forms are also called Clifforms in the literature~\cite{CF1, CF2}. 

In an similar fashion, the other gauge field 1-forms related to the gauge symmetries $SU(3)_{C} \times SU(2)_{WL} \times U(1)_{WR} \times U(1)_{B-L}$ can be defined as
\begin{align}
\label{eq:oneform}
&\hat{G} = \hat{G}_{\mu}dx^{\mu}, \\
&\hat{W}_{L} =\hat{W}_{L\mu}dx^{\mu}, \\
&\hat{W}_{R} = \hat{W}_{R\mu}dx^{\mu}, \\
&\hat{A}_{BL} =\hat{A}_{BL\mu}dx^{\mu}, 
\end{align}
where the strong interaction gauge field $\hat{G}_{\mu}$, the left-handed weak interaction gauge field $\hat{W}_{L\mu}$, the right-handed weak interaction gauge field $\hat{W}_{R\mu}$, and the BL interaction gauge field $\hat{A}_{BL\mu}$ are valued in the gauge generator space of $SU(3)_{C}$, $SU(2)_{WL}$, $U(1)_{WR}$, and $U(1)_{B-L}$ (Eqs.~\eqref{eq:gaugeB}, \eqref{eq:gaugeC}, \eqref{eq:SU3} and \eqref{eq:J}), respectively. As an example, according to the gauge group generators specified in Eqs.~\eqref{eq:gaugeB}, the left-handed weak interaction gauge field $\hat{W}_{L\mu}$ can be defined as 
\begin{align}
\label{eq:BL2}
&\hat{W}_{L\mu}= \frac{1}{2}(W^1_{L\mu}\Gamma_2\Gamma_3 + W^2_{L\mu}\Gamma_3\Gamma_1 + W^3_{L\mu}\Gamma_1\Gamma_2), 
\end{align}
where $W^1_{L\mu}$, $W^2_{L\mu}$, and $W^3_{L\mu}$ are real-valued. The other gauge fields $\hat{G}_{\mu}$, $\hat{W}_{R}$, and $\hat{A}_{BL\mu}$ can be defined in a similar manner using the corresponding gauge group generators specified in Eqs.~ \eqref{eq:gaugeC}, \eqref{eq:SU3} and \eqref{eq:J}, respectively. We adopt the notation convention (e.g. $\hat{G}$ rather than $G$) to highlight the fact that these gauge fields are {\it Clifford-valued} 1-forms, i.e. Clifforms. 

The chiral gauge-covariant derivatives of the chiral spinor fields are defined as
\begin{align}
&D_L\psi_L = (d + \hat{\omega} + \hat{W}_{L})\psi_L + \psi_L (\hat{G} + \hat{A}_{BL} ), \\
&D_R\psi_R = (d + \hat{\omega} + \hat{W}_{R})\psi_R + \psi_R (\hat{G} + \hat{A}_{BL} ),
\end{align}
where $d$ is the exterior derivative $d=dx^{\mu}\partial_{\mu}$. We follow the convention of putting the gauge coupling constant in the coefficient of the Yang-Mills Lagrangian (see e.g. Eq.~\eqref{eq:YM2}), rather than in the gauge-covariant derivatives above (via re-scaling the gauge fields). The gauge-covariance property of  the covariant derivatives is ensured by the gauge transformation rules of the gauge fields involved~\cite{WL1,WL3}.  It is essential that the gauge fields should appear on the proper side of the spinor, which is dictated by how the gauge symmetries are defined in Section~\ref{subsec:Gauge}. Specifically, $\hat{\omega}$ and $\hat{W}_{L/R}$ correspond to left-sided gauge symmetries, thus they should act on the left side of the spinor. On the other hand, $\hat{G}$ and $\hat{A}_{BL}$ correspond to right-sided gauge symmetries, hence they should be applied to the right side. 

Note that there is no extra imaginary number $i$ in front of the gauge fields (such as $i{W}_{L}$ in the conventional QFT formalism) in the definition of the gauge-covariant derivatives. This conforms with the way the gauge transformations are defined in Subsection~\ref{subsec:Gauge} where we choose the mathematicians' convention without the extra $i$ in the definition of gauge transformations. It is also consistent with the general rule stated earlier that we should eschew the quantum imaginary number $i$ in the definition of classical fields and their corresponding gauge-covariant derivatives.

The spin connection $\hat{\omega}$, as the gauge field of the Lorentz gauge group, is crucial in maintaining the {\it local } Lorentz gauge covariance of $D_{L}\psi_{L}$ and $D_{R}\psi_{R}$. On the other hand, given that the vierbein $\hat{e}$ is not a gauge field, $\hat{e}$ is conspicuously absent in the gauge-covariant derivatives of the spinor fields $\psi_{L/R}(x)$. Nonetheless, as we will learn below, $\hat{e}$ shows up in other parts of the Lagrangian and it plays a pivotal role in the model building.  

The gauge interactions are formulated as curvature $2$-forms, namely $\hat{R}$, $\hat{F}_{G}$, $\hat{F}_{WL}$, $\hat{F}_{WR}$, and $\hat{F}_{BL}$. For instance, the spin connection curvature $2$-form $\hat{R}$ and the left-handed weak interaction curvature 2-form $\hat{F}_{WL}$ are expressed as
\begin{align}
\label{eq:R}
&\hat{R} = d\hat{\omega} + \hat{\omega}\wedge\hat{\omega}, \\
&\hat{F}_{WL} = d\hat{W}_{L} +\hat{W}_{L}\wedge\hat{W}_{L}  =\frac{1}{2} \hat{F}_{WL\mu\nu} dx^{\mu}\wedge dx^{\nu} \\
&= \frac{1}{4} (F_{WL\mu\nu}^1\Gamma_2\Gamma_3 + F_{WL\mu\nu}^2\Gamma_3\Gamma_1 + F_{WL\mu\nu}^3\Gamma_1\Gamma_2)dx^{\mu}\wedge dx^{\nu},
\end{align}
where $\wedge$ stands for outer product between differential forms. For later usage we have expanded $\hat{F}_{WL}$ in more details. The other curvature $2$-forms $\hat{F}_{WR}$, $\hat{F}_{G}$, and $\hat{F}_{BL}$ can be defined in a similar way. Note that the outer product term vanishes for abelian interactions such as $\hat{F}_{WR}$ and $\hat{F}_{BL}$.

Now we are ready to write down the local gauge- and diffeomorphism-invariant Lagrangian of the world
\begin{align}
\label{eq:world}
\mathcal{L}_{World} =  &\mathcal{L}_{Fermion}  \\
+ &\mathcal{L}_{Gravity} + \mathcal{L}_{CC} \\
+ &\mathcal{L}_{YM-Color} + \mathcal{L}_{YM-Weak-Left} + \mathcal{L}_{YM-Weak-Right} + \mathcal{L}_{YM-BL} \label{SYM} \\
+ &\mathcal{L}_{Higgs-Majorana} + \mathcal{L}_{Higgs-Electroweak}.
\end{align}
The fermion Lagrangian in curved space-time is of the form
\begin{align}
\label{eq:fermion}
&\mathcal{L}_{Fermion} \sim i\left\langle I\hat{e}\wedge \hat{e}\wedge \hat{e} \wedge  (\psi_L \overline{D_L\psi_L}) \right\rangle + i\left\langle I\hat{e}\wedge \hat{e}\wedge \hat{e} \wedge  (\psi_R \overline{D_R\psi_R}) \right\rangle,
\end{align}
where $\overline{D_{L/R}\psi_{L/R}}$ is the Dirac conjugate of $D_{L/R}\psi_{L/R}$ defined as $\overline{D_{L/R}\psi_{L/R}} =(D_{L/R}\psi_{L/R})^{\dagger}\gamma_0$. The gravity plus cosmological constant Lagrangian terms are of the form 
\begin{align}
\label{eq:gravity}
\mathcal{L}_{Gravity} + \mathcal{L}_{CC} &= \frac{1}{8\pi G}\left\langle I(\hat{e}\wedge \hat{e} \wedge \hat{R} - \frac{\Lambda}{4!}\hat{e}\wedge \hat{e} \wedge \hat{e}\wedge \hat{e}) \right\rangle,
\end{align}
where $\hat{R}$ is the spin connection curvature $2$-form~\eqref{eq:R}, $G$ is gravitational constant, and $\Lambda$ is the cosmological constant. The left-handed weak interaction Yang-Mills Lagrangian term in curved space-time is of the form 
\begin{align}
\label{eq:YM}
&\mathcal{L}_{YM-Weak-Left} \sim \frac{\left\langle (I\hat{e}\wedge \hat{e} \wedge \hat{F}_{WL})(I\hat{e}\wedge \hat{e} \wedge \hat{F}_{WL})\right\rangle} {\left\langle I\hat{e}\wedge \hat{e} \wedge \hat{e}\wedge \hat{e}  \right\rangle}.
\end{align}
The Yang-Mills-type Lagrangian terms for the other gauge interactions such as $\hat{F}_{G}$, $\hat{F}_{WR}$, and $\hat{F}_{BL}$ take a similar form, which for brevity sake we will not write out explicitly. Here we adopted the generic ``Yang-Mills-type'' label, albeit historically the term Yang-Mills is reserved for non-abelian gauge fields only. Note that $\left\langle \ldots\right\rangle$ denotes the Clifford-scalar part of the enclosed expression, as defined earlier. 

The Higgs field-related Lagrangian terms $\mathcal{L}_{Higgs-Majorana}$ and $\mathcal{L}_{Higgs-Electroweak}$ will be outlined in Sections \ref{subsec:Majorana} and \ref{subsec:Electroweak} when we discuss the Higgs mechanism.  Upon symmetry breaking, the Majorana and Dirac mass terms will emerge from  $\mathcal{L}_{Higgs-Majorana}$ and $\mathcal{L}_{Higgs-Electroweak}$ via the Higgs mechanism. For later reference, we write down the  Dirac mass term for a given standard model fermion $\psi_f$ in curved space-time as
\begin{align}
\label{eq:diracMass}
&\mathcal{L}_{Dirac-Mass} \sim im_f\left\langle I\hat{e}\wedge \hat{e}\wedge \hat{e} \wedge  \hat{e}(\psi_f I \bar{\psi}_f) \right\rangle,
\end{align}
where $m_f$ is the Dirac mass of fermion $\psi_f$. 

It shall be reminded that in the Yang-Mills Lagrangian term, the 4-form factor $d^4x = dx^{0}\wedge dx^{1}\wedge dx^{2}\wedge dx^{3}$  from one of the $I\hat{e}\wedge \hat{e} \wedge \hat{F}_{WL}$ should be canceled out by the similar 4-form factor from the denominator $I\hat{e} \wedge \hat{e}\wedge \hat{e} \wedge \hat{e}$ before multiplication with the other $I\hat{e}\wedge \hat{e} \wedge \hat{F}_{WL}$. As such, the Yang-Mills Lagrangian terms, along with the other Lagrangian terms of the world, are diffeomorphism-invariant 4-forms on the 4-dimensional space-time manifold. 

Also note that the Clifford algebra elements of the vierbein $\hat{e}$ and the curvature $2$-forms such as $\hat{F}_{G}$ formally commute with each other in the Yang-Mills Lagrangian, since they transform under commuting gauge groups. Since the vierbein $\hat{e}$ transforms as a vector under Lorentz gauge transformation~\eqref{eq:vector} and is invariant under the other gauge transformations, all the terms of the Lagrangian of the world can be proved to be invariant under the local gauge symmetry transformations~\eqref{eq:symmetry1}. 

If we look at the fermion Lagrangian term $\mathcal{L}_{Fermion}$ in isolation, it can potentially accommodate Pati-Salam's left-right symmetrical $ SU(2)_L\times SU(2)_R\times SU(4)$, since no quark/lepton projections or $P_{\pm}$ projections (which would otherwise spoil the Pati-Salam symmetries) are applied to the fermions $\psi_L$ and $\psi_R$ in $\mathcal{L}_{Fermion}$. However, as we will learn later in section~\ref{sec:SSB}, the Higgs Lagrangian terms $\mathcal{L}_{Higgs-Majorana}$ and $\mathcal{L}_{Higgs-Electroweak}$ do involve the Pati-Salam-violating quark/lepton projections (on the Majorana-Higgs field and fermion fields) and $P_{\pm}$ projections (on the standard model Higgs field and right-handed fermion fields). Therefore, the Lagrangian of the world as a whole does not enjoy the Pati-Salam symmetries. 

The action of the world is
\begin{align}
\label{eq:Sworld}
S_{world}  = \int \mathcal{L}_{World}, 
\end{align}
where the integration over $d^4x = dx^{0}\wedge dx^{1}\wedge dx^{2}\wedge dx^{3}$ is already embedded in the definition of the Lagrangian as 4-form. We know that the space-time metric $g_{\mu\nu}$ can be derived from the vierbein (see Eq.~\eqref{eq:metricG}). Thus the metric tensor $g_{\mu\nu}$-related quantities  in the conventional metric gravity can be constructed using various combinations/transformations of the vierbein. For instance, the 4-form $\frac{1}{4!} I \hat{e} \wedge \hat{e}\wedge \hat{e} \wedge \hat{e}$ plays the role of the metric volume form $\sqrt{|g|}d^4x$. Another example is that the conventional Hodge star $\star$ is replaced by the specific configuration of vierbeins in the Yang-Mills Lagrangian (\ref{eq:YM}). 

We subscribe to the general notion of effective field theory~\cite{EFFE0,EFFE2}, which states that all the terms allowed by the symmetry requirements should be included in the Lagrangian of the world. Since one goal of this paper is to treat gravity and Yang-Mills interactions on an equal footing, we should consider Lagrangian terms that is linear in Yang-Mills curvature 2-forms as well, such as
\begin{align}
\label{eq:YMb}
\left\langle I\hat{e}\wedge \hat{e} \wedge \hat{F}_{WL} \right\rangle.
\end{align}
The above term is analogs to the gravity Lagrangian~\eqref{eq:gravity} which is linear in the spin connection curvature $2$-form $\hat{R}$. However, it can be verified that such linear Lagrangian terms for the Yang-Mills curvature 2-forms are identically zero. The only allowable linear term other than the gravity Lagrangian is the Holst term~\cite{HOLS} 
\begin{align}
\label{eq:gravity2}
\left\langle \hat{e}\wedge \hat{e} \wedge \hat{R}\right\rangle,
\end{align}
which differs from the gravity Lagrangian~\eqref{eq:gravity} by removing the pseudoscalar $I$.

When it comes to the Lagrangian terms with two or more gauge curvature 2-forms, there is a plethora of allowable forms besides the Yang-Mills-type Lagrangian.  Some examples are the topological CP-violating terms for the Yang-Mills interactions, the topological Gauss-Bonnet term and Nieh-Yan term~\cite{NIEH}, and the higher-derivative gravity terms~\cite{EFFE2}. While all these higher-order terms should in principle be included in the Lagrangian, the key for model building is to recognize that the practical predictions of any model must be made within the context of separation of energy scales. The gravity and Yang-Mills Lagrangian terms happen to be amongst the first few order terms that are relevant at the energy scale accessible to experiments. 

The attentive readers may have noticed the presence of the imaginary number $i$ in the fermion kinetic Lagrangian~\eqref{eq:fermion} and Dirac mass Lagrangian~\eqref{eq:diracMass}. As explained in the introduction section, there are two kinds of imaginary numbers. One is the genuine $i$, and the other can be replaced by the pseudoscalar $I$. We have demonstrated that the algebraic spinors and the gauge group generators are valued in the real Clifford space. Hence we have managed to stay away from the imaginary number $i$. So why do we need $i$ in the fermion Lagrangian terms? It has to do with the requirement that the classical action of the world should be real
\begin{align}
\label{eq:super}
&S^*_{world}  = S_{world}.
\end{align}
We know that $\left\langle \ldots\right\rangle$ is employed in each Lagrangian term. By definition, it is the Clifford-scalar part of the expression, which means $\left\langle \ldots\right\rangle$ is real as long as the related fields in the expression are valued in the real Clifford space.  Since the bosonic fields (gauge fields, vierbein, and Higgs fields) are valued in the real Clifford space, the reality condition is automatically satisfied for all the bosonic field-related Lagrangian terms, such as the gravity and Yang-mills Lagrangian terms.

On the other hand, since spinors are valued in the real Clifford space with Grassmann-odd coefficients, the fermion action can essentially be reduced to a sum/integral of terms like $i\psi_n\psi_m$, where $\psi_n$ and $\psi_m$ are Grassmann-odd  coefficients of the spinors. Given that $\psi_n$ and $\psi_m$ are defined as real $\psi^*_n =\psi_n$ and $\psi^*_m =\psi_m$, the complex conjugate of $i\psi_n\psi_m$ can be calculated as
\begin{align}
\label{eq:super3}
&(i\psi_n\psi_m)^* =i^*\psi^*_m\psi^*_n = -i\psi_m\psi_n = i\psi_n\psi_m.
\end{align}
Hence the reality condition is satisfied, which is otherwise violated if $i$ in the expression is removed. Note that the complex conjugate of the multiplication of Grassmann numbers is defined as
\begin{align}
\label{eq:super4}
&(\psi_n\psi_m)^* =\psi^*_m\psi^*_n,
\end{align}
with no extra minus sign, even though both $\psi_n$ and $\psi_m$ are Grassmann-odd.

In summary, we are compelled to include the imaginary number $i$ in the definition of the fermion Lagrangian to enforce the reality condition. That said, as mentioned earlier, the imaginary number $i$ is intimately related to the quantum theory. It will be shown in Section~\ref{sec:quantization} that the appearance of $i$ in the fermion Lagrangian is the tip of the iceberg of quantum essence of almost everything. 

\section{Spontaneous Symmetry Breaking}
\label{sec:SSB}
In this section, we investigate spontaneous symmetry breaking (SSB) driven by the non-degenerate vacuum expectation values (VEVs) of various bosonic fields, such as vierbein $\hat{e}$, Majorana-Higgs field $\phi_M$, the standard model Higgs field $\phi$, antisymmetric-tensor Higgs field $\phi_{AT}$, and $\Phi$ fields.   Among these bosonic fields, the vierbein $\hat{e}$ field and the antisymmetric-tensor Higgs field $\phi_{AT}$ are Lorentz vector and Lorentz sextet respectively, while the rest are Lorentz scalars. Along the way, we also study the fermion mass hierarchies and the lepton number-conserving Majorana mass.

Consistent with Section~\ref{sec:Clifford}, we regard these bosonic fields as classical fields {\it prior to field quantization}. Therefore, the quantum imaginary number $i$ should not be allowed in the definition of these classical fields such as the classical vierbein and Higgs fields. In the current section, all these symmetry-breaking bosonic fields are expressed as {\it real} Clifford algebra $Cl(0,6)$ multivectors. The important topic of field quantization via the Clifford functional integral formalism and the formal introduction of imaginary number $i$ will be discussed in detail in Section~\ref{sec:quantization}. 

There are two common threads running through these bosonic fields. The first common thread is that all these bosonic fields develop non-zero VEVs and consequently lead to SSB.  Note that while the Majorana-Higgs field $\phi_M$, the standard model Higgs field $\phi$, and the antisymmetric-tensor Higgs field $\phi_{AT}$ break gauge symmetries only, the vierbein $\hat{e}$ field breaks both Lorentz gauge symmetry and diffeomorphism symmetry given that vierbein $\hat{e}$ is a differential 1-form unlike the 0-form Higgs fields. The $\Phi$ fields are unique in that they are gauge invariant and hence they break global symmetries and generate (pseudo) Nambu-Goldstone bosons. 

The second common thread is that all these symmetry-breaking fields are composite fields of quantum condensation origin, albeit in the current section we treat these symmetry-breaking fields as fundamental fields only. The compositeness of these bosonic fields will be investigated in Section~\ref{sec:quantization}. Note that all the standard model fermion fields and all the gauge fields examined in Section~\ref{sec:Clifford} are still fundamental non-composite fields.

\subsection{Vierbein-induced SSB and the residual global Lorentz symmetry}
\label{subsec:Lorentz}
The SSB saga of the universe starts with the vierbein field $\hat{e}$ acquiring a nonzero VEV. As a consequence, the local Lorentz gauge symmetry $Spin(1,3)$ and the diffeomorphism symmetry are violated. The remaining local gauge symmetries are $SU(3)_{C} \times SU(2)_{WL} \times U(1)_{WR} \times U(1)_{B-L}$ plus a residual {\it global} Lorentz symmetry. 

The ``ground state'' of the vierbein $\hat{e}$ and spin connection $\hat{\omega}$ should satisfy the field equations, which are obtained by varying the world action $S_{world}$ with the fields $\hat{e}$ and $\hat{\omega}$ independently. The resultant Einstein-Cartan equations read
\begin{align}
&\frac{1}{8\pi G}(\hat{R}\wedge \hat{e} + \hat{e} \wedge \hat{R} - \frac{\Lambda}{3!}\hat{e}\wedge \hat{e} \wedge \hat{e})I = \mathbb{T}, \label{eq:EC:1}\\
&\frac{1}{8\pi G}(\hat{T}\wedge \hat{e} - \hat{e}\wedge \hat{T})I = \mathbb{S}, \label{eq:EC:2}
\end{align}
where the energy-momentum current $3$-form $\mathbb{T}$ and the spin current $3$-form $\mathbb{S}$ arise from the matter sector, such as the fermion, Yang-Mills, and Higgs action terms. Note that $\hat{R}$ is the spin connection curvature $2$-form~\eqref{eq:R}, and $\hat{T}$ is the torsion $2$-form 
\begin{align}
\label{eq:torsion}
\hat{T} &= d\hat{e} + \hat{\omega} \wedge \hat{e} + \hat{e} \wedge \hat{\omega}.
\end{align}

When the spin-current $\mathbb{S}$ is zero, the second Einstein-Cartan equation~\eqref{eq:EC:2} amounts to enforcing the zero-torsion condition
\begin{align}
\label{eq:zeroT}
\hat{T} &=  0,
\end{align}
which can be used to express the spin connection $\hat{\omega}$ in terms of the vierbein $\hat{e}$. In this case, the remaining (first) Einstein-Cartan equation~\eqref{eq:EC:1} can be shown to be equivalent to the regular Einstein field equations of gravity plus a cosmological constant term. 

Upon SSB, the vierbein field develops a nonzero VEV 
\begin{align}
\label{eq:vierbeinVEVflat}
\hat{e} = \delta_{\mu}^a \gamma_a dx^{\mu} = \gamma_{\mu} dx^{\mu},
\end{align}
while the spin connection remains zero 
\begin{align}
\hat{\omega} = 0.
\end{align}
It can be verified that the above $\hat{e}$ and $\hat{\omega}$ satisfy the Einstein-Cartan equations, provided that $\Lambda = 0$, $\mathbb{T} = 0$, and $\mathbb{S} = 0$. Subsequently, the space-time metric $g_{\mu\nu}= \left\langle \hat{e}_{\mu}\hat{e}_{\nu} \right\rangle$ reduces to
\begin{equation}
g_{\mu\nu} = \left\langle \gamma_{\mu}\gamma_{\nu} \right\rangle= \eta_{\mu\nu},
\end{equation}
which is the Minkowski flat space-time metric. 

With the substitution of $\hat{e}$ and $\hat{\omega}$ by the VEVs, the fermion action in the flat Minkowski space-time reduces to
\begin{align}
\label{eq:fermion2}
S_{fermion} = \int i{\left\langle \bar{\psi}_L \gamma^{\mu} D_{L\mu} \psi_L
+ \bar{\psi}_R \gamma^{\mu} D_{R\mu} \psi_R \right\rangle d^{4}x}, 
\end{align}
where
\begin{align}
\label{eq:partial}
D_{L \mu}\psi_{L} = (\partial_{\mu} + \hat{W}_{L\mu})\psi_{L} + \psi_{L} (\hat{G}_{\mu} + \hat{A}_{BL\mu}), \\
D_{R \mu}\psi_{R} = (\partial_{\mu} + \hat{W}_{R\mu})\psi_{R} + \psi_{R} (\hat{G}_{\mu} + \hat{A}_{BL\mu}).
\end{align}
Similarly, the Yang-Mills action(\ref{eq:YM}) of the left-handed weak interaction $\hat{F}_{WL}$ can be rewritten as
\begin{align}
\label{eq:YM2}
S_{YM-Weak-Left}  =  -\frac{1}{4g_{WL}}\int {F_{WL\mu\nu}^iF_{WL}^{i, \mu\nu} d^{4}x},
\end{align}
where $g_{WL}$ is the dimensionless coupling constant of the left-handed weak interaction. The Yang-Mills-type action terms of the right-handed weak, strong, and BL interactions take a similar form, with coupling constants $g_{WR}$, $g_{G}$, and $g_{BL}$, respectively. Note that
\begin{align}
\label{eq:reciprocal}
&\gamma^{\mu}\eta_{\mu\nu} = \gamma_{\nu}, \\
&F_{WL}^{i, \mu\nu}\eta_{\mu\alpha}\eta_{\nu\beta} = F_{WL\alpha\beta}^i,
\end{align}
where $\{\gamma^{\mu}\}$ is the reciprocal frame of  $\{\gamma_a\}$. The reciprocal frame $\{\gamma^{\mu}\}$ is the avatar of the vierbein 3-form $I\hat{e}\wedge \hat{e}\wedge \hat{e}$ from the original fermion Lagrangian~\eqref{eq:fermion} when the vierbein field $\hat{e}$ acquires the nonzero VEV in Eq.~\eqref{eq:vierbeinVEVflat}. Therefore, when the fermion action in the flat Minkowski space-time~\eqref{eq:fermion2} is employed to derive the massless Dirac equation (the Dirac equation with nonzero mass will be discussed in Section~\ref{subsec:Electroweak})
\begin{align}
\label{eq:Dirac0}
&\gamma^{\mu} D_{L, \mu} \psi_L = 0,\\
&\gamma^{\mu} D_{R, \mu} \psi_R = 0,
\end{align}
what shows up in the Dirac equation is the reciprocal frame. As such, it's the reciprocal frame  $\{\gamma^{\mu}\}$ that corresponds to the gamma matrices used in the conventional formalism. This fact also explains the minus sign we mentioned earlier: $-I$  plays the role of the conventional pseudoscalar when it is applied to the left side of a spinor such as $-I\psi$. The conventional pseudoscalar is defined using the reciprocal frame $\{\gamma^{\mu}\}$, whereas we define the pseudoscalar using the original Clifford algebra basis $\{\gamma_{a}\}$. Hence there is a minus sign. 

It's worth mentioning that when we derive the flat space-time fermion action from the curved space-time counterpart, we have leveraged $(\gamma^{\mu})^\dagger = \gamma_0\gamma^{\mu}\gamma_0$ and the following properties 
\begin{subequations}\label{eq:CliffordCommute}
\begin{align}
&\left\langle FG\right\rangle = -\left\langle GF\right\rangle, \\
&\left\langle (FG)^\dagger \right\rangle = -\left\langle FG \right\rangle,
\end{align}
\end{subequations}
where the Grassmann-odd $F$ and $G$ are functionals of odd multiples of $\psi(x)$ and $\bar{\psi}(x)$. The minus sign arises from the Grassmann-odd nature of $\psi(x)$ and $\bar{\psi}(x)$. 

Upon SSB induced by the nonzero VEV of the vierbein $\hat{e}$,  both the local Lorentz gauge symmetry $Spin(1,3)$ and the diffeomorphism symmetry are violated. That said, it can be verified that the fermion~\eqref{eq:fermion2} and the Yang-Mills~\eqref{eq:YM2} actions in flat Minkowski space-time is invariant under the residual {\it global} Lorentz transformations
\begin{align}
\label{eq:Lorentz1}
x^\mu & \quad\rightarrow\quad (\Lambda^{-1})^\mu_\nu x^\nu, \\
W^i_{L\mu}(x) & \quad\rightarrow\quad (\Lambda^{-1})_\mu^{\nu}W^i_{L\nu}(\Lambda^{-1}x), \\
\psi(x) & \quad\rightarrow\quad e^{\frac{1}{4}\theta^{ab}\gamma_a\gamma_b}\psi(\Lambda^{-1}x), 
\end{align}
where the Lorentz transformation parameters $\theta^{ab} = -\theta^{ba}$ and $\Lambda^\mu_\nu$ are related via the equation
\begin{align}
\label{eq:Lorentz2}
e^{\frac{1}{4}\theta^{ab}\gamma_a\gamma_b}\gamma^\mu e^{-\frac{1}{4}\theta^{ab}\gamma_a\gamma_b} = \Lambda^\mu_\nu \gamma^\nu.
\end{align}
Note that the global Lorentz transformation parameters $\theta^{ab}$ above are independent of position, as opposed to the position-dependent $\theta^{ab}(x)$ for the local Lorentz gauge transformations. 


The situation here parallels the Higgs mechanism where there remains a global $SU(2)$ custodial symmetry~\cite{CUSTOD} after the electroweak symmetry breaking. In the literature~\cite{Percacci1991,Percacci2023,Sardanashvily}, comparisons have been made between the vierbein-induced gravitational symmetry breaking and the Higgs-induced electroweak symmetry breaking. 

In the case of the vierbein-induced SSB, the vestigial global Lorentz symmetry is a synchronization (enforced by Eq. \eqref{eq:Lorentz2}) between the global  portion of the local Lorentz gauge transformation ($e^{\frac{1}{4}\theta^{ab}\gamma_a\gamma_b}$) for spinors and the global volume-preserving portion of the diffeomorphism transformation ($\Lambda_\mu^{\nu}$) for space-time coordinates. The local Lorentz gauge transformation involves Clifford algebraic elements such as $\gamma_a\gamma_b$ labeled by Roman indices, while the diffeomorphism transformation involves coordinates $x^\mu$ and gauge fields  (e.g. the electromagnetic field $A_\mu$ in Eq.~\eqref{eq:EM}) labeled by Greek indices. At the nexus is the VEV of the vierbein $e_{\mu}^a = \delta_{\mu}^a$ which acts as a soldering form gluing together the Roman and Greek propers.

\subsection{Majorana mass and absence of neutrinoless double beta decay}
\label{subsec:Majorana}
The second stage of SSB is triggered by the Majorana-Higgs field $\phi_M$ that couples to the right-handed neutrinos only. It is a Higgs-like bosonic field in addition to the well-known electroweak symmetry-breaking Higgs field $\phi$ of the standard model. The VEV of $\phi_M$ generates Majorana mass for the right-handed neutrino. This is why we call it the Majorana-Higgs field. It breaks the gauge symmetries from $SU(3)_{C} \times SU(2)_{WL} \times U(1)_{WR} \times U(1)_{B-L}$ down to the standard model symmetries. This subsection also presents one major thesis of our paper: The neutrino Majorana mass preserves lepton number and therefore it does not lead to the neutrinoless double beta decay. 

The Majorana-Higgs field $\phi_M$ is valued in the {\it real} Clifford algebraic subspace spanned by two multivectors
\begin{align}
\label{eq:phiM}
&\phi_M =  (\phi_{M1} + \phi_{M2} I) \gamma_0 P_l,
\end{align}
where $\phi_{M1}$ and $\phi_{M2}$ are two real numbers, and $P_l$ is the lepton projection operator~\eqref{eq:color}.  The Majorana-Higgs field obeys gauge transformation rules
\begin{align}
\label{eq:phiMgauge}
&\phi_M \quad\rightarrow\quad  e^{- \frac{1}{2}\theta_{WR}I - \frac{1}{2}\epsilon_{BL}J} \; \phi_M \; e^{ \frac{1}{2}\theta_{WR}I + \frac{1}{2}\epsilon_{BL}J}, 
\end{align}
where $\theta_{WR}$ and $\epsilon_{BL}$ are the right-handed weak $U(1)_{WR}$ and BL $U(1)_{B-L}$ gauge transformation parameters. As such, $\phi_M$ is invariant under the Lorentz $Spin(1,3)$ (Lorentz scalar), left-handed weak $SU(2)_{WL}$ (weak singlet), and color $SU(3)_{C}$ (color singlet) gauge transformations. Given that $JP_l = -IP_l$, one can replace $J$ with $-I$ in the above transformation. We keep $J$ to highlight its relevance to the BL symmetry. Note that $\phi_M$ is Clifford-odd, different from the Clifford-even electroweak Higgs field which will be investigated in Section~\ref{subsec:Electroweak}. 

The Majorana-Higgs Lagrangian reads
\begin{align}
\mathcal{L}_{Higgs-Majorana} =  \left\langle (D^{\mu} \phi_M)^\dagger(  D_{\mu} \phi_M) - V_M + (y_MiI)\bar{\nu}_R(I\Gamma_2\Gamma_3){\nu}_R \phi_M\right\rangle,
\end{align}
where ${\nu}_R$ is the right-handed neutrino, $y_M$ is the Majorana-Yukawa coupling constant, and the Majorana-Higgs potential  $V_M$ is 
\begin{align}
&V_M = -\mu_{M}^2 \phi_M^\dagger\phi_M + \lambda_{M}(\phi_M^\dagger\phi_M)^2.
\end{align}
The gauge-covariant derivative of $\phi_M$  is defined as
\begin{align}
\label{eq:partialM}
&D_{\mu} \phi_M = (\partial_{\mu} - \frac{1}{2}W_{R\mu}I - \frac{1}{2}A_{BL\mu}J)\phi_M + \phi_M (\frac{1}{2}W_{R\mu}I + \frac{1}{2}A_{BL\mu}J),
\end{align}
which involves the right-handed weak gauge field $\hat{W}_{R\mu}$ and the BL gauge field $\hat{A}_{BL\mu}$.
Note that the bi-vector $\Gamma_2\Gamma_3$ in the Yukawa term can be replaced by an arbitrary combination of $\Gamma_2\Gamma_3$ and $\Gamma_1\Gamma_3$. But it does not change the overall picture.  The imaginary number $i$ shows up in the Yukawa term. This is in compliance with the reality condition for the Majorana-Higgs Lagrangian. It's worth mentioning that for flat space-time we follow the convention of not including $d^4x$ in the definition of the Lagrangian, as opposed to the curved space-time case where $d^4x$ is embedded in the definition of the Lagrangian as 4-form. 

Given that $\phi_M$ is a weak singlet, the Majorana-Yukawa term couples with the right-handed neutrinos only, as opposed to the electroweak Higgs Yukawa term which couples with both the left- and right-handed fermions. The Majorana-Higgs Lagrangian $\mathcal{L}_{Higgs-Majorana}$ can be verified to be $SU(3)_{C} \times SU(2)_{WL} \times U(1)_{WR} \times U(1)_{B-L}$ gauge invariant. Due to the gauge transformation properties of $\phi_M$~\eqref{eq:phiMgauge}, it can be shown that a similar Majorana-Yukawa term coupling to the right-handed electrons is prohibited since such a term violates the gauge symmetries. Therefore, the Majorana-Yukawa term couples to the right-handed neutrinos exclusively. 

By virtue of the Mexican hat-shaped potential  $V_M$, the Majorana-Higgs field $\phi_M$ acquires a nonzero VEV
\begin{align}
\label{eq:MajoranaScale}
&\phi_M = \frac{1}{\sqrt{2}}\upsilon_{M}\gamma_0P_l,
\end{align}
where $\upsilon_{M} = \frac{\mu_{M}}{\sqrt{\lambda_{M}}}$ is called the Majorana scale (or the seesaw scale).

As a result, the gauge symmetry related to the gauge field $\hat{Z}'_{\mu}$
\begin{align}
&\hat{Z}'_{\mu} = W_{R\mu}I + A_{BL\mu}J,
\end{align}
is spontaneously broken. The would-be Nambu-Goldstone boson is "eaten''  by the gauge field $\hat{Z}'_{\mu}$ which gains a mass as a consequence. The local gauge symmetries are broken down to the standard model symmetries $SU(3)_{C}  \times SU(2)_{WL} \times U(1)_Y$, with the hypercharge gauge symmetry $U(1)_Y$ specified by the synchronized double-sided gauge transformation~\eqref{eq:symmetryHyper}. The $U(1)_Y$ gauge field $\hat{A}_Y$ remains massless and has an effective coupling constant of~\cite{WL3}
\begin{align}
&g_{Y} = \frac{g_{WR}g_{BL}}{\sqrt{g_{WR}^2 + g_{BL}^2}},
\end{align}
where $g_{WR}$ and $g_{BL}$ are the right-handed weak and BL coupling constants, respectively. 

After replacement of $\phi_M$ with its VEV, the Majorana-Yukawa term reduces to the Majorana mass term of the right-handed neutrino
\begin{align}
\label{eq:MM}
&Mi\left\langle I\bar{\nu}_{R}(I\Gamma_2\Gamma_3)\nu_{R}\gamma_0\right\rangle = Mi\left\langle I\bar{\nu}_{R}(\nu_{R})_{C'}\right\rangle,
\end{align}
where $(\nu_{R})_{C'}$ is the weaker form charge conjugation $C'$~\eqref{eq:weakerC} of $\nu_{R}$ and the Majorana mass $M$ is
\begin{align}
&M = \frac{1}{\sqrt{2}}y_{M}\upsilon_{M}.
\end{align}
It can be verified that the Majorana mass term respects all the standard model symmetries. This kind of mass is allowed for a standard model singlet such as $\nu_{R}$. It can also be shown that the Majorana mass term is permitted only if $\nu_{R}$ is valued in the Clifford algebraic space with real {\it Grassmann} coefficients. The Majorana mass term would be identically zero if $\nu_{R}$ were valued in the Clifford algebraic space with real coefficients. 

If we juxtapose the Majorana mass term with a typical Dirac mass term between neutrinos
\begin{align}
\label{eq:DM}
&m_\nu i\left\langle I\bar{\nu}\nu \right\rangle =  m_\nu i\left\langle I\bar{\nu}_{L}\nu_{R} + I\bar{\nu}_{R}\nu_{L}\right\rangle,
\end{align}
we can see that the former couples $\bar{\nu}_{R}$ with $(\nu_{R})_{C'}$, while the later couples $\bar{\nu}_{R}$ with $\nu_{L}$. We know that the weaker form of charge conjugation $C'$~\eqref{eq:weakerC} is a Clifford-odd operation. It converts the Clifford-even $\nu_{R}$ to Clifford-odd $(\nu_{R})_{C'}$. Consequently, $(\nu_{R})_{C'}$ is effectively left-handed and could be coupled to the right-handed $\bar{\nu}_{R}$ in a similar fashion as the Dirac mass term. 

The observation of neutrino oscillations\cite{FUK, AHM, EGU}  indicates that neutrinos have nonzero masses which are much smaller than that of the other standard model fermions.  If we assume that the neutrino Majorana mass $M$ is much heavier than the neutrino Dirac mass $m_\nu$, a very small effective mass of the order of $m_\nu^2/M$ can thus be generated for the neutrino. This appealing explanation for the tiny neutrino mass is called the seesaw mechanism~\cite{NEUT}. If we could experimentally detect the Majorana nature of the neutrino mass, it would lend support to the seesaw mechanism. 

Given that the traditional definition of the Majorana mass involves the charge conjugation $C$~\eqref{eq:chargeC} that converts a particle into its corresponding antiparticle, the traditional Majorana mass term violates the conservation of lepton number and could be confirmed by the lepton number-violating process of the neutrinoless double beta decay. Therefore, it's widely believed that the observation of the neutrinoless double beta decay could be a confirmation of the Majorana mass. The lepton number-violating process can also be used to explain the origin of matter in the universe via the leptogenesis mechanism~\cite{LEPTO}. A slew of experiments have been commissioned to search for the neutrinoless double beta decay. As of yet, no evidence of such decay has ever been found~\cite{NDBD1,NDBD2}. 

On the other hand, the weaker form of charge conjugation $C'$~\eqref{eq:weakerC} does not invoke complex conjugate, and thus there is no particle-antiparticle interchange. Consequently, the Majorana mass term as shown in Eq.~\eqref{eq:MM} conserves lepton number, which is dissimilar to the  traditional Majorana mass term that invokes the stronger form of charge conjugation $C$. This suggests that the absence of the neutrinoless double beta decay does not disapprove the Majorana mass  as defined in Eq.~\eqref{eq:MM}. We shall seek other means of the Majorana mass detection.

\subsection{Scalar and antisymmetric-tensor Higgs fields}
\label{subsec:Electroweak}
The third step of SSB concerns the well-known electroweak symmetry-breaking Higgs field $\phi$ of the standard model which couples to both the left-handed and the right-handed fermions. The VEV of the scalar $\phi$ breaks the standard model symmetries down to $SU(3)_{C} \times U(1)_{EM}$. The SSB pattern outlined in this subsection is in many ways similar to the conventional Higgs mechanism, only that it's transposed onto the Clifford algebraic landscape. That said, toward the end of this subsection we will touch upon a non-scalar Higgs field which could have cosmological implications. 

The Higgs field $\phi$ is valued in the {\it real} Clifford algebraic subspace spanned by four Clifford-even multivectors
\begin{align}
\label{eq:phi}
\phi &=  (\phi_{0} + \phi_{1} \Gamma_2\Gamma_3 + \phi_{2} \Gamma_3\Gamma_1 + \phi_{3} \Gamma_1\Gamma_2) P_{+}, 
\end{align}
where $P_+$ is the projection operator~\eqref{IDEM6} with the property $\Gamma_1\Gamma_2P_+ = - IP_+$.  The 4 real coefficients $\{\phi_{a}; a = 0, 1, 2, 3\}$ correspond to the 2 complex components (i.e. 4 real degrees of freedom) of the traditional Higgs doublet. The Higgs field obeys gauge transformation rules
\begin{align}
\label{eq:phigauge}
&\phi \quad\rightarrow\quad  e^{\frac{1}{2}\theta_{WL}^1\Gamma_2\Gamma_3 + \frac{1}{2}\theta_{WL}^2\Gamma_3\Gamma_1 + \frac{1}{2}\theta_{WL}^3\Gamma_1\Gamma_2} \; \phi \; e^{- \frac{1}{2}\epsilon_Y\Gamma_1\Gamma_2}, 
\end{align}
where $\{ \theta_{WL}^1, \theta_{WL}^2, \theta_{WL}^3 \}$ and $\epsilon_Y$ are the left-handed weak $SU(2)_{WL}$ and the hypercharge $U(1)_Y$ gauge transformation parameters. As such, $\phi$ is a weak doublet and is invariant under the Lorentz $Spin(1,3)$ (Lorentz scalar) and the color $SU(3)_{C}$ (color singlet) gauge transformations. Given that $P_+\Gamma_1\Gamma_2 = -P_+ I$, one can replace $-\frac{1}{2}\epsilon_Y\Gamma_1\Gamma_2$ with $\frac{1}{2}\epsilon_YI$ in the above transformations and therefore the Higgs field has Hypercharge $1$ (or $\frac{1}{2}$ depending on the Hypercharge definition convention). We keep $-\frac{1}{2}\epsilon_Y\Gamma_1\Gamma_2$ to highlight its relevance to the Hypercharge symmetry~\eqref{eq:symmetryHyper}. 

Note that the Higgs field could take values in a complimentary Clifford-even subspace
\begin{align}
\label{eq:phi2}
\phi' &=  (\phi'_{0} + \phi'_{1} \Gamma_2\Gamma_3 + \phi'_{2} \Gamma_3\Gamma_1 + \phi'_{3} \Gamma_1\Gamma_2) P_{-}, 
\end{align}
where the projection operator is changed from $P_{+}$ to $P_{-}$. We call the original $\phi$~\eqref{eq:phi} and the complimentary $\phi'$~\eqref{eq:phi2} the $\phi_{+}$-type  and $\phi_{-}$-type Higgs fields, respectively.  They have opposite Hypercharges $P_{\pm}\Gamma_1\Gamma_2 = \mp P_{\pm} I$ and thus their coupling patterns with the isospin up-type and down-type fermions differ from each other. 

It can be verified that the 4 degrees of freedom of $\phi_{+}$-type Higgs field plus the 4 degrees of freedom of $\phi_{-}$-type Higgs field amount to the 8 degrees of freedom of the Clifford subspace $\{1, \Gamma_i\Gamma_j, I, I \Gamma_i\Gamma_j \}$. The additional $\phi_{-}$-type Higgs field can be exploited in various extensions of the standard model such as the two-Higgs-doublet model (2HDM)~\cite{TDL,2HDM} or the three-Higgs-doublet model (3HDM)~\cite{3HDM}. We will circle back to this point in Section~\ref{subsec:Hierarchy}. But for now, let's focus on the $\phi_{+}$-type Higgs field only. 

The standard model gauge-invariant Higgs Lagrangian reads
\begin{align}
\label{eq:HiggsLag}
\mathcal{L}_{Higgs-Electroweak} =  \left\langle (D^{\mu} \phi)^\dagger(D_{\mu} \phi) - V + (y_eiI\bar{l}_L\phi e_R + h.c.)\right\rangle,
\end{align}
where $l_L = \nu_L + e_L$ is the left-handed lepton doublet, $e_R$ is the right-handed electron, $y_e$ is the Yukawa coupling constant, and Higgs potential  $V$ is 
\begin{align}
&V = -\mu^2 \phi^\dagger\phi + \lambda(\phi^\dagger\phi)^2.
\end{align}
The gauge-covariant derivative is defined as
\begin{align}
\label{eq:partial}
&D_{\mu} \phi = (\partial_{\mu} + \frac{1}{2}W^1_{L\mu}\Gamma_2\Gamma_3 + \frac{1}{2}W^2_{L\mu}\Gamma_3\Gamma_1 + \frac{1}{2}W^3_{L\mu}\Gamma_1\Gamma_2)\phi - \phi (\frac{1}{2}A_{Y\mu}\Gamma_1\Gamma_2),
\end{align}
which involves the left-handed weak gauge field $\hat{W}_{L\mu}$ and the Hypercharge gauge field $\hat{A}_{Y\mu}$. We omit Yukawa terms for non-electron fermions which will be investigated in more detail in Section~\ref{subsec:Hierarchy} when we tackle the issue of the fermion mass hierarchies. Note that the imaginary number $i$ shows up in the Yukawa term. This is in compliance with the reality condition. 

By virtue of the Mexican hat-shaped potential  $V$, the Higgs field acquires a nonzero VEV
\begin{align}
&\phi = \frac{1}{\sqrt{2}}\upsilon_{EW} P_+,
\end{align}
where $\upsilon_{EW} = \frac{\mu}{\sqrt{\lambda}}$ is usually called the electroweak scale. As a result, the gauge fields $\hat{W}^{\pm}_{L\mu}$ and $\hat{Z}_{\mu}$ gain masses. The standard model symmetries are broken down to $SU(3)_{C} \times U(1)_{EM}$. The electromagnetic $U(1)_{EM}$ gauge field $\hat{A}$ remains massless and has an effective coupling constant of~\cite{WL3}
\begin{align}
&g_{EM} = \frac{g_{WL}g_{WR}g_{BL}}{\sqrt{g_{WL}^2g_{WR}^2 + g_{WL}^2g_{BL}^2 + g_{WR}^2g_{BL}^2}},
\end{align}
where $g_{WL}$, $g_{WR}$ and $g_{BL}$ are the left-handed weak, the right-handed weak and the BL coupling constants, respectively. The electromagnetic field $\hat{A}$-related gauge-covariant derivative of the algebraic spinor $\psi$ can be cast into the form
\begin{align}
&D_\mu\psi = (\partial_{\mu} + \frac{1}{2}A_\mu\Gamma_1\Gamma_2)\psi + \psi (\frac{1}{2}A_\mu J) = \partial_{\mu}\psi + q A_\mu\psi I,
\end{align}
where $q$ is the electric charge. 

With replacement of $\phi$ with its VEV, the Yukawa term reduces to the Dirac mass term of electron
\begin{align}
\label{eq:DMe}
&m_ei\left\langle I\bar{e}e \right\rangle =  m_ei\left\langle I\bar{e}_{L}e_{R} + I\bar{e}_{R}e_{L}\right\rangle,
\end{align}
where the electron mass $m_e$ is 
\begin{align}
&m_e = \frac{1}{\sqrt{2}}y_{e}\upsilon_{EW}.
\end{align}

At this final stage of SSB, we are ready to write down the action of the electron
\begin{align}
\label{eq:electron} 
S_{electron} = \int \mathcal{L}\; d^{4}x= i\int {\left\langle \bar{\psi} \gamma^{\mu} (\partial_{\mu}\psi +qA_\mu \psi I) + m\bar{\psi}\psi I\right\rangle d^{4}x},
\end{align}
 where  $q= -1$ and we relabel $e$ as $\psi$ and $m_e$ as $m$. The Clifford algebraic Dirac equation can be readily derived
\begin{align}
\label{eq:dirac}
 & \gamma^{\mu} (\partial_{\mu}\psi I-qA_\mu \psi) - m \psi = 0.
\end{align}
It's similar to the conventional Dirac equation, provided that $i$ is replaced with $I$ positioned on the right side of $\psi$. It can be used to derive the equation for the $C'$ charge conjugation~\eqref{eq:weakerC} counterpart $\psi_{c'}$ 
\begin{align}
\label{eq:dirac2}
 & \gamma^{\mu} (\partial_{\mu}\psi_{c'}I +qA_\mu \psi_{c'} ) - m \psi_{c'} = 0.
\end{align}
The above equation demonstrates that the weaker form of charge conjugation $C'$ indeed changes the sign of the electric charge.

We ought to emphasize that the Clifford algebraic Dirac equation~\eqref{eq:dirac} without the quantum imaginary number $i$ is a {\it classical} field equation for the {\it classical} spinor field $\psi$, whereas the Clifford functional-differential Schwinger-Dyson equation~\eqref{eq:SD} (which will be derived in Section~\ref{sec:quantization} on field quantization) is the true QFT equation for the quantized spinor field. This is different from the view of regarding Dirac equation as a relativistic version of quantum mechanical equation. As defined in Section~\ref{sec:Clifford}, the classical spinor field $\psi$~\eqref{eq:psi} is a {\it Grassmann-valued} Clifford algebraic multivector, which is obviously not a {\it complex-valued} quantum wave function. Hence the Clifford algebraic Dirac equation~\eqref{eq:dirac} could not be regarded as a quantum mechanical equation for quantum wave function. In Section~\ref{sec:quantization} on field quantization, we will formally introduce the quantum imaginary number $i$ via the Clifford functional integral formalism and further dispel the notion of ``first quantization'' usually ascribed to Dirac equation. 

The electroweak symmetry breaking process delineated in this section bears close resemblance to the traditional Higgs mechanism. Curiously, the Clifford algebra framework allows for a  non-scalar Higgs field $\phi_{AT}$ which could potentially break both the electroweak and Lorentz symmetries. The non-scalar Higgs field is valued in the real Clifford algebraic subspace spanned by $4*6 = 24$ Clifford-even multivectors\cite{WL3}
\begin{align}
\label{eq:AT16}
&\gamma_a\gamma_b, \quad \gamma_a\gamma_b\Gamma_i\Gamma_j,
\end{align}
where $i, j = 1, 2, 3, i>j,  a, b = 0, 1, 2, 3, a>b$. We have the flexibility of only considering the projected portion $\phi_{AT{\pm}} = \phi_{AT}P_{\pm}$ with each half having 12 independent Clifford algebraic components. The non-scalar Higgs field obeys the transformation rules
\begin{align}
\label{eq:phigauge2}
&\phi_{AT} \quad\rightarrow\quad  e^{\frac{1}{4}\theta^{ab}\gamma_a\gamma_b+\frac{1}{2}\theta_{WL}^1\Gamma_2\Gamma_3 + \frac{1}{2}\theta_{WL}^2\Gamma_3\Gamma_1 + \frac{1}{2}\theta_{WL}^3\Gamma_1\Gamma_2} \; \phi_{AT}  \; e^{- \frac{1}{2}\epsilon_Y\Gamma_1\Gamma_2 -\frac{1}{4}\theta^{ab}\gamma_a\gamma_b}, 
\end{align}
where $\theta^{ab}\gamma_a\gamma_b$ are Lorentz transformations. As such, $\phi_{AT}$ is a weak doublet as well as a Lorentz sextet (an antisymmetric tensor rather than a scalar). If $\phi_{AT}$ acquires a nonzero VEV
\begin{align}
&\phi_{AT} =  \frac{1}{\sqrt{2}} \upsilon_{AT}\gamma_0\gamma_3, \label{LAT}
\end{align}
it would break the electroweak and Lorentz symmetries at the same time, since it singles out a specific space-time direction via $\gamma_0\gamma_3$. The magnitude of this VEV could be extremely small compared with the electroweak scale $\upsilon_{AT}\ll \upsilon_{EW}$, rendering the $\upsilon_{AT}$-related effects unobservable in laboratories. We hypothesize that the ethereal VEV of the  antisymmetric-tensor Higgs field might manifest itself as the large-scale anisotropies of the universe~\cite{CMB1, CMB2, CMB3, CMB4, ANISO1,ANISO2}.

\subsection{The 3HDM and the fermion mass hierarchy problem}
\label{subsec:Hierarchy}
Dimensionless ratios between physical constants appearing in a physical theory cannot be accidentally small. The technical naturalness principle is elegantly defined by 't Hooft~\cite{TH}: A quantity should be small only if the underlying theory becomes more symmetric as that quantity tends to zero. The weakly broken symmetry ensures that the smallness of a parameter is preserved against possible large quantum corrections. 


\begin{subequations}\label{eq:UV}
For the application of the naturalness principle, let's examine two global symmetries related to the vector $U_V(1)$ phase transformation
\begin{align}
\psi_L &\quad\rightarrow\quad \psi_Le^{\theta_V I}, \quad \quad \psi_R \quad\rightarrow\quad \psi_Re^{\theta_V I}, \\
\phi_M &\quad\rightarrow\quad \phi_Me^{2\theta_V I}, \\
\phi &\quad\rightarrow\quad \phi,
\end{align}
\end{subequations}
and the axial $U_A(1)$ phase transformation \begin{subequations}\label{eq:UA}
\begin{align}
\psi_L &\quad\rightarrow\quad  \psi_Le^{-\theta_A I}, \quad \psi_R \quad \rightarrow\quad  \psi_Re^{\theta_A I}, \\
\phi_M &\quad\rightarrow\quad  \phi_Me^{2\theta_A I}, \\
\phi &\quad\rightarrow\quad  \phi e^{2\theta_A I},
\end{align}
\end{subequations}
where $\psi_L$ is the left-handed spinor,  $\psi_R$ is the right-handed spinor, $\phi_M$ is the Majorana-Higgs field, and $\phi$ is the regular Higgs field.  The phase transformation rules for $\phi_M$ and $\phi$ may not seem intuitive. But when we consider $\phi_M$ and $\phi$ as multi-fermion condensations in Section~\ref{subsec:Higgs}, the reason for these transformation rules will become clear. 

For later discussion, let's also introduce a $U_\alpha(1)$ phase transformation for the right-handed spinors \begin{subequations}\label{eq:UA}
\begin{align}
\psi_L &\quad\rightarrow\quad \psi_L,\quad \quad \quad  \psi_R \quad \rightarrow\quad \psi_Re^{\alpha I}, \\
\phi_M &\quad\rightarrow\quad \phi_M e^{2\alpha I}, \\
\phi &\quad\rightarrow\quad \phi e^{\alpha I},
\end{align}
\end{subequations}
which is basically a combination of  $U_V(1)$  and $U_A(1)$  .

It can be checked that, when $\phi_M$ and $\phi$ are replaced by their VEVs, the Majorana~\eqref{eq:MM} and Dirac~\eqref{eq:DMe} mass terms violate the  $U_V(1)$/$U_A(1)$/$U_\alpha(1)$  and $U_A(1)$/$U_\alpha(1)$ symmetries, respectively. Hence the Majorana and Dirac masses are technically natural, given that these global symmetries can be restored if the Majorana and Dirac masses are set to zero. In other words, the smallness of the Majorana and Dirac masses  are protected by the global symmetries against possible large quantum corrections. 

Prior to the SSB induced by $\phi_M$ and $\phi$, these two Higgs fields would transform according to the aforementioned global phase transformation rules. It can be shown that all the terms of the Lagrangian of the world~\eqref{eq:world} respect the $U_V(1)$, $U_A(1)$, and $U_\alpha(1)$ global symmetries, with only one exception which is the $U_A(1)$/$U_\alpha(1)$-violating electron Yukawa term in Eq.~\eqref{eq:HiggsLag}. It would be nice if we can tinker with the $U_A(1)$/$U_\alpha(1)$ transformation rule for $\phi$, so that the entire Lagrangian of the world is invariant. Contrary to our expectation, it's not achievable. Within the confines of a single standard model Higgs field $\phi$, it is impossible to make {\it both} the isospin up-type and down-type fermion Yukawa terms $U_A(1)$/$U_\alpha(1)$-invariant. 

The seemingly worrisome symmetry-violating Yukawa terms can be turned into our advantage. In the spirit of the technical naturalness principle, we can exploit the global symmetry  properties to explain the vast range of fermion masses which span five orders of magnitude between the heaviest top quark and the lightest electron. The key for solving the fermion mass hierarchy problem is to realize that some of the Yukawa coupling constants are actually not constants at all~\cite{WL4}. Embedded in the Yukawa couplings, there are six Clifford-valued bosonic scalar fields 
\begin{align}
\label{eq:Phis}
\Phi_{\alpha t} &= \Phi_{\alpha t 1} + \Phi_{\alpha t 2}I, \quad \quad  \Phi_{\beta t} = \Phi_{\beta t 1} + \Phi_{\beta t 2}I, \\
\Phi_{\alpha {\nu_\tau}} &= \Phi_{\alpha {\nu_\tau} 1} + \Phi_{\alpha {\nu_\tau}2}I, \quad \Phi_{\beta {\nu_\tau}} = \Phi_{\beta {\nu_\tau} 1} + \Phi_{\beta {\nu_\tau}2}I, \\
\Phi_{\alpha {\tau}} &= \Phi_{\alpha {\tau} 1} + \Phi_{\alpha {\tau}2}I,\quad \quad  \Phi_{\beta {\tau}} = \Phi_{\beta {\tau} 1} + \Phi_{\beta {\tau}2}I,
\end{align}
where all the coefficients, such as $\Phi_{\alpha t 1}$ and $\Phi_{\alpha t 2}$, are real numbers. 

The three $\alpha$-type $\Phi$ fields are tied to the $U(1)_\alpha$ symmetry, while the  three $\beta$-type $\Phi$ fields are tied to a novel $U(1)_\beta$  symmetry (see table~\ref{globalS}). The global symmetry-violating nature of the effective Yukawa terms originates from these six $\Phi$ fields acquiring nonzero VEVs via the SSB mechanism. Note that these $\Phi$  fields are gauge singlets, i.e. they are invariant under all the gauge transformations. This is in contrast to the regular electroweak Higgs field which is a weak $SU(2)_{WL}$ doublet. 

To account for the masses of the three families of fermions, we adopt three Higgs fields in our model (a.k.a. 3HDM), namely, the top-quark Higgs field $\phi_t$, the tau-neutrino Higgs field $\phi_{\nu_\tau}$, and the tau-lepton Higgs field $\phi_{\tau}$. The naming convention of the three Higgs fields and the six $\Phi$ fields will become clear when we related them to their corresponding quantum condensations in Section~\ref{subsec:Higgs}. Among these Higgs fields, $\phi_t$ and $\phi_{\nu_\tau}$ are $\phi_{+}$-type Higgs field~\eqref{eq:phi}, while $\phi_{\tau}$ is $\phi_{-}$-type Higgs field~\eqref{eq:phi2}. All the three Higgs doublets obey the usual gauge transformation rules for the electroweak Higgs field~\eqref{eq:phigauge}.  

We introduce one more global symmetry $U(1)_\beta$ which, like $U(1)_\alpha$, is related to the phases of the right-handed fermions. However, the $U(1)_\beta$  charge is not uniformly assigned to fermions. While $U_\alpha(1)$ transforms all the right-handed fermions by the same phase $e^{\alpha I}$, $U_\beta(1)$ transforms the isospin up-type quarks ($u_R$, $c_R$, $t_R$) and down-type leptons ($e_R$, $\mu_R$, $\tau_R$)  by the phase $e^{\beta I}$ and it transforms the down-type quarks ($d_R$, $s_R$, $b_R$) and up-type leptons ($\nu_{eR}$, $\nu_{{\mu}R}$, $\nu_{{\tau}R}$) by the opposite phase  $e^{-\beta I}$. The $U(1)_\alpha$/$U(1)_\beta$ charge assignments are summarized in table~\ref{globalS}.
\begin{table}[ht]
\caption{The $U(1)_\alpha$ and $U(1)_\beta$ charge}
\centering
{\begin{tabular}{|c|c|c|c|c|c|c|c|}
\hline
 & ${u_R}$,${c_R}$,${t_R}$,            & ${d_R}$,${s_R}$,${b_R}$,                                  & $\phi_t$  & $\phi_{\nu_\tau}$  & $\phi_{\tau}$  & $\Phi_{\alpha t}$, $\Phi_{\alpha {\nu_\tau}}$,  & $\Phi_{\beta t}$, $\Phi_{\beta {\nu_\tau}}$, \\ 
 & ${e_R}$,${\mu_R}$,${\tau_R}$   & ${\nu_{eR}}$,${\nu_{\mu R}}$,${\nu_{\tau R}}$ &              &                           &                    &                 $\Phi_{\alpha {\tau}}$       & $\Phi_{\beta {\tau}}$\\ 
\hline
$U(1)_\alpha$ & 1 & 1& 1 &1& 1&-2&0 \\ 
\hline
$U(1)_\beta$ & 1 & -1& 1 &-1& 1&0& -2\\ 
\hline
\end{tabular} \label{globalS}}
\end{table}

It will be shown later in this subsection that the two global symmetries $U(1)_\alpha$ and $U(1)_\beta$ are instrumental in determining the relative magnitudes of the effective Yukawa coupling constants, and consequently establishing the fermion mass hierarchies. As the $U(1)_\alpha$ charge assignment is analogous to that of the Peccei-Quinn $U(1)_{PQ}$ symmetry~\cite{AXN1}, we will use the term $U(1)_\alpha$ and $U(1)_{PQ}$ interchangeably henceforth. 

\begin{subequations}\label{eq:Yukawa}
Now we are ready to write down the Yukawa coupling terms for all three generations of the standard model fermions (plus right-handed neutrinos)
\begin{align}
&i\left\langle 
g_tI\bar{q}^3_{L}\tilde{\phi}_t{t}_{R} +   
g_{\nu_{e}}{\Phi}^\dagger_{\beta t}I\bar{l}^1_{L}\tilde{\phi}_t{\nu}_{eR}  +  
g_b\Phi_{\alpha t}I\bar{q}^3_{L}{\phi}_t{b}_{R} +  
g_e\Phi_{\alpha t}\Phi_{\beta t} I\bar{l}^1_{L}{\phi}_t{e}_{R}\right\rangle + h.c. \\
+&i\left\langle 
g_{{\nu}_{\tau}}I\bar{l}^3_{L}\tilde{\phi}_{\nu_{\tau}}{\nu}_{\tau R} +   
g_c\Phi_{\beta {\nu_\tau}}I\bar{q}^2_{L}\tilde{\phi}_{\nu_{\tau}}{c}_{R} +  
g_{\mu}\Phi_{\alpha {\nu_\tau}}I\bar{l}^2_{L}{\phi}_{\nu_{\tau}}{\mu}_{R} +  
g_d\Phi_{\alpha {\nu_\tau}}{\Phi}^\dagger_{\beta {\nu_\tau}}I\bar{q}^1_{L}{\phi}_{\nu_{\tau}}{d}_{R}\right\rangle \nonumber \\
 &+ h.c. \\
+&i\left\langle  
g_{\tau}I\bar{l}^3_{L}\tilde{\phi}_{\tau}{\tau}_{R}   +  
g_{s}{\Phi}^\dagger_{\beta {\tau}}I\bar{q}^2_{L}\tilde{\phi}_{\tau}{s}_{R} +  
g_{\nu_{\mu}}\Phi_{\alpha {\tau}}I\bar{l}^2_{L}{\phi}_{\tau}\nu_{\mu R} +  
g_u\Phi_{\alpha {\tau}}\Phi_{\beta {\tau}}I\bar{q}^1_{L}{\phi}_{\tau}{u}_{R}\right\rangle + h.c.,
\end{align}
\end{subequations}
where $g_t, g_{{\nu}_{\tau}}, \cdots$ are the bare Yukawa coupling constants which are dimensionless parameters of order $O(1)$. The left-handed doublets  are 
\begin{align}
&l^1_{L} = \nu_{eL} + e_L, \quad q^1_{L} = u_L + d_L , \\
&l^2_{L} = \nu_{\mu L} + \mu_L, \quad q^2_{L} = c_L + s_L , \\
&l^3_{L} = \nu_{\tau L} + \tau _L, \quad q^3_{L} = t_L + b_L , 
\end{align}
where quarks stand for color triplets, such as $u_L  = u_{rL}+u_{gL}+ u_{bL}$. From the Yukawa coupling pattern we can tell that the $\Phi$ singlets are of mass dimension zero, different from the traditional mass dimension-one scalar fields. Alternatively, we can rewrite these singlets as the conventional mass dimension-one scalar fields, as long as they show up in the Yukawa terms as $\Phi/M$ with $M$ being an unknown energy scale. 

As we have mentioned before, ${\phi_t}$ is a $\phi_+$-type Higgs field, which means that it can only couple to the isospin down-type fermions such as ${b}_{R}$ and ${e}_{R}$, whereas direct coupling to the up-type fermions is prohibited. Only a transformed form of  ${\phi_t}$
\begin{align}
\label{eq:HiggT1}
&\tilde{\phi_t} = \frac{1}{4}\gamma^\mu \phi_t \gamma_\mu,
\end{align}
can couple to the up-type fermions such as ${t}_{R}$ and ${\nu}_{eR}$. Note that $\gamma_\mu$ flips the sign of $P_{\pm}$: $P_{\pm}\gamma_\mu = \gamma_\mu P_{\mp}$. Therefore, $\tilde{\phi_t}$ effectively turns a $\phi_+$-type Higgs field $\phi_t$ into a  $\phi_-$-type Higgs field. Similar logic applies to ${\phi}_{\nu_{\tau}}$ and ${\phi}_{\tau}$, which are $\phi_+$-type and $\phi_-$-type, respectively. 

The conventional QFT formalism leverages a different transformation of the standard model Higgs field
\begin{align}
\label{eq:HiggT}
&\tilde{\phi} = i\sigma_2\phi^*,
\end{align}
so that $\tilde{\phi}$ can be coupled to up-type fermions. It involves the complex conjugate, whereas the Clifford algebra version~\eqref{eq:HiggT1} doesn't. 

The Yukawa coupling scheme~\eqref{eq:Yukawa} partitions fermions into three cohorts, namely
\begin{align}
\label{eq:chorts}
{\phi}_t \quad cohort &: t, {\nu}_{e}, b, e, \\
{\phi}_{\nu_{\tau}} \quad cohort &: \nu_{\tau}, c, {\mu}, d, \\
{\phi}_{\tau} \quad cohort &: {\tau}, s, \nu_{\mu}, u.
\end{align}
The right-handed fermions in each cohort only couple to the designated Higgs field, thus preventing the flavor-changing neutral currents (FCNCs). Within each of the Higgs field cohort, only one out of the four Yukawa terms is free from the $\Phi$ coupling. This is because that in each of the Higgs field cohort, three out of the four Yukawa terms violate the $U(1)_\alpha$/$U(1)_\beta$ symmetries without including the additional $\Phi$ singlets. After inserting these $\Phi$ singlets in the Yukawa couplings, it can be verified that all the Yukawa terms respect the $U(1)_\alpha$  and $U(1)_\beta$ global symmetries.

The coupling patterns~\eqref{eq:Yukawa} of the three Higgs fields are predicated on an alternative generation/family assignment,  
\begin{align}
Generation \quad 0 &: \quad t, b, \nu_e, e,  \label{Z0}\\
Generation \quad + &: \quad c, s, \nu_{\tau}, \tau,  \label{ZP}\\
Generation \quad - &: \quad u, d, \nu_{\mu}, \mu, \label{ZN}
\end{align}
which are tied to the flavor projection operators \{$\zeta^{0}$, $\zeta^{+}$, $\zeta^{-}$\}~\cite{WL3,WL4} associated with ternary Clifford algebra~\cite{Abramov,Azcarraga}. These flavor projection operators dictate the $\zeta^{+}$/$\zeta^{-}$ mixing between the ${\phi}_{\nu_{\tau}}$ cohort and the ${\phi}_{\tau}$ cohort. The projection operators can also be applied to the Majorana-type Yukawa coupling and the resultant Majorana mass can directly mix the $\nu_{\mu R}$ and $\nu_{\tau R}$ neutrinos~\cite{WL3,WL4} which is evidenced in the observation of neutrino oscillations\cite{FUK, AHM, EGU}. Other flavor mixing phenomena might also be accommodated by the framework, which we defer to future research. 

We assume that the Lagrangians of the three Higgs doublets and six $\Phi$ singlets are analogous to the one specified for the regular Higgs mechanism~\eqref{eq:HiggsLag}, albeit the potential $V$ could be more complicated, such as involving cross terms mixing different scalar fields~\cite{2HDM, 3HDM}. The study of the exact form of $V$ is beyond the scope of this paper. For our purpose here, we just postulate that these fields acquire the following nonzero VEVs
\begin{align}
&\phi_t = \frac{1}{\sqrt{2}}\upsilon_{t}P_+, \quad {\phi}_{\nu_{\tau}}  = \frac{1}{\sqrt{2}}\upsilon_{\nu_{\tau}}P_+, \quad {\phi}_{\tau} = \frac{1}{\sqrt{2}}\upsilon_{\tau}P_-, \\
& \Phi_{\alpha t} = \upsilon_{\alpha t}, \quad \Phi_{\alpha {\nu_\tau}} = \upsilon_{\alpha {\nu_\tau}}, \quad \Phi_{\alpha {\tau}} = \upsilon_{\alpha {\tau}}, \\
& \Phi_{\beta t} = \upsilon_{\beta t},\quad  \Phi_{\beta {\nu_\tau}} = \upsilon_{\beta {\nu_\tau}},\quad  \Phi_{\beta {\tau}} = \upsilon_{\beta {\tau}}.
\end{align}

The three $\Phi_{\alpha}$-type fields break the global $U(1)_\alpha$/$U(1)_{PQ}$ symmetry, while the three $\Phi_{\beta}$-type fields break the global $U(1)_\beta$ symmetry. Note that the $\Phi$ fields do not break any local gauge symmetry since they are gauge singlets. Post the SSB, there will be six massive sigma modes and six Nambu-Goldstone modes. As opposed to the Higgs mechanism, the Nambu-Goldstone modes are not ``eaten'' by the gauge field. Due to the explicit symmetry breaking originated from the quantum anomaly and instanton effects, the otherwise massless Nambu-Goldstone bosons of the three $\Phi_\alpha$ fields acquire masses and turn into pseudo-Nambu-Goldstone bosons in a similar fashion as the axions~\cite{AXN1, AXN2, AXN3}. Since the $\Phi_\alpha$ fields are local gauge (especially electroweak) singlets, they are more in line with the invisible axions~\cite{INV1, INV2, INV3, INV4}. The axions have historically been proposed as dark matter candidates as well as a possible solution to the strong CP problem. We leave the investigation of the $U(1)_\beta$-type (pseudo-)Nambu-Goldstone boson's role as dark matter to future study. 

After replacement of the three Higgs doublets and six $\Phi$ singlets with their VEVs, the Yukawa terms reduce to the Dirac mass terms 
\begin{subequations}\label{eq:Dmass}
\begin{align}
			 &g_{t}\upsilon_{t}i\left\langle I\bar{t}t\right\rangle
					+ g_{\nu_{e}}\upsilon_{\beta t}\upsilon_{t} i\left\langle I\bar{\nu}_e\nu_e \right\rangle
					+ g_{b}\upsilon_{\alpha t}\upsilon_{t} i\left\langle I\bar{b}b\right\rangle
					+ g_{e}\upsilon_{\alpha t}\upsilon_{\beta t}\upsilon_{t} i\left\langle I\bar{e}e\right\rangle \\
			+ &g_{\nu_{\tau}}\upsilon_{\nu_{\tau}} i\left\langle I\bar{\nu}_{\tau}\nu_{\tau}\right\rangle
					+ g_{c}\upsilon_{\beta {\nu_\tau}}\upsilon_{\nu_{\tau}} i\left\langle I\bar{c}c \right\rangle
					+ g_{\mu}\upsilon_{\alpha {\nu_\tau}}\upsilon_{\nu_{\tau}} i\left\langle I\bar{\mu}\mu \right\rangle
					+ g_{d}\upsilon_{\alpha {\nu_\tau}}\upsilon_{\beta {\nu_\tau}}\upsilon_{\nu_{\tau}} i\left\langle I\bar{d}d\right\rangle \\
			+ &g_{{\tau}}\upsilon_{{\tau}} i\left\langle I\bar{{\tau}}{\tau}\right\rangle
					+g _{s}\upsilon_{\beta {\tau}}\upsilon_{{\tau}} i\left\langle I\bar{s}s \right\rangle
					+g _{\nu_{\mu}}\upsilon_{\alpha {\tau}}\upsilon_{{\tau}} i\left\langle I\bar{\nu}_{\mu}\nu_{\mu} \right\rangle
					+g _{u}\upsilon_{\alpha {\tau}}\upsilon_{\beta {\tau}}\upsilon_{{\tau}} i\left\langle I\bar{u}u\right\rangle,
\end{align}
\end{subequations}
where for the sake of brevity, all terms are re-scaled by $\sqrt{2}$.

Before making contact with the experimental results, we have to identify which mode of the three Higgs doublets corresponds to the 125 GeV boson observed at the Large Hadron Collider~\cite{H125A,H125C}. Generally speaking, a Higgs boson can be defined as a linear combination of the Clifford-scalar sector  (with VEVs subtracted) of the three Higgs fields. For an order-of-magnitude analysis, let's assume that the $125\; GeV$ Higgs boson is aligned with the top-quark Higgs field $\phi_t$. Therefore the VEV of $\phi_t$ is approximately
\begin{align}
\label{eq:VEVt}
\upsilon_{t} \approx  246 GeV,
\end{align}
and the bare Yukawa coupling constant $g_t$ can be identified as the top quark Yukawa constant $y_t = g_t$. For the sake of estimation, we make the further assumption that the  bare Yukawa coupling constants are almost uniform
\begin{align}
\label{eq:UniY}
y_t = g_t \approx g_{\nu_{e}} \approx \cdots \approx g _{u}.
\end{align}
The standard model Yukawa constants, except  $y_t$, can be regarded as effective coupling constants. For example, the bottom quark's effective Yukawa constant is $y_b =  y_t\upsilon_{\alpha t}$, and tau neutrino's effective Yukawa constant is $y_{\nu_{\tau}}= y_t\upsilon_{\nu_{\tau}}/\upsilon_{t}$. 

Aided by the above assumptions and the mass formula~\eqref{eq:Dmass}, we arrive at an estimation of the Higgs/$\Phi$ VEVs and the neutrino Dirac masses as shown in table~\ref{hierarchy}, where the known fermion masses are also included for comparison.
\begin{table}[ht]
\caption{Higgs VEVs (MeV), $\Phi$ VEVs, and Dirac masses (MeV)}
\centering
{\begin{tabular}{|l|lr|lr|lr|}
\hline
 & \multicolumn{2}{|c|}{$\phi_t$ Cohort} & \multicolumn{2}{c|}{${\phi}_{\nu_{\tau}}$ Cohort} & \multicolumn{2}{c|}{${\phi}_{\tau}$ Cohort} \\ 
\hline
&&&&&& \\
Higgs VEVs & $\upsilon_{t}$ &246,000 & $\upsilon_{\nu_{\tau}}$ &41,900 & $\upsilon_{\tau}$ &2,530 \\ 
\hline
&&&&&& \\
$\Phi_\alpha$ VEVs & $\upsilon_{\alpha t}$ &$1/41$ & $\upsilon_{\alpha {\nu_\tau}}$ &$1/278$ & $\upsilon_{\alpha {\tau}}$ &$1/44$ \\ 
&&&&&& \\
$\Phi_\beta$ VEVs & $\upsilon_{\beta t}$ &$1/8200$ & $\upsilon_{\beta {\nu_\tau}}$ &$1/23$ & $\upsilon_{\beta {\tau}}$ &$1/19$ \\ 
\hline
&&&&&& \\
& $t$ &173,000 & $\nu_{\tau}$ &29,500 & ${\tau}$ &1,780 \\ 
&&&&&& \\
Dirac Masses& ${{\nu_e}}$ &21 & ${{c}}$ &1,280 & ${{s}}$ &96 \\ 
&&&&&& \\
& ${b}$ &4,180 & ${\mu}$ &106 & ${{\nu_\mu}}$ &40\\ 
&&&&&& \\
& ${e}$ &0.51 & ${d}$ &4.6 & ${u}$ &2.2 \\
\hline
\end{tabular} \label{hierarchy}}
\end{table}

There are a couple of takeaways from the above estimations. First of all, the magnitudes of $\Phi$ VEVs ($\upsilon_{\alpha t}, \upsilon_{\beta t}, \cdots$) are all small, albeit to varying degrees. In accordance with the technical naturalness principle of 't Hooft~\cite{TH}, the weakly broken symmetries of $U(1)_\alpha$/$U(1)_\beta$ ensure that the smallness of the VEVs is preserved against possible quantum corrections. We have mentioned earlier that the $\Phi$ singlets can be considered as traditional mass dimension-one scalar fields characterized by an  energy scale $M$. We assume that the energy scale $M$ is higher than the Majorana scale $\upsilon_{M}$~\eqref{eq:MajoranaScale}. This is to ensure that the $\Phi$ field-induced $U(1)_\alpha$/$U(1)_\beta$ global symmetry breaking process is decoupled from the Higgs(-like) symmetry breaking mechanism triggered by either the Majorana-Higgs field $\phi_M$ or the electroweak Higgs fields $\phi_t$, ${\phi}_{\nu_{\tau}}$, and ${\phi}_{\tau}$. 

The $\Phi$ VEVs play a crucial role in determining the magnitudes of the effective Yukawa constants and thus establishing the fermion mass hierarchies within each of the $\phi_t$, ${\phi}_{\nu_{\tau}}$, and ${\phi}_{\tau}$ cohorts. On the other hand, the relative fermion mass sizes between different $\phi$ cohorts are controlled by the Higgs VEVs ($\upsilon_{t},\upsilon_{\nu_{\tau}}, \upsilon_{\tau}$). 

The fermion masses within a given ${\phi}$ cohort are mostly in a descending order in each column of table ~\ref{hierarchy}. The only exception is the reversed order between  the ${\nu_e}$ and ${b}$ masses due to the abnormally small magnitude of the $\Phi_{\beta t}$ VEV ($\upsilon_{\beta t} \sim 1/8200$) compared with the other $\Phi$ VEVs. Note that the estimated neutrino masses are meant to be the {\it Dirac} masses, as opposed to the much smaller seesaw effective masses or the vastly larger Majorana masses discussed in Section~\ref{subsec:Majorana}.  Interestingly, according to our estimation, the Dirac mass of the $\nu_{\tau}$ neutrino ($m_{\nu_{\tau}} \sim 29,500 MeV$) is considerably larger than those of the ${{\nu_\mu}}$ and ${{\nu_e}}$ neutrinos ($m_{{\nu_\mu}} \sim 40 MeV$ and $m_{{\nu_e}} \sim 21 MeV$). 

Secondly, assuming that there is no cross term between the kinetic part of the three Higgs Lagrangians, the masses of the $W^{\pm}$ and $Z^0$ bosons can be calculated as
\begin{align}
\label{eq:Wboson}
m_{W^{\pm}} &= \frac{1}{2} \upsilon_{total} g_{WL}, \\
m_{Z^0} &= \frac{1}{2} \upsilon_{total} \sqrt{g_{WL}^2 + g_{Y}^2}, 
\end{align}
where $g_{WL}$ and $g_{Y}$ are the weak and Hypercharge gauge coupling constants. The total electroweak scale $\upsilon_{total}$ is dependent on all three Higgs VEVs
\begin{align}
\label{eq:upsilon}
\upsilon_{total} = \sqrt{\upsilon_{t}^2 + \upsilon_{\nu_{\tau}}^2 + \upsilon_{{\tau}}^2}.
\end{align}
According to table~\ref{hierarchy}, the estimated three Higgs VEVs \{$\upsilon_{t}$, $\upsilon_{\nu_{\tau}}$, $\upsilon_{{\tau}}$ \} have a hierarchical structure
\begin{align}
&246 \; GeV \gg 42\;  GeV \gg 2.5\;  GeV,
\end{align}
where the $\phi_t$ Higgs VEV is significantly larger than the other two. The $\phi_{\nu_{\tau}}$ Higgs VEV plays a non-negligible role in the electroweak scale saturation. 

The total electroweak scale $\upsilon_{total}$ is dominated by the $\phi_t$ Higgs VEV. The ratio between them is given by
\begin{align}
&\frac{\upsilon_{total}}{\upsilon_{t}} = 1. 014.
\end{align}
Given the assumption that the $125\; GeV$ Higgs boson is attributed to the top-quark Higgs field $\phi_t$, this $1.4\%$ discrepancy might be the underlying reason for the deviation of the measured W-boson mass from the standard model prediction~\cite{CDF}. If we tweak the uniform  Yukawa coupling assumption by proposing that the bare Yukawa couplings of the $\phi_{\nu_{\tau}}$ cohort are 5 times larger than the other bare Yukawa couplings, then the ${\upsilon_{total}}/{\upsilon_{t}}$ difference is around $0.06\%$, in the neighborhood of what is observed by the CDF Collaboration~\cite{CDF}. 

And lastly, according to the Yukawa coupling scheme~\eqref{eq:Yukawa}, the muon belongs to the tau-neutrino Higgs field $h_{\nu_{\tau}}$ cohort. Given the intrinsic connection between the muon and the $h_{\nu_{\tau}}$ Higgs field, it is worthwhile to investigate the $h_{\nu_{\tau}}$ Higgs field's contribution to the muon anomalous magnetic moment, especially in light of the recent muon $g-2$ measurement with improved accuracy which confirms a deviation from the standard model prediction~\cite{G2}. 

\section{Quantum Condensation and Naturalness Problems}
\label{sec:quantization}
In this section, we investigate the extended gauge symmetries and contemplate naturalness problems through the lens of quantum condensations. While in the previous section the symmetry-breaking fields are regarded as fundamental fields, in the current section we propose that each and every symmetry-breaking bosonic field is an effective representation of a unique multi-fermion quantum condensation via the dynamical symmetry breaking mechanism. All the mass scales of the universe, including the Planck scale $M_\mathrm{pl}$, are quantum emergent. Each mass scale  is associated with a particular symmetry and the corresponding symmetry breaking process via quantum condensation. 

According to the QFT renormalization procedure, a divergent Feynman  integral $A$ needs to be regularized at the intermediate stage. After renormalization, a finite and regularization scheme-independent result $<A>_R$ can be obtained, so that it can be compared with measurable quantities. As detailed in this section, we advocate a paradigm shift from the conventional renormalization procedure: We stipulate that the relationship involving the renormalized values of the {\it multiplications} of divergent Feynman integrals such as $<A^4>_R = (<A^2>_R)^2$ shall be avoided.  This means that even though we have seemingly only one divergent Feynman integral $A$, there could be multiple unrelated scales, in stark contrast to the conventional QFT renormalization procedure (or the Wilsonian renormalization group approach) characterized by a single renormalization scale.  

In the case of the cosmological constant $\Lambda$ , the implication is that $\Lambda$ (which is linked to $<A^4 >_R$) should be deemed as a parameter completely decoupled from the Planck scale $M_\mathrm{pl}$ (which is linked to $(<A^2>_R)^2$). The magnitudes of these two scales could differ vastly from each other. Hence the cosmological constant problem can be evaded.

\subsection{Quantization via the Clifford functional integral formalism}
\label{subsec:functional}
As mentioned in the introduction section, there are two kinds of imaginary numbers. One is the genuine $i$ which is central to the quantum theory. The other one can be replaced by the pseudoscalar $I$ which shows up in the definition of spinors and gauge fields. Intriguingly, the imaginary number $i$ is inextricably embedded in the { \it classical } fermion Lagrangians \eqref{eq:fermion} and \eqref{eq:electron}, which suggests that there might be quantum phenomenon lurking beneath the veneer of the classical Lagrangian terms. It eventually leads us to the epiphany that quantum condensations may hold the golden key to various sorts of naturalness problems. 

But first, let's examine how to quantize the classical spinor fields. Historically, field quantization is also referred to  as ``second quantization''. However, ``second quantization'' would be a misnomer in our case here, since there is no separate ``first quantization'' in the first place when we define the {\it classical} spinor, gauge, vierbein, and Higgs fields in Section~\ref{sec:Clifford} and Section~\ref{sec:SSB}. 

Different Clifford algebraic field quantization methods have been proposed in the literature~\cite{Pavsic,Borstnik}. With the goal of quantizing the classical Grassmann-odd spinor fields valued in the Clifford algebraic space, we have developed the Clifford functional integral formalism in our earlier paper~\cite{WL5}, whereby the generating functional $Z[\eta]$ for the spinors can be represented by the Clifford functional integral
\begin{align}
\label{eq:Z}
Z[\eta] &= \int \mathcal{D}\psi e^{\frac{1}{2}\int d^4x\{i\mathcal{L}[\psi] + \left\langle I\bar{\eta}(x)\psi(x) + I\bar{\psi}(x)\eta(x)\right\rangle\}},
\end{align}
where we formally introduce the imaginary number $i$ in front of the Lagrangian $i\mathcal{L[\psi]}$ which is essential for quantization. The real Grassmann-odd sources $\eta(x)$ and $\bar{\eta}(x)$ are valued in the same Clifford space as $\psi(x)$ and $\bar{\psi}(x)$. It is understood that $Z[\eta]$ satisfies the normalization condition $Z[0] = 1$. 

Note that we have adopted the natural units $c = \hbar = 1$ in this paper. If we write out the Planck constant $\hbar$ explicitly, the  Lagrangian $\mathcal{L}[\psi] $ in the Clifford functional integral~\eqref{eq:Z} shall be replace by $\mathcal{L}/\hbar$. As a universal rule, we should always deem $\mathcal{L}/\hbar$ as an inseparable quantity whenever we use the Lagrangian with or without quantization. One implication is that the Planck constant $\hbar$ from the Dirac derivative $\gamma^{\mu}i\hbar\partial_{\mu}$ in the spinor Lagrangian $\mathcal{L}[\psi]$ is canceled out by the denominator $\hbar$ in $\mathcal{L}/\hbar$. Therefore, there is no explicit  $\hbar$ in the spinor $\mathcal{L}/\hbar$, which dovetails nicely with our earlier argument about no ``first quantization''. The generalized rule of treating ${\mathcal{L}}/{\hbar}$ as de facto Lagrangian could lead to an alternative view on the choice of physics units~\cite{Barut,Ralston}.

We regard $\psi(x)$ and $\bar{\psi}(x)$ as dependent variables, as opposed to the traditional way of treating them as independent variables (as such, the functional integration is over $\mathcal{D}\psi$ rather than $\mathcal{D}\psi\mathcal{D}\bar{\psi}$). The same logic applies to $\eta(x)$ and $\bar{\eta}(x)$. Therefore, an extra ${1}/{2}$ factor in front of the action is required to keep the calculated quantities, such as the fermion propagators, consistent with those of the conventional formalism. Note that the single-source format $\left\langle I\bar{\eta}(x)\psi(x) + I\bar{\psi}(x)\eta(x)\right\rangle$ is employed here. Alternatively, we can adopt the bilocal-source format $\left\langle \bar{\eta}(x)\psi(x)\right\rangle \left\langle \eta(y)\bar{\psi}(y)\right\rangle$~\cite{WL5} which is well-suited for the non-perturbative approximations analogous to the two-particle irreducible (2PI) effective action approach~\cite{CJT}.  

One feature of the Clifford functional integral formalism is that we don't need to literally perform the functional integration for most cases. Rather, we resort to the property that the functional integration of a total functional derivative is zero
\begin{align}
\label{eq:prop}
\int \mathcal{D}\psi \frac{\delta}{\delta \psi(x)} e^{\frac{1}{2}\int d^4x\{i\mathcal{L}[\psi] +\left\langle I\bar{\eta}(x)\psi(x) + I\bar{\psi}(x)\eta(x)\right\rangle\}} &= 0.
\end{align}
Similar property holds for $\bar{\psi}(x)$. It means that the functional integral is invariant under a shift of $\psi(x)$. 

In our earlier paper on the Clifford functional integral formalism~\cite{WL5}, we have provided the specific definition of the Clifford functional derivatives ${\delta}/{\delta \psi(x)}$ and ${\delta}/{\delta \bar{\psi}(x)}$. For our purpose here, we only need to know the basic Leibniz rule
\begin{align*}
\frac{\delta}{\delta \psi(x)} \left\langle \psi(y) F[\psi]\right\rangle &= \delta(x-y) F[\psi] + \frac{\delta}{\delta \psi(x)} \left\langle \psi(y) \dot{F}[\psi]\right\rangle,
\end{align*}
where the dot on $\dot{F}[\psi]$ denotes functional derivative performed on $F[\psi]$ only.  Similar Leibniz rules can be applied to $\bar{\psi}(x)$, $\eta(x)$ and $\bar{\eta}(x)$. The Leibniz rules, coupled with the other two Clifford algebra properties~\eqref{eq:CliffordCommute}, enable us to perform the relevant Clifford functional derivatives in this paper. 

As an exercise, let's apply the property ~\eqref{eq:prop} to the fermion Lagrangian~\eqref{eq:electron}  absent the electromagnetic coupling. We arrive at the Clifford functional-differential version of the Schwinger-Dyson (SD) equation 
\begin{align}
\label{eq:SD}
&\gamma^\mu{\partial}_{\mu} \{ \frac{\delta}{\delta \bar{\eta}(x)} Z[\eta] I\} - m \frac{\delta}{\delta \bar{\eta}(x)} Z[\eta] + \eta(x)I Z[\eta] = 0.
\end{align}
The solution to the SD equation can be readily obtained as
\begin{align}
\label{eq:Z0}
Z[\eta] = e^{-\frac{1}{2}\int \frac{d^4p}{(2\pi)^4}\left\langle I\bar{\eta}(p)S(p)\eta(p)\right\rangle},
\end{align}
where $\eta(p) = \int d^4x \eta(x)e^{Ip\cdot x}$ and $p\cdot x = p_\mu x^\mu$.  Note that the ``Fourier transformation'' of $ \eta(x)$ is in terms of $e^{Ip\cdot x}$ rather than $e^{ip\cdot x}$. Hence $\eta(x)e^{Ip\cdot x} \neq e^{Ip\cdot x}\eta(x)$, since Clifford-odd part of $\eta(x)$ anticommutes with pseudoscalar $I$. The  Feynman propagator $S(p)$ is given by
\begin{align}
\label{eq:S}
S(p) = \frac{1}{\slashed{p} - m + {i}\epsilon},
\end{align}
where $\slashed{p} = p_\mu\gamma^\mu$. 

A few comments are in order. First of all, the imaginary number $i$ does not explicitly show up in the DS equation~\eqref{eq:SD}. It is because the $i$ in the Clifford functional integral~\eqref{eq:Z} and the $i$ in the fermion Lagrangian~\eqref{eq:electron} cancel out. When gauge fields are included in the Lagrangian, there is no such cancellation due to the absence of $i$ in the Yang-Mills-type Lagrangian terms. 

Secondly, the Feynman propagator $S(p)$ has poles at $p^2 = m^2$. The propagator is not properly defined without a prescription on the $p_0$-related integral in the vicinity of the poles. A well-defined Lorentz-invariant Feynman propagator hinges on the contour integral on the $p_0$ complex plane prescribed by $ i \epsilon$. As will be shown later in this paper, Feynman's $ i \epsilon$ trick has profound implications for quantum condensations and quantum loop integrals, ultimately contributing to the mysterious appearance of the quantum imaginary number $i$ in the { \it classical } fermion Lagrangians \eqref{eq:fermion} and \eqref{eq:electron}.

In the subsequent subsections, the Feynman propagator will be used extensively in various calculations of quantum loop effects. For the sake of brevity, going forward we will not explicitly write down $i \epsilon$ in the propagators.

\subsection{Composite Higgs and the Higgs mass naturalness problem}
\label{subsec:Higgs} 
The discovery of the 125 GeV Higgs boson~\cite{H125A,H125C} has renewed the interest in the possible explanation for the Higgs mass naturalness problem~\cite{GIUD,DINE,CRAIG}, notwithstanding differing views~\cite{Bianchi,Hossenfelder} on the merits of naturalness and fine-tuning arguments in particle physics and cosmology. 

The 125 GeV Higgs mass is technically unnatural according to 't Hooft~\cite{TH}, since even if one takes the massless Higgs boson limit, the symmetry of the standard model is not enhanced. The perturbative quantum corrections tend to draw the Higgs mass towards higher scale. This is in contrast to the case of the fermion mass, which is protected by the $U_A(1)$ global symmetry against possible large quantum corrections. 


One way of addressing the Higgs mass naturalness problem is to replace the fundamental Higgs boson with a fermion-antifermion condensation, such as in the technicolor\cite{TC1, TC2, TC3} and the (extended) top condensation models~\cite{TOP1,TOP2,TOP3,TOP4,TOP5,NEU1,NEU2,NEU3,NEU4,NEU5,WL4}. In these models, the Higgs sector is an effective description of the low energy physics represented by the composite Higgs field. The condensation is induced via the dynamical symmetry breaking (DSB) mechanism, which is a profound concept in physics. It is introduced into the relativistic QFT by Nambu and Jona-Lasinio (NJL)\cite{NJL}, inspired by the earlier Bardeen-Cooper-Schriefer (BCS) theory of superconductivity\cite{BCS}.

Motivated by the proximity of top quark mass scale and the electroweak symmetry breaking scale, the top condensation model has been extensively studied. The simplest version of the top condensation model assumes the top quark-antiquark condensation only. With a view toward explaining the fermion mass hierarchies in the context of composite electroweak Higgs bosons, we have proposed the extended top condensation model in our previous work~\cite{WL4}.  In addition to the top quark condensation, the extended top condensation model involves the tau neutrino and tau lepton condensations as well. The 3HDM in Section~\ref{subsec:Hierarchy} is essentially an effective representation of these three condensations. 

The top condensation model in its original form is based on the NJL-like four-fermion interactions. For example, the top quark interaction term takes the form
\begin{align}
\label{eq:tt4f}
V_{top-quark} & \sim g\left\langle I \bar{q}^3_{L}\gamma^{\mu}{q}^3_{L}I\bar{t}_{R}\gamma_{\mu}{t}_{R} \right\rangle, 
\end{align}
where g is the four-fermion coupling constant and $q^3_{L} = t_L + b_L$. Comparing above with the $\phi_t$ Higgs field Yukawa coupling term~\eqref{eq:Yukawa}, one can see that $\phi_t$ is an effective representation of the fermion-antifermion pair 
\begin{align}
i\phi_{t} & \sim g {q}^3_{L}I\bar{t}_{R}, \label{phit} 
\end{align}
where the $i$ multiplier will be explained later in this subsection. With the replacement of the effective Higgs field, the top quark interaction term~\eqref{eq:tt4f} turns into the top quark Yukawa term
\begin{align}
\label{eq:ttYukawa}
V_{top-quark} & \sim i\left\langle I \bar{q}^3_{L}\gamma^{\mu}\phi_{t}\gamma_{\mu}{t}_{R} \right\rangle \sim i\left\langle I \bar{q}^3_{L}\tilde{\phi}_{t}{t}_{R} \right\rangle, 
\end{align}
Now let's investigate what kind of Clifford algebraic value the effective Higgs field $\phi_t$ can take. First of all,  $\phi_t$ is Clifford-even, given that ${q}^3_{L}$ is Clifford-odd and ${t}_{R}$ is Clifford-even (meaning that $\bar{t}_{R}={t}_{R}^\dagger \gamma_0$ is Clifford-odd). And since ${t}_{R} = P_- {t}_{R}$ (meaning that $\bar{t}_{R}={(P_- t)}^\dagger_{R}\gamma_0 =  \bar{t}_{R}P_+$),  $\phi_t = \phi_t P_+$ should be the $\phi_+$-type Higgs field. Therefore, there are $32/2=16$ components which correspond to the combination of the 4-component  $\phi_+$-type scalar Higgs field~\eqref{eq:phi} and the 12-component $\phi_{AT+}$-type antisymmetric-tensor Higgs field~\eqref{eq:AT16}. 

For a free fermion, it's straightforward to obtain the solution~\eqref{eq:Z0} to the SD equation~\eqref{eq:SD}.  In the presence of interactions such as~\eqref{eq:tt4f}, solving the corresponding SD equation is notoriously hard. The path well-trodden is to find a perturbative solution, under the assumption that a certain coupling constant is small. In our previous paper~\cite{WL5}, we follow a non-perturbative scheme dubbed as the bilocal-source approximation~\cite{Roch,Roch2}, which effectively treats the bilocal-source term as a series expansion parameter. The zeroth-order approximation of the Clifford functional SD equation~\cite{WL5} is equivalent to the self-consistent Hartree mean-field approximation (a.k.a. rainbow approximation). According to the DSB mechanism, when the four-fermion interaction~\eqref{eq:tt4f} is strong enough, it will trigger a quantum condensation
\begin{align}
\label{eq:gap}
i\upsilon_t & \sim  g\int \frac{d^4p}{(2\pi)^4}\frac{m}{p^2 - m^2},
\end{align}
where $\upsilon_t$ is the magnitude of the condensation and $m$ is the emergent top quark mass. 

We can see that the above integral is quadratically divergent. The integral is seemingly a real number. However, as we mentioned in Section~\ref{subsec:functional}, the fermion propagator has poles at $p^2 = m^2$. Feynman's $i \epsilon$ trick ensures that the integral on $p_0$ is well-defined. The proper contour integral on the complex plane of $p_0$ (or equivalently the Wick rotation of time axis) would pick up an imaginary number $i$, thus making the quadratically divergent Feynman integral~\eqref{eq:gap} imaginary valued. Therefore, even if there is no  $i$ in the original four-fermion interaction~\eqref{eq:tt4f}, the imaginary number shows up explicitly in effective Higgs field definition~\eqref{phit} and in the Higgs Yukawa coupling term~\eqref{eq:ttYukawa}.

It is well know that the calculations of QFT are plagued by divergent Feynman integrals, which need to be regularized at the intermediate stage. After renormalization, a finite and regularization scheme-independent result can be obtained for the renormalizable theories. The top condensation model's four-fermion interaction is nonrenormalizable in the conventional sense, since the four-fermion interaction is a dimension six operator. That said, as mentioned earlier, we subscribe to the general notion of effective field theory~\cite{EFFE0,EFFE2}, according to which the seemingly nonrenormalizable models, including a viable quantum theory of gravity~\cite{EFFE,Donoghue2017,Burgess}, are nonetheless manageable renormalization-wise and predictive quantum effect-wise, insofar as there is a separation of low energy physics from the high energy quantum perturbations.

Historically the NJL model has been  presented with the energy cutoff schemes~\cite{Klev}, which usually break the Lorentz invariance. In the presence of a cutoff scale, the four-fermion interaction coupling constant has to be fine-tuned in order to establish the hierarchy between the putative large cutoff scale and the much smaller fermion mass scale. Thus the naturalness problem seems to haunt us again in the Lorentz symmetry-violating cutoff approach. 

On the other hand, there is a Lorentz symmetry-preserving implicit regularization framework~\cite{IMPL,Batt,Batt2} (IR) wherein the divergent parts of Feynman integrals could be isolated in a few Lorentz-invariant primitive integrals that are independent of the external momentum, whereas the remaining external momentum-dependent integrals are convergent. Because the convergent integrals are separated from the divergent ones, the finite parts can be integrated free from the effects of regularization. 

In our earlier paper~\cite{WL5}, we applied the IR technique to the NJL-type model. Granted that the divergent primitive integrals are independent of the external momentum, they can be treated as finite quantities as a result of unspecified (implicit) regularization. The central tenet of the IR approach is that no attempt whatsoever shall be made to  calculate these divergent primitive integrals via  explicit regularization. 
The external momentum-independent divergent primitive integrals are regarded as free parameters of the model that shall be determined by comparing with measurable quantities, such as the emergent fermion mass, the composite boson mass, and the vacuum energy. 

Given that  explicit regularization is eschewed in the calculation, the traditional notion of a cutoff scale and a fine-tuned coupling constant are of no relevance in the IR approach. Instead, the smallness of the symmetry breaking scale of the fermion mass $m$ is an {\it a priori} assumption. Once a small scale of $m$ is assumed at the lower order of approximation,  it's ensured that the smallness of $m$ is preserved against possible higher order disturbances due to the protection from the weakly broken axial $U_A(1)$ symmetry, which is in accordance with the technical naturalness principle. 

Before proceeding with examining the bosonic bound state properties of the composite Higgs model, we would like to mention some open questions. One question is how to properly calculate the vacuum energy shift due to the quantum condensation. And the other is the long-standing issue of the momentum routing ambiguity associated with the fermion bubble diagram~\cite{Will}. With the goal of tackling these issues, we propose an improved version of the IR methodology by adding two supplementary rules below~\cite{WL5}. 

Supplementary rule No. 1: The original IR approach sets forth the rule that if an external momentum-independent primitive divergent Feynman integral $A$ such as Eq.~\eqref{eq:gap} is isolated, at the final stage of calculation it should be replaced with a finite renormalized value $<A>_R$ so that it can be compared with the measurable quantities. We denote this renormalization procedure (R procedure) as 
\begin{align}
\label{eq:renor}
& A \quad \rightarrow \quad <A>_R.
\end{align}
Rule No. 1 stipulates that if divergent Feynman integrals $A$ and $B$ are related to the same physical process, then
\begin{align}
\label{eq:multi}
& <AB>_R \neq <A>_R<B>_R.
\end{align}
In other words, the relationship involving the multiplication of divergent Feynman integrals such as $<AB>_R = <A>_R<B>_R$ shall be avoided if $A$ and $B$ are related to the same physical process. The value of $<AB>_R$ should be treated as independent of $<A>_R$ or $<B>_R$. On the other hand, the R procedure is allowed to be applied recursively to the multiplication of two primitive divergent Feynman integrals if $A$ and $B$ are linked to independent physical processes, such as two independent condensations. 

Supplementary rule No. 2: When a Feynman integral is convergent or logarithmically divergent, the integral is independent of the momentum routing parameter, because the parameter can be shifted away by a translation of the integration variable. When it comes to integrals that are more than logarithmically divergent, one should proceed with caution. For example, the quadratically divergent Feynman integral corresponding to the fermion bubble diagram~\cite{NJL,Will} in the scalar (Higgs boson) channel is
\begin{align}
\label{eq:Pi}
\Pi_s(q) &= i\int \frac{d^4p}{(2\pi)^4}\left\langle S(p+(1-\alpha)q)S(p-\alpha q)\right\rangle, 
\end{align}
where $S$ is the fermion propagator and $\alpha$ is an arbitrary parameter controlling the momentum shifting~\cite{Will}. Unlike the case of convergent or logarithmically divergent Feynman integrals, the seemingly innocuous momentum shifting changes the integral values. Rule No. 2 stipulates that for the quadratically (or higher order) divergent Feynman integrals with momentum routing ambiguities, the momentum routing parameter $\alpha$ shall be set at the symmetrical value. For the above instance, the momentum routing parameter should be set at $\alpha=\frac{1}{2}$, so that $(1-\alpha)q = \alpha q$ is symmetrical. Note that a related ambiguity problem is the triangle diagrams of the Adler-Bell-Jackiw (ABJ) anomaly~\cite{ABJ1,ABJ2}, where the integrals are linearly divergent. This ambiguity is fixed by enforcing the vector Ward identity, at the expense of the axial Ward identity. 

With these two supplementary rules specified, let's investigate the bosonic bound state properties of the composite Higgs model. To this end, we go beyond the zeroth-order bilocal-source approximation and turn to the  first-order approximation of the Clifford functional  SD equation~\cite{WL5}. The collective mode of the composite Higgs boson can be determined via the pole of the composite boson propagator (a.k.a. the fermion-antifermion channel T-matrix) in the scalar channel
\begin{align}
\label{eq:D}
 D_s(p) &\sim \frac{1}{g^{-1} - \Pi_s(p)}, 
\end{align}
where $\Pi_s(p)$ is the bubble function~\eqref{eq:Pi}. The composite boson propagators $D_s(p)$ is the re-summation of the infinite order chain of the fermion bubble diagrams. Similar leading order calculation in the context of contact interactions goes by various names, such as the random-phase approximation, ladder approximation, Bethe-Salpeter T-matrix equation, and 1/N expansion. 

After setting the momentum routing parameter in Eq.~\eqref{eq:Pi} to $\alpha=\frac{1}{2}$, the pole (i.e. the Higgs boson mass) of the composite boson propagator $D_s(p)$ can be calculated as~\cite{WL5}
\begin{align}
\label{eq:Hmass}
 m_h &= \frac{2}{\sqrt{1 + |\Delta|}} m,
\end{align}
where $m$ is the dynamically generated fermion mass and $\Delta$ is defined by the renormalized logarithmically divergent Feynman integral
\begin{align}
\label{eq:Delta}
\Delta^{-1} = 64\pi^2 <\int \frac{d^4p}{(2\pi)^4}\frac{1}{(p^2-m^2)^2}>_R.
\end{align}
Because of the $\sqrt{1 + |\Delta|}$ factor, the composite Higgs boson mass $m_h$ is less than $2m$, which deviates from the typical first-order approximation prediction $m_h =2m$~\cite{NJL}. In other words, $2m$ serves as an upper bound of the Higgs boson mass, which implies that the Higgs boson mass is also protected by the weakly broken axial symmetry, given that the Higgs boson mass and the fermion mass are simultaneously generated by the same DSB mechanism. And additionally, at the electroweak scale there is no elementary Higgs mass term to be modified by any higher order quantum perturbation from external sources. Therefore, the composite Higgs mass is naturally small. 

Historically, when it comes to the top quark condensation model, one phenomenological problem is related to the  prediction of the Higgs-top mass ratio. Since the 2012 discovery~\cite{H125A,H125C}, the Higgs boson is known to be lighter than the top quark. According to the traditional way of Higgs mass calculation, the top condensation model appears to fail since it gives too heavy Higgs mass compared with the top quark mass. However, in our calculation the Higgs mass and the top mass relation involves an extra primitive divergent Feynman integral~\eqref{eq:Delta}. According to the central rule of the IR approach, the value of such integral should not be explicitly calculated. Rather, it is determined by the experimental measurements. Therefore, the observed Higgs-top mass ratio does not falsify the top condensation model. Instead, the ratio fixes the dimensionless parameter $|\Delta|$ of the model. Based on the measured top quark mass ($173$Gev) and Higgs mass ($125$Gev), we arrive at an estimation of $|\Delta|=6.66$ from Eq.~\eqref{eq:Hmass}.

In the same vein as the composite electroweak Higgs field, the Majorana-Higgs field $\phi_{M}$ can be regarded as a composite field representing the condensation of a right-handed neutrino-antineutrino pair~\cite{WL4}
\begin{align}
i\phi_M \sim I\bar{\nu}_R(I\Gamma_2\Gamma_3){\nu}_R. \label{phiM} 
\end{align}
Similarly, the $\Phi$ singlets can be linked to the condensation of fermion-antifermion double pairs~\cite{WL4} such as
\begin{align}
\Phi_{\alpha t} &\sim \bar{b}_{R}{t}_{R}\bar{q}^3_{L}{q}^3_{L}, \label{phialpha} \\
\Phi_{\beta t} &\sim \bar{l}^1_{L}\gamma^\mu {q}^3_{L}\bar{{t}}_{R}\gamma_\mu \nu_{ eR}. \label{phibeta} 
\end{align}
The quantum condensation details of the $\phi_{M}$ and $\Phi$ composite fields can be worked out along the lines of the aforementioned composite electroweak Higgs field. Note that  there is no imaginary number $i$ in the definition of the effective $\Phi$ field, since even numbers of $i$ (two fermion-antifermion pairs) cancel out. 

In summary, the Higgs mass could be naturally small and we have demystified the imaginary number $i$ in the Yukawa/mass term as a vestige of the quantum condensation. Emboldened by these achievements, one might wonder whether we can also surmount the naturalness problem of vacuum energy and decipher the origin of the imaginary number $i$ in the fermion kinetic Lagrangian term.  That is the subject of the next subsection. 

\subsection{Composite vierbein and the cosmological constant problem}
\label{subsec:CC}
Quantum fluctuations of the vacuum contribute to the cosmological constant $\Lambda$. The calculated vacuum energy is extremely large compared with the commonly accepted estimation of $\Lambda$~\cite{Riess1,Perl,Peeb,Cop}. The vacuum energy is $10^{120}$ times too large according to the zero-point energy calculation, or $10^{55}$ times too large according to  the electroweak symmetry breaking calculation. The cosmological constant problem is perceived as the most severe naturalness problem in physics~\cite{Wein,SOLA,BURG}. 

Inspired by the composite Higgs model investigated in Section~\ref{subsec:Higgs}, we turn to the composite vierbein field~\cite{AKA,VOL1,WET1,WET2,DIAK1,DIAK2,OBU} as a possible solution to the cosmological constant problem. Paralleling the dynamical symmetry breaking (DSB) mechanism of the composite Higgs approach, the vierbein field ${\hat{e}} $ can be considered as an effective representation of the the standard model fermion-antifermion condensation 
\begin{align}
i{\hat{e}} & \sim E = \psi d\bar{\psi } \label{veir},
\end{align}
where $d$ is the exterior derivative, hence $E_\mu =\psi \partial_\mu \bar{\psi }$.  Consequently, quantum gravity is realized indirectly via the quantized the standard model spinor field which underlies the composite vierbein field.

Unlike the previous composite vierbein approaches, the Clifford-valued composite vierbein field above is not restricted to the vector space $\gamma_a$, albeit its VEV will congeal around  $\gamma_a$. This is analogous to the Higgs mechanism where the Higgs VEV settles around the subspace $\phi_0 P_+$ out of the full Higgs doublet space of $(\phi_0 + \phi_1\Gamma_2\Gamma_3 + \phi_2\Gamma_3\Gamma_1 + \phi_3\Gamma_1\Gamma_2)P_+$. We will delve into more details about the extended vierbein space in Section~\ref{subsec:chiral}, where a more accurate definition of the chiral composite vierbein fields will be provided when we examine the gauge-covariant chiral vierbeins. 

The composite vierbein field is to be compared with a generalized version of the composite Higgs field 
\begin{align}
i\phi & \sim H = \psi I\bar{\psi } \label{higgs}.
\end{align}
For simplicity reasons, we consider a generic spinor field $\psi$ with both chirality and ignore gauge field coupling. 

There are a couple of similarities and dissimilarities between the spinor bilinears $H$ and $E$. Given that there is no $\left\langle \ldots\right\rangle$ operation in the definition of $H$ and $E$, both $H$ and $E$ are allowed to take values in the general Clifford algebraic space. The Higgs spinor bilinear $H$ is a 0-form, whereas the vierbein spinor bilinear $E= E_{\mu}dx^{\mu}$ is a 1-form  which conforms with fact that the vierbein field $\hat{e}$ is a 1-form. The Higgs spinor bilinear $H$ acquires a Clifford-even VEV, which implies that the Higgs field $\phi$ describes the condensation of an opposite-handed fermion-antifermion pair ($\psi_L I \bar{\psi}_R$ or $\psi_R I \bar{\psi}_L$). On the other hand, the vierbein spinor bilinear $E$ acquires Clifford-odd VEVs valued in \{$\gamma_a$\},  which suggests that the vierbein field $\hat{e}$ describes the condensation of a like-handed fermion-antifermion pair ($\psi_L d \bar{\psi}_L$ or $\psi_R d \bar{\psi}_R$). 

In the above $H$ and $E$ definitions, we follow the tradition~\cite{WL1,VOL2} of regarding the spinor field $\psi$ as dimensionless, a.k.a. bare spinor field. Consequently, the Higgs spinor bilinear $H$   is also dimensionless. The vierbein spinor bilinear $E_\mu =\psi \partial_\mu \bar{\psi }$ is endowed with mass dimension one from the partial derivative. As such, a proper differential form would remain dimensionless. For the example of $E = E_\mu dx^{\mu}$, the mass dimension one of $E_\mu$ is canceled out by the mass dimension minus one of $dx^{\mu}$. The same logic applies to any 1-form gauge field (such as $\hat{A}=\hat{A}_\mu dx^{\mu}$), provided that $\hat{A}_\mu$ is assigned mass dimension one. If we construct a Lagrangian term using the proper differential forms, the mass dimension assignment convention implies that the coefficient in front of the Lagrangian term should be of mass dimension zero. The conventional Lagrangian parameter mass dimensions can be recovered when we re-scale the bare spinor field which will be discussed later in this subsection. 

\begin{subequations}\label{eq:Lags1}
Leveraging these spinor bilinears $E$ and $H$, we can write down the diffeomorphism-invariant Lagrangian terms of the pre-condensation primordial world
\begin{align}
\mathcal{L}_{Fermion+CC} &\sim \left\langle I E \wedge E \wedge E \wedge E \right\rangle, \\
\mathcal{L}_{Yukawa+CC} &\sim \left\langle I E \wedge E \wedge E \wedge E H^2\right\rangle, \\
\mathcal{L}_{Yang-Mills} &\sim \frac{\left\langle (IE\wedge E \wedge \hat{F})(IE\wedge E \wedge \hat{F})\right\rangle} {\left\langle I E \wedge E \wedge E \wedge E \right\rangle}, \label{eq:Lags1YM}\\
\mathcal{L}_{Gravity} &\sim \left\langle IE \wedge E \wedge \hat{R} \right\rangle,
\end{align}
\end{subequations}
where $\hat{F}$ stands for any Yang-Mills-type gauge field curvature $2$-form and $\hat{R}$ is the spin connection curvature $2$-form~\eqref{eq:R}. Note that $\mathcal{L}_{Yukawa+CC}$ could have some variations, such as $\left\langle I E \wedge E \wedge E \wedge H E H\right\rangle$ which corresponds to the top quark-type Yukawa interaction term~\eqref{eq:tt4f}. 

Diffeomorphism-invariance is guaranteed since all the Lagrangian terms are 4-forms on the 4-dimensional space-time manifold. As mentioned earlier, the coefficients in front of the Lagrangian terms (for brevity sake not explicitly written out) are all of mass dimension zero. And we further assume that these dimensionless coefficients should be of order $O(1)$. In other words, there shouldn't be any unnaturally small or large coefficients. 

It's worth mentioning that there are even numbers of fermion-antifermion pairs in each Lagrangian term. It's mandated by the two imperatives: The pre-condensation Lagrangian shall be real and there shall be no explicit imaginary number $i$ in the Lagrangian.  Note that a bare cosmological constant term is not allowed, since diffeomorphism-invariance demands that the Lagrangian terms must be 4-forms and the available gauge-covariant differential forms are either the fermion field-related $E$ spinor bilinear 1-form or the gauge field-related curvature 2-forms. 

The fermion $\mathcal{L}_{Fermion+CC}$ and the Yukawa $\mathcal{L}_{Yukawa+CC}$ Lagrangian terms are comprised of 8 and 12 Grassmann-odd fermion fields respectively. This is very different from the typical fermion Lagrangian. Upon quantum condensation, they will give rise to the conventional fermion kinetic and mass terms as well as the effective cosmological constant term. More specifically, when the three $E$ spinor bilinears in $\mathcal{L}_{Fermion+CC}$ are replaced by their condensation values, the Lagrangian term $\mathcal{L}_{Fermion+CC}$ is left with one $E$ spinor bilinear and is turned into the effective fermion kinetic term. Similarly, when the four $E$ spinor bilinears and one $H $  spinor bilinear in $\mathcal{L}_{Yukawa+CC}$ are replaced by their condensation values, the Lagrangian term $\mathcal{L}_{Yukawa+CC}$ is left with one $H$ spinor bilinear and is turned into the effective Dirac mass term. Lastly, when all spinor bilinears in $\mathcal{L}_{Fermion+CC}$ and $\mathcal{L}_{Yukawa+CC}$ are replaced by their condensation values, these two Lagrangian terms are left with no spinor bilinear and are turned into the effective cosmological constant term. 

Upon DSB-generated quantum condensation, the diffeomorphism-invariance is broken and the effective fermion propagator $S(p)$ assumes the flat space-time form
\begin{align}
\label{eq:S2}
S(p) = \frac{a_0}{\slashed{p} - m_0},
\end{align}
where the emergent mass $m_0$ arises from the $\mathcal{L}_{Yukawa+CC}$ term, and the parameter $a_0$ comes from the $\mathcal{L}_{Fermion+CC}$ term.  The parameter $a_0^{-1}$ is of mass dimension three, since it's related to the condensations of three mass dimension-one $E_\mu$ spinor bilinears. The parameters $a_0$ and $m_0$ can be determined self-consistently via their respective mean-field ``gap'' equations in a similar fashion as the NJL-type model~\cite{NJL,WL5}. 

Leveraging the Clifford generating functional $Z[\eta]$ (Eq.~\eqref{eq:Z0}), the mean-field VEVs of $E_\mu$ and $H$ can be calculated as
\begin{subequations}\label{eq:gap2}
\begin{align}
&E_\mu \sim iM_0\gamma_\mu =  a_0 \gamma_\mu\int \frac{d^4p}{(2\pi)^4}\frac{p^2}{p^2 - m_0^2}, \label{eq:gap2a}\\
&H  \sim i\upsilon_0   =  a_0\int \frac{d^4p}{(2\pi)^4}\frac{m_0}{p^2 - m_0^2}, \label{eq:gap2b}
\end{align}
\end{subequations}
where the VEV magnitudes $M_0$ and $\upsilon_0$ are of mass dimension one and zero, respectively. The VEV of $H$ takes value in the Clifford-scalar space,  while the VEV of $E_\mu$ takes value in the Clifford-vector space \{$\gamma_\mu$\} as expected for an effective vierbein field.  We can see that the above primitive quantum loop integrals for $E_\mu$ and $H$ are quartically and quadratically divergent, respectively. According to the contour integral rule on the complex plane of $p_0$, these integrals pick up an imaginary number $i$ factor. The VEV magnitudes $M_0$ and $\upsilon_0$ are subject to the renormalization procedure $<\cdots>_R$ as delineated in Section~\ref{subsec:Higgs}. 

As we know, $E_\mu$ is of mass dimension one. To be consistent with the conventional formalism of the dimensionless vierbein,  the correspondence between the spinor bilinear $E_\mu$ and the dimensionless vierbein $\hat{e}_\mu$ should be
\begin{align}
\label{eq:gap3}
E_\mu \sim  i M_0\hat{e}_\mu,
\end{align}
which means $\hat{e}_\mu = \gamma_\mu$ according to~\eqref{eq:gap2a}, as expected for the flat space-time. The above calculation of VEVs are based on the flat space-time fermion propagator $S(p)$~\eqref{eq:S2}. In the following discussion, we will assume that the assignment of $E_\mu \sim  i M_0\hat{e}_\mu$ is applicable for the general cases of curved space-time as well. 

\begin{subequations}\label{eq:Lags2}
After replacing  the condensated $E_\mu$ and $H$ with $i M_0\hat{e}_\mu$ and $i\upsilon_0$ respectively and retaining the lowest order terms in the non-condensated $E$ and $H$, the effective  Lagrangian terms take the form
\begin{align}
\mathcal{L}_{Fermion-Kinetic} &\sim i <M_0^3>_R \;\; \left\langle I \hat{e} \wedge \hat{e}\wedge \hat{e} \wedge E \right\rangle, \label{eq:Lags2fermion} \\
\mathcal{L}_{Fermion-Mass} &\sim i <M_0^4 \upsilon_0>_R \;\; \left\langle I \hat{e} \wedge \hat{e}\wedge \hat{e} \wedge \hat{e}  H \right\rangle, \label{eq:Lags2mass}\\
\mathcal{L}_{Yang-Mills} &\sim \frac{<M_0^4 >_R}{<M_0^4 >_R}\;\; \frac{\left\langle (I\hat{e}\wedge \hat{e} \wedge \hat{F})(I\hat{e}\wedge \hat{e} \wedge \hat{F})\right\rangle} {\left\langle I \hat{e} \wedge \hat{e}\wedge \hat{e} \wedge \hat{e} \right\rangle},\label{eq:Lags2YM}\\
\mathcal{L}_{Gravity} &\sim <M_0^2>_R \;\; \left\langle I\hat{e} \wedge \hat{e} \wedge \hat{R} \right\rangle,\label{eq:Lags2Gravity}\\
\mathcal{L}_{CC-Fermion} &\sim <M_0^4 >_R  \;\; \left\langle I \hat{e} \wedge  \hat{e}\wedge \hat{e} \wedge \hat{e}  \right\rangle, \label{eq:Lags2CC1}\\
\mathcal{L}_{CC-Yukawa} &\sim <M_0^4 \upsilon_0^2>_R  \;\; \left\langle I \hat{e} \wedge \hat{e}\wedge \hat{e} \wedge \hat{e}  \right\rangle, \label{eq:Lags2CC2}
\end{align}
where the $\mathcal{L}_{Fermion-Kinetic}$ and $\mathcal{L}_{CC-Fermion}$ terms are derived from $\mathcal{L}_{Fermion+CC}$, while the $\mathcal{L}_{Fermion-Mass}$ and $\mathcal{L}_{CC-Yukawa}$ terms are derived from $\mathcal{L}_{Yukawa+CC}$.
\end{subequations}

Now we can trace the origin of the imaginary number $i$ in the ``classical'' Lagrangian terms. The imaginary number $i$ stems from the primitive divergent integrals~\eqref{eq:gap2} related to the quantum condensations of the $E_\mu$ and $H$ spinor bilinears. If there are odd number of condensations (or equivalently odd number of quantum loop integrals), there is an $i$ in the coefficient of the effective ``classical'' Lagrangian, such as the fermion kinetic~\eqref{eq:Lags2fermion} and mass~\eqref{eq:Lags2mass} terms. On the other hand, if there are even number of condensations (or equivalently even number of quantum loop integrals), there is no $i$ in the coefficient of the effective ``classical'' Lagrangian since $i$ squares to $-1$, such as the Yang-Mills~\eqref{eq:Lags2YM}, gravity~\eqref{eq:Lags2Gravity}, and cosmological constant~\eqref{eq:Lags2CC1}~\eqref{eq:Lags2CC2} terms. 

We can verify that the fermion kinetic~\eqref{eq:Lags2fermion} and mass~\eqref{eq:Lags2mass} terms conform with the corresponding terms specified in the Lagrangian of the world in Section~\ref{subsec:Lagrangian} (see Eq. \eqref{eq:fermion} and \eqref{eq:diracMass}). The only difference is in the coefficients, given that we have adopted the dimensionless bare spinor field  in this subsection. To map to the traditional mass dimension $3/2$ spinor field (a.k.a. dressed spinor field) in Section \ref{subsec:Lagrangian}, we can leverage the field renormalization relationship
\begin{align}
\label{eq:dressed}
&\psi_{dressed} = \sqrt{<M_0^3>_R}\;\; \psi_{bare},
\end{align}
where $\psi_{bare}$ stands for the dimensionless bare spinor field and $\psi_{dressed}$ stands for the mass dimension $3/2$ dressed spinor field. With the substitution of $\psi_{bare}$ with $\psi_{dressed}$,  we can see that the fermion kinetic term~\eqref{eq:Lags2fermion} regains the conventional form~\eqref{eq:fermion} with the coefficient normalized to one. The same substitution in the fermion mass~\eqref{eq:Lags2mass} term  implies that the fermion mass for the dressed spinor field is
\begin{align}
\label{eq:dmass}
m_0 \sim \frac{<M_0^4 \upsilon_0>_R}{<M_0^3>_R}.
\end{align}

For the effective Yang-Mills-type Lagrangian~\eqref{eq:Lags2YM}, the $<M_0^4 >_R$ factors from the numerator and the denominator cancel out. As we mentioned earlier, the dimensionless coefficient in front of the original Yang-Mills-type Lagrangian term~\eqref{eq:Lags1YM} is assumed to be of order $O(1)$. Therefore we expect that the effective Yang-Mills-type coupling constants should not be far from being of order $O(1)$ as well. As a verification, the QED's fine-structure constant is $\alpha \approx 1/137$, which meets our expectation. 

For the effective gravity Lagrangian~\eqref{eq:Lags2Gravity}, the coefficient $<M_0^2>_R$ is of mass dimension two and can be identified with the Planck mass $M_\mathrm{pl}$
\begin{align}
\label{eq:planck}
<M_0^2>_R \;  \approx \; M_\mathrm{pl}^2.
\end{align}

The effective cosmological constant $\Lambda$ is of mass-dimension two. It is defined by the ratio between the $\mathcal{L}_{CC-Fermion}$/$\mathcal{L}_{CC-Yukawa}$ and $\mathcal{L}_{Gravity}$ Lagrangian coefficients
\begin{align}
\label{eq:CC}
\Lambda \sim \frac{<M_0^4 >_R + <M_0^4 \upsilon_0^2>_R}{<M_0^2>_R}  \approx \;  \frac{<M_0^4 >_R + <M_0^4 \upsilon_0^2>_R}{M_\mathrm{pl}^2},
\end{align}
where  $<M_0^4 >_R$ is the vacuum energy contribution from the fermion Lagrangian, and $<M_0^4 \upsilon_0^2>_R$ is the Higgs ground state energy contribution from the Yukawa Lagrangian.

According to the conventional wisdom, each $M_0$ factor in the above equations can be identified with the Planck mass $M_\mathrm{pl}$. Resultantly, the effective fermion mass $m_0$~\eqref{eq:dmass} is estimated as 
\begin{align}
\label{eq:wrongmass}
m_0 \sim \upsilon_0 M_\mathrm{pl}.
\end{align}
Using the top quark mass as an example, $\upsilon_0$ is calculated as of order $\upsilon_0 \approx 10^{-17}$. Similarly, $\Lambda$ is estimated as of order
\begin{align}
\label{eq:wrongLambda}
\Lambda \sim (1 + \upsilon_0^2)M_\mathrm{pl}^2 \approx M_\mathrm{pl}^2,
\end{align}
which is astronomically larger than the commonly accepted estimation of $\Lambda \sim 10^{-120} M_\mathrm{pl}^2 $. 

However, there is a loophole in the above reasoning. According to the supplementary rule No. 1 of the implicit regularization in Section~\ref{subsec:Higgs}, the renormalization procedure can not be applied recursively to the multiplication of primitive divergent integrals
\begin{align}
\label{eq:correct}
& <M_0^4 >_R \; \neq \; <M_0^2 >_R\;<M_0^2 >_R \; \approx \; M_\mathrm{pl}^4.
\end{align}
As such, $<M_0^4 >_R$ should be deemed as a parameter completely decoupled from the scale of $M_\mathrm{pl}^4$. The magnitudes of these two could differ vastly from each other. By the same token,  each of the renormalized primitive divergent integrals $<M_0^3>_R$, $<M_0^4 \upsilon_0>_R$, and $<M_0^4 \upsilon_0^2>_R$ should be regarded as an individual parameter which can only be determined by comparing with measurable results. This means that even though we have seemingly only one divergent  Feynman integral $M_0$, there could be multiple unrelated scales such as the Planck mass $M_\mathrm{pl}$ and the cosmological constant $\Lambda$, in stark contrast to the conventional QFT renormalization procedure (or the Wilsonian renormalization group approach) characterized by a single renormalization scale.  

Therefore, Eq.~\eqref{eq:CC} can not be used to predict the size of the cosmological constant. Rather, one should use the measured magnitude of $\Lambda$ to impute that $<M_0^4 >_R + <M_0^4 \upsilon_0^2>_R\;\sim 10^{-120} M_\mathrm{pl}^4$. Hence the cosmological constant problem can be evaded. 

It's also worth mentioning that there are other contributions to the cosmological constant, such as the Majorana-Higgs-induced phase transitions. In principal, these phase transitions can be treated in a similar way as delineated above.

In the last part of this subsection, we turn our attention to the possible experimental evidences of the composite vierbein. When we write out the effective Lagrangian~\eqref{eq:Lags2}, we only retain the lowest order terms of the non-condensated $E$/$H$ spinor bilinears, under the assumption that the other terms are negligible at low energies. At an elevated energy level, additional terms could become relevant. As an example, let's examine the fermion Lagrangian term with two non-condensated $E$ spinor bilinears. Therefore there are four remaining spinor fields in the effective Lagrangian
\begin{align}
\label{eq:compo}
 \frac{<M_0^2 >_R}{(<M_0^3 >_R)^2}\left\langle \gamma^{\mu}\gamma^{\nu}\psi \partial_\mu \bar{\psi}\;  \psi  \partial_\nu \bar{\psi}\right\rangle,
\end{align}
where $\mu \neq \nu$ and $\psi$ denotes the dressed spinor field with the field renormalization~\eqref{eq:dressed}. We specifically write down the Lagrangian term in flat space-time to highlight the fact that the partial derivatives $\partial_\mu$ and $\partial_\nu$ are orthogonal as opposed to being aligned, which is very different from a typical scalar field Lagrangian that involves two partial derivatives. 

The Lagrangian term could be considered as two perpendicular fermion currents interacting with each other. If such an event is detected experimentally, it would be a telltale sign that one of the composite $\hat{e}$ fields in the effective fermion kinetic Lagrangian term~\eqref{eq:Lags2fermion} is broken down into the underlying fermion fields. In other words, such an event exposes the fermion compositeness of the space-time metric. The coefficient of the above four-fermion term is of mass dimension $-4$. It implies an energy scale
\begin{align}
\label{eq:composcale}
M_{comp} \sim  \left(\frac{(<M_0^3 >_R)^2}{<M_0^2 >_R}\right)^{\frac{1}{4}} \approx \left(\frac{<M_0^3 >_R}{M_\mathrm{pl}}\right)^{\frac{1}{2}},
\end{align}
above which the space-time fabric disintegrates into the underlying fermionic components. Note that we have no theoretical recourse to pinpoint the exact compositeness scale $M_{comp}$, since it involves the primitive divergent integral $<M_0^3 >_R$ which could only be ascertained via empirical means as per the IR tenet. Of particular interest is the fact that the compositeness scale $M_{comp}$ is different from the Planck scale $M_\mathrm{pl}$. They are two unrelated scales. The compositeness scale is the scale above which there could be measurable evidences of the composite vierbein broken down into the fermionic components, whereas the Planck scale is the scale at which the higher-order gravitational Lagrangian terms become relevant. Therefore, if  $M_{comp}< M_\mathrm{pl}$, we could have a chance of probing  the so-called Planck-scale physics at an energy level below the Planck scale.

\subsection{Extended symmetries and gravi-weak interaction}
\label{subsec:chiral}
In the case of the composite Higgs fields, we have benefited from various clues guiding us towards the conclusion that there are three specific fermions catalyzing the electroweak symmetry breaking process, namely, the top quark, tau neutrino, and tau lepton condensations~\cite{WL4}. When it comes to the composite vierbeins, due to lack of evidences we are not able to speculate which of the standard model fermions are involved in the vierbein-related condensations. Nonetheless, we can still make progress by investigating the general symmetry properties of the composite vierbeins. 

Considering the gauge transformation characteristics of the standard model fermions, we cast the effective vierbeins into three categories
\begin{align}
i{\hat{e}_L} & \sim E_L = \psi_L \overline{D_L{\psi_L }} \label{veirL}, \\
i{\hat{e}_{Ru}} & \sim E_{u} = \psi_{Ru} \overline{D_R{\psi_{Ru} }} \label{veirR1}, \\
i{\hat{e}_{Rd}} & \sim E_{d} = \psi_{Rd} \overline{D_R{\psi_{Rd}}} \label{veirR2}, 
\end{align}
where $D_L$ and $D_R$ are the left-handed and right-handed gauge-covariant derivatives, $\psi_L$ is any left-handed doublet such as $u_L + d_L$, $\psi_{Ru}$ is any right-handed up-type singlet such as $u_R$, and $\psi_{Rd}$ is any right-handed down-type singlet such as $d_R$.

Given the freedom afforded by the above chiral vierbeins unconstrained by the Clifford subspace \{$\gamma_a$\}, the gauge symmetry groups~\eqref{eq:symmetry1} we studied earlier can be expanded to (but still a subset of the enveloping symmetries \eqref{eq:largest} and \eqref{eq:largest2})
\begin{align}
\label{eq:symmetry3}
 &Spin(1,3)_{L} \times Spin(1,3)_{R} \times Spin(1,3)_{WL} \times Spin(1,1)_{WR} \times U(1)_{WR} \nonumber \\
 &\times SU(3)_{C}\times U(1)_{B-L},
\end{align} 
where $Spin(1,3)_{L}$ and $Spin(1,3)_{R}$ are the left-handed and right-handed Lorentz gauge groups respectively. The extended left-handed weak group $Spin(1,3)_{WL}$ generators are
\begin{align}
\label{eq:symmetry4}
 &\Gamma_2\Gamma_3, \; \Gamma_3\Gamma_1, \; \Gamma_1\Gamma_2, \;  \Gamma_0\Gamma_1, \; \Gamma_0\Gamma_2, \; \Gamma_0\Gamma_3, 
\end{align} 
where $\Gamma_0 = \gamma_1\gamma_2\gamma_3$ and the pseudo-weak portion of the extended right-handed $Spin(1,1)_{WR}$ group  generator is
\begin{align}
\label{eq:symmetry4}
 &\Gamma_0\Gamma_3.
\end{align} 
The left-handed and right-handed fermions transform independently under the chiral left-sided gauge transformations $Spin(1,3)_{L} \times Spin(1,3)_{WL}$ and $Spin(1,3)_{R}\times Spin(1,1)_{WR} \times U(1)_{WR}$ respectively, whereas the left-handed and right-handed fermions transform in unison under the right-sided gauge transformation  $SU(3)_{C}\times U(1)_{B-L}$.  

Note that the  extended left-handed weak group $Spin(1,3)_{WL}$  comprises the regular weak group $SU(2)_{WL}$ generated by \{$\Gamma_i\Gamma_j$\}  as well as the weak-boosts generated by \{$\Gamma_0\Gamma_i$\}. These are the counterparts of the spacial rotations generated by \{$\gamma_i\gamma_j$\} and the Lorentz boosts generated by \{$\gamma_0\gamma_i$\}. The weak-boost is not a group on its own, it's rather the coset $Spin(1,3)_{WL}/SU(2)_{WL}$. Given the relationships such as $\Gamma_0\Gamma_3 = I\Gamma_1\Gamma_2$, we also call \{$\Gamma_0\Gamma_i$\} the pseudo-weak generators. Henceforth, we will use the terms weak-boost and pseudo-weak interchangeably. We would like to highlight the fact that both the local Lorentz symmetry and the extended  left-handed weak symmetry are $Spin(1,3)$, hinting at an interesting duality between Lorentz gravity and weak interactions. Historically, there have been various attempts~\cite{Nesti2008,Alexander,Woit2021,Woit2023} trying to explore the possible connection between the gravitational and weak interactions.

Given that both the vierbeins and the left-side gauge transformations $V$ are acting on the left side of a spinor, the left-handed and right-handed vierbeins should transform as vectors ($\hat{e} \rightarrow V  \hat{e}   V^{-1}$) under the gauge transformations $Spin(1,3)_{L} \times Spin(1,3)_{WL}$ and $Spin(1,3)_{R} \times Spin(1,1)_{WR}$ respectively to ensure the gauge-invariance of the spinor Lagrangian. Consequently, the left-handed ${\hat{e}_L}$ ought to be valued in the extended Clifford algebraic subspace spanned by the $4*4=16$ multivectors
\begin{align}
\label{eq:VLspace}
&\gamma_a, \quad \gamma_a \Gamma_0\Gamma_i,
\end{align}
where $a= 0,1,2,3$ and $i=1, 2, 3$.  The right-handed ${\hat{e}_{Ru}}$ is valued in the Clifford algebraic subspace spanned by the $4$ multivectors
\begin{align}
\label{eq:VRu}
&\gamma_a P_{+},
\end{align}
while the right-handed ${\hat{e}_{Rd}}$ is valued in the Clifford algebraic subspace spanned by the $4$ multivectors
\begin{align}
\label{eq:VRd}
&\gamma_a P_{-},
\end{align}
where $P_{\pm}$ are the projection operators~\eqref{IDEM6}.  

Alternatively,  ${\hat{e}_L}$ may take values in the complimentary Clifford algebraic subspace spanned by the $4*4=16$ multivectors
\begin{align}
\label{eq:VLspace2}
&\gamma_a I, \quad \gamma_a \Gamma_i\Gamma_j.
\end{align}
As such, ${\hat{e}_L}$ could develop VEVs valued in the pseudo-vector  subspace \{$\gamma_a I$\}, rather than the regular vector subspace \{$\gamma_a$\}. The same logic goes for ${\hat{e}_{Ru}}$ and ${\hat{e}_{Rd}}$. Nevertheless, our discussion in this paper is concentrated on the regular vector-type ${\hat{e}_L}$, ${\hat{e}_{Ru}}$, and ${\hat{e}_{Rd}}$ vierbeins. 

Note that if the left-handed ${\hat{e}_L}$ were restricted to the vector space $\gamma_a$, it could not accommodate the left-handed weak-boost transformation $\hat{e}_L  \rightarrow e^{\frac{1}{2}\theta^{0i}\Gamma_0\Gamma_i}  \; \hat{e}_L \;   e^{-\frac{1}{2}\theta^{0i}\Gamma_0\Gamma_i}$. The  left-handed ${\hat{e}_L}$ has $16*4 = 64$ individual components ${e}^a_{L\mu}$ with $a = 1\dots 16$ and $\mu = 0,1,2,3$. We keep using the term ``vierbein'' (4-legs) for $\hat{e}_L$ given the 4-dimensional space-time indexed by $\mu= 0,1,2,3$. If we want to accentuate the 16-dimensional Clifford algebraic subspace indexed by $a = 1\dots 16$, we could refer to $\hat{e}_L$ as ``vielbein'' (multi-legs). The  metric $g_{\mu\nu}$, defined as $g_{\mu\nu}= \left\langle \hat{e}_{\mu}\hat{e}_{\nu} \right\rangle$, is still a $4\times 4$ tensor regardless of the dimension of the Clifford algebraic subspace of the vierbeins, since the underlying space-time (indexed by $\mu$ and $\nu$ ) is 4-dimensional. It's worth mentioning that historically various proposals~\cite{Percacci,NestiPercacci,Chamseddine,KrasnovPercacci,MaiezzaNesti,Konitopoulos,Volovik2022,Obukhov} have been made to extend the vierbein (vielbein) space while keeping the 4-dimensional space-time. 


With the extended symmetries, the chiral gauge-covariant derivatives of the left- and right-handed spinor fields $\psi_{L/R}(x)$ are defined by
\begin{align}
&D_L\psi_L = (d + \hat{\omega}_L + \hat{W}_{L} + \hat{W'}_{L})\psi_L + \psi_L (\hat{G} + \hat{A}_{BL} ), \\
&D_R\psi_R = (d + \hat{\omega}_R + \hat{W}_{R}+ \hat{W'}_{R})\psi_R + \psi_R (\hat{G} + \hat{A}_{BL} ),
\end{align}
where the left-hand weak $SU(2)_{WL}$ gauge field $\hat{W}_{L}$, the right-hand weak $U(1)_{WR}$ gauge field $\hat{W}_{R}$, the color $SU(3)_{C}$ gauge field $\hat{G}$, and the BL $U(1)_{B-L}$ gauge field $\hat{A}_{BL}$ follow the same definition as specified previously~\eqref{eq:oneform}. The newly introduced gauge fields are the left- and right-handed spin connections of the $Spin(1,3)_{L}$ and $Spin(1,3)_{R}$ Lorentz gauge groups
\begin{align}
\label{eq:chiralw}
&\hat{\omega}_{L} = \frac{1}{4}\omega_{L\mu}^{ab}\gamma_{a}\gamma_{b}dx^\mu, \\
&\hat{\omega}_{R} = \frac{1}{4}\omega_{R\mu}^{ab}\gamma_{a}\gamma_{b}dx^\mu,
\end{align}
the pseudo-weak portion of the extended left-handed weak gauge field
\begin{align}
\label{eq:boostL}
&\hat{W'}_{L}= \frac{1}{2}(W'^1_{L\mu}\Gamma_0\Gamma_1 + W'^2_{L\mu}\Gamma_0\Gamma_3 + W'^3_{L\mu}\Gamma_0\Gamma_3)dx^\mu, 
\end{align}
and the pseudo-weak portion of the extended right-handed weak gauge field
\begin{align}
\label{eq:boostR}
&\hat{W'}_{R}= \frac{1}{2}W'^3_{R\mu}\Gamma_0\Gamma_3dx^\mu. 
\end{align}
The combination of the regular weak $\hat{W}_{L}$ and the pseudo-weak $\hat{W'}_{L}$ constitutes the overall gauge field of $Spin(1,3)_{WL}$
\begin{align}
\label{eq:boostR}
&\hat{\omega}_{Iso-L}= \hat{W}_{L} + \hat{W'}_{L}. 
\end{align}
We call $\hat{\omega}_{Iso-L}$ the isospin connection since it is in many ways analogous to the regular spin connection~\eqref{eq:chiralw} of the Lorentz group. 

The chiral spin connections $\hat{\omega}_L$ and $\hat{\omega}_R$ are crucial in maintaining the  chiral Lorentz gauge covariance of $D_L\psi_{L}$ and $D_R\psi_{R}$, which are leveraged in conjunction with  the chiral vierbeins to ensure the chiral Lorentz gauge invariance of the Lagrangian terms. 

The gauge interaction curvature $2$-forms for $\hat{\omega}_{L}$, $\hat{\omega}_{R}$, $\hat{W'}_{L}$, and $\hat{W'}_{R}$ are expressed as
\begin{align}
\label{eq:NewForce}
&\hat{R}_L = d\hat{\omega}_L  + \hat{\omega}_L \wedge\hat{\omega}_L , \\
&\hat{R}_R = d\hat{\omega}_R  + \hat{\omega}_R \wedge\hat{\omega}_R , \\
&\hat{F}_{W'L} = d\hat{W'}_{L} +\hat{W'}_{L}\wedge\hat{W}_{L} + \hat{W}_{L}\wedge\hat{W'}_{L}, \\
&\hat{F}_{W'R} = d\hat{W'}_{R},
\end{align}
where the outer product between gauge fields vanishes for the abelian interaction $\hat{F}_{W'R}$. The regular left-handed weak force is appended with an additional cross product term of $\hat{W'}_{L}$
\begin{align}
\label{eq:NewWeak}
&\hat{F}_{WL} = d\hat{W}_{L} +\hat{W}_{L}\wedge\hat{W}_{L} + \hat{W'}_{L}\wedge\hat{W'}_{L}. 
\end{align}
The combination of  the weak $\hat{F}_{WL}$ and the pseudo-weak $\hat{F}_{W'L}$ constitutes the overall gauge curvature 2-form of the weak $Spin(1,3)_{WL}$
\begin{align}
\label{eq:boostL}
&\hat{R}_{Iso-L}= \hat{F}_{WL} + \hat{F}_{W'L}=d\hat{\omega}_{Iso-L} + \hat{\omega}_{Iso-L}\wedge \hat{\omega}_{Iso-L}. 
\end{align}
We call $\hat{R}_{Iso-L}$ the isospin connection curvature 2-form (or the extended weak force) in parallel with the spin connection curvature 2-form $\hat{R}_L$~\eqref{eq:NewForce} of the Lorentz group.

The local gauge- and diffeomorphism-invariant Lagrangian terms of the world are similar to the ones we inspected earlier in Section~\ref{subsec:Lagrangian}, provided that the chirality and isospin conjugations are taken care of. The following are some examples 
\begin{subequations}\label{eq:world3}
\begin{align}
\mathcal{L}_{Fermion} \sim &\;i\left\langle I\hat{e}_L\wedge \hat{e}_L\wedge \hat{e}_L \wedge  (\psi_L \overline{D_L\psi_L}) \right\rangle  \\
&+ \;i\left\langle I\hat{e}_{Rd}\wedge \hat{e}_{Ru}\wedge \hat{e}_{Rd} \wedge  (\psi_{Ru} \overline{D_R\psi_{Ru}}) \right\rangle \\
&+ \;i\left\langle I\hat{e}_{Ru}\wedge \hat{e}_{Rd}\wedge \hat{e}_{Ru} \wedge  (\psi_{Rd} \overline{D_R\psi_{Rd}}) \right\rangle, \\
\mathcal{L}_{Gravity-Left}  \sim &\;\left\langle I\hat{e}_L\wedge \hat{e}_L \wedge \hat{R}_L \right\rangle, \label{eq:world3gravity}\\
\mathcal{L}_{Gravity-Right}  \sim &\;\left\langle I(\hat{e}_{Ru}\wedge \hat{e}_{Rd} + \hat{e}_{Rd}\wedge \hat{e}_{Ru})\wedge \hat{R}_R \right\rangle, \label{eq:world3gravity2}\\
\mathcal{L}_{CC-Left} \sim &\;\left\langle I\hat{e}_L\wedge \hat{e}_L \wedge \hat{e}_L\wedge \hat{e}_L \right\rangle, \label{eq:world3CC1}\\
\mathcal{L}_{CC-Right} \sim &\;\left\langle I\hat{e}_{Ru}\wedge \hat{e}_{Rd} \wedge \hat{e}_{Ru}\wedge \hat{e}_{Rd} \right\rangle,\label{eq:world3CC2}
\end{align}
\end{subequations}
where the alternation between $\hat{e}_{Ru}$ and $\hat{e}_{Rd}$ is because of the properties $\hat{e}_{Ru} = P_{-}\hat{e}_{Ru}P_{+}$ and $\hat{e}_{Rd} = P_{+}\hat{e}_{Rd}P_{-}$.

In view of the extended symmetries~\eqref{eq:symmetry3} of the Lagrangian of the world, let's revisit the diffeomorphism and Lorentz gauge symmetry breaking triggered by the nonzero VEVs of the vierbeins. It can be checked that the flat space-time VEV~\eqref{eq:vierbeinVEVflat} of the vierbein violates the gauge symmetries $Spin(1,3)_{L} \times Spin(1,3)_{R} \times Spin(1,1)_{WR}$ and the coset $Spin(1,3)_{WL}/SU(2)_{WL}$. The remaining gauge symmetries are $SU(3)_{C} \times SU(2)_{WL} \times U(1)_{WR} \times U(1)_{B-L}$ plus the residual global Lorentz symmetry as the combined remnant of the local Lorentz and diffeomorphism symmetries.  We would like to highlight the fact that our model has no conflict with the Coleman-Mandula theorem~\cite{Coleman}, since neither the local Lorentz symmetry nor the residual global Lorentz symmetry  has nontrivial mixing with the  internal gauge symmetries. The interested reads are encouraged to read the discussion in the literature~\cite{Nesti2008} regarding why the residual global Lorentz symmetry, rather than the broken local Lorentz symmetry, is the symmetry relevant to the Coleman-Mandula theorem. 

Note that the VEV magnitudes and orientations of the three vierbeins $\hat{e}_L$, $\hat{e}_{Ru}$, and $\hat{e}_{Rd}$ may not align with each other. That said, we have the freedom to re-scale and re-orientate (via the global Lorentz rotations) the corresponding fermions, so that every vierbein takes the same flat space-time VEV everywhere all at once. Hence it is ensured that all fermions, regardless of their chirality and isospin types, share the universal Minkowski flat space-time metric.  

However, there are some factors that can not be re-scaled away. These are the differences between the coefficients of the chiral gravity~\eqref{eq:world3gravity}~\eqref{eq:world3gravity2} Lagrangian terms and the differences between the coefficients of the chiral cosmological constant~\eqref{eq:world3CC1}~\eqref{eq:world3CC2} Lagrangian terms. As a result, the left- and right-handed fermions experience different strengths of gravitational interactions. Under normal conditions, this is not a problem since the left- and right-handed matters are usually commensurate with each other. Thus we can't really discern which chiral gravity is stronger since we routinely observe the collective gravitational interaction governed by a combined effective gravitational constant. 

The only exception is when there is imbalance between the left- and right-handed matters. For example, let's assume that the left-handed gravitational constant is larger than the right-handed counterpart. We could observe an unexplainable drop of gravitational force compared with expectation if there is an excess of right-handed matter. In this regard, we would like to draw attention to the right-handed neutrinos, since they are endowed with extremely large Majorana masses. If there is a large concentration of the right-handed neutrinos in certain parts of the universe, the discrepancies between the chiral gravitational forces would possibly be revealed. 

In the last part of this subsection, we turn to a novel kind of gravi-weak Lagrangian terms
\begin{subequations}\label{eq:world4}
\begin{align}
\mathcal{L}_{Gravity-Weak-Left}  \sim &\;\left\langle I\hat{e}_L\wedge \hat{e}_L \wedge \hat{R}_{Iso-L}\right\rangle, \label{eq:world4gravity}\\
\mathcal{L}_{Holst-Weak-Left} \sim &\;\left\langle \hat{e}_L\wedge \hat{e}_L \wedge \hat{R}_{Iso-L}\right\rangle, 
\end{align}
\end{subequations}
where $\hat{R}_{Iso-L}$ is the left-handed isospin connection curvature 2-form~\eqref{eq:boostL}. As indicated by the Lagrangian names, these terms bear close resemblance to the regular gravity~\eqref{eq:gravity} and Holst~\eqref{eq:gravity2} Lagrangian terms. We mentioned earlier that under normal circumstances a Lagrangian term with a single Yang-Mills field curvature 2-form is identically zero. This is because the traditional vierbein is {\it invariant} under the Yang-Mills field-related gauge transformations. However, the extended left-handed vierbein ${\hat{e}_L}$ {\it transforms as a vector} under the extended left-handed weak symmetry $Spin(1,3)_{WL}$, which makes the above single-curvature gravi-weak terms possible. 

Let's derive the field equations for the left-handed gravity Lagrangian~\eqref{eq:world3gravity} and the left-handed  gravi-weak Lagrangian~\eqref{eq:world4gravity} by varying with  $\hat{e}_L$, $\hat{\omega}_L$, and $\hat{\omega}_{Iso-L}$, respectively. The resultant extended Einstein-Cartan equations read (for brevity sake we drop the $L$ subscripts and quantities are implicitly left-handed)
\begin{align}
&\frac{1}{8\pi G}(\hat{R}\wedge \hat{e} + \hat{e} \wedge \hat{R})I + \frac{1}{8\pi G_{Iso}}(\hat{R}_{Iso}\wedge \hat{e} + \hat{e} \wedge \hat{R}_{Iso})I= \mathbb{T}, \label{eq:EC:b1}\\
&\frac{1}{8\pi G}(\hat{T}\wedge \hat{e} - \hat{e}\wedge \hat{T})I = \mathbb{S}, \label{eq:EC:b2} \\
&\frac{1}{8\pi G_{Iso}}(\hat{T}_{Iso}\wedge \hat{e} - \hat{e}\wedge \hat{T}_{Iso})I = \mathbb{S}_{Iso}, \label{eq:EC:b3}
\end{align}
where $G$ is the gravitational constant, $G_{Iso}$ is the iso-gravitation constant for the gravi-weak Lagrangian~\eqref{eq:world4gravity}, $\mathbb{T}$ is the energy-momentum current $3$-form, $\mathbb{S}$ is the  spin current $3$-form, $\mathbb{S}_{Iso}$ is the isospin current $3$-form, $\hat{T}$ is the torsion $2$-form~\eqref{eq:torsion}, and $\hat{T}_{Iso}$ is the iso-torsion $2$-form
\begin{align}
\label{eq:Isotorsion}
\hat{T}_{Iso} &= d\hat{e} + \hat{\omega}_{Iso} \wedge \hat{e} + \hat{e} \wedge \hat{\omega}_{Iso}.
\end{align}
Compared with the regular Einstein-Cartan equations, we have an additional equation for the iso-torsion $\hat{T}_{Iso}$. 
Given that the weak gauge field $\hat{W}_{L}$ (as part of the isospin connection $\hat{\omega}_{Iso-L}$) is susceptible to the SSB effects from the electroweak Higgs mechanism, we expect that the isospin connection $\hat{\omega}_{Iso-L}$  would have a different kind of impact on gravity than the regular spin connection $\hat{\omega}_{L}$. 

This sort of gravi-weak modification to the gravitational equations could have cosmological implications. As we know, the concordance $\Lambda$CDM cosmological model is besieged with a multitude of  discordances~\cite{Bull,DISCORD1,DISCORD2}, with the most acute one being the Hubble tension~\cite{Riess2,Planck,Planck2}. Various modified gravity models~\cite{SF,CDL,HCK} have been proposed to remediate the shortcomings of the $\Lambda$CDM model. There are two noteworthy modified gravity theories invoking two different scales: a characteristic Hubble scale $h_0$ and a characteristic acceleration scale $a_0$. In cosmology, the characteristic Hubble scale $h_0$ marks the boundary between the validity domains of Friedmann equation and modified Friedmann equation (MOFE) which could explain the late-time accelerated expansion of the universe {\it without dark energy} \cite{WL2,WL5}, whereas in galactic dynamics the characteristic acceleration scale $a_0$ marks the boundary between the validity domains of Newtonian dynamics and modified Newtonian dynamics (MOND) which could explain the galactic rotation curves {\it without dark matter} \cite{MOND}. We hope that the gravi-weak interplay delineated above could possibly shed some light on the dark side of the universe. 

\section{Conclusions}
\label{sec:concl}
The naturalness problems have been front and center in physics researches~\cite{Wein,SOLA,BURG,GIUD,DINE,CRAIG}. The cosmological constant problem is arguably the most severe naturalness problem in physics, with the runner-up being the Higgs mass/electroweak hierarchy problem. 
With the goal of addressing the naturalness problems, we propose that each and every symmetry-breaking bosonic field is an effective representation of a unique multi-fermion quantum condensation via the dynamical symmetry breaking mechanism. Note that while these symmetry-breaking fields such as the vierbein and  Higgs fields are composite fields, all the standard model fermion  fields and all the gauge fields in our model are still fundamental non-composite fields.

Our research originated from drawing a previously unappreciated distinction between two imaginary numbers~\cite{WL4,WL5}: The first one is the bona fide imaginary number $i$ which governs the quantum world, whereas the other one is the unit pseudoscalar $I$ of Clifford algebra $Cl(0,6)$ masquerading as imaginary number which shows up in the definition of spinors, gauge fields, and their transformations. In the Clifford algebra approach, we can manage to completely avoid imaginary number $i$ in classical field equations with classical spinors and gauge fields defined as Clifford multivectors. This is demonstrated by the Clifford algebraic Dirac equation~\eqref{eq:dirac}, where the conventional $i$ is replaced by pseudoscalar $I$ as long as we stick to the regime of applying pseudoscalar $I$ to the right side of the algebraic spinor. 

However, when it comes to the fermion Lagrangian, the imaginary number $i$ is irreplaceable in both the kinetic and mass terms. The conundrum of the {\it quantum} $i$ enmeshed in the { \it classical } spinor Lagrangian indicates that the regular classical Lagrangian terms might be of quantum origin. We propound that the imaginary number in the fermion Lagrangian stems from the odd number of quantum loop integrals related to the odd number of fermion-antifermion quantum condensations. On the other hand, if there are even fermion-antifermion pairs involved in the quantum condensations with even number of quantum loop integrals, there is no $i$ in the Lagrangian term, such as the Yang-Mills, gravity, and cosmological constant terms.   

We would like to underscore the fact that there is no fixed mass/energy scale in the pre-symmetry breaking  world, since the original Lagrangian coefficients are dimensionless. In other words, the  pre-condensation Lagrangian terms are scale-invariant. Each mass scale of the universe, including the Planck scale, is associated with a particular symmetry and the corresponding symmetry breaking process via quantum condensation. This pan-emergence paradigm of mass scales parallels Landau’s symmetry breaking scheme in condensed matter physics characterized by the nonzero order parameter. One exception to this rule might be the QCD scale $\Lambda_{QCD}$ which is possibly generated by a topological order in quark-gluon plasma and nucleons akin to the fractional quantum Hall effect~\cite{WL6}. 

Let's take stock of various symmetry-breaking quantum condensations examined in this paper. The first category of quantum condensation involves a standard model fermion-antifermion pair of the same chirality, with a gauge-covariant derivative sandwiched in between. The effective representation of this sort of condensation corresponds to the vierbein field $\hat{e}$ in the Lorentz gauge  theory of gravity. In the composite vierbein scenario, the standard model fermions play the dual role of interacting with the space-time metric as well as being the metric. Consequently, quantum gravity is realized indirectly via the quantized spinor fields which underlie the composite space-time metric. Allowed by the extended vierbein valued in the 16-dimensional multivector subspace spanned by \{ $\gamma_a \Gamma_0\Gamma_i$ \} in addition to  \{ $\gamma_a$\}, our model accommodates an extended left-handed weak gauge group $Spin(1,3)_{WL}$ which encompasses the standard model left-handed weak gauge group $SU(2)_{WL}$. A gravi-weak interplay is thus permitted between the extended vierbein field and the extended weak gauge field. 

The local Lorentz and pseudo-weak symmetries are broken when the vierbeins acquire nonzero VEVs via the dynamical symmetry breaking mechanism. One interesting implication is that there are two different energy scales related to this symmetry breaking process. One is the compositeness scale above which there are measurable evidences of the composite vierbeins broken down into the underlying fermionic components, while the other is the Planck scale at which the higher-order gravitational Lagrangian terms become relevant in the effective field theory of quantum gravity. The second implication is that the coefficients of the effective cosmological constant and gravity Lagrangian terms are dictated by the divergent Feynman integrals of quantum condensations. We advocate a paradigm shift from the conventional single-scale renormalization procedure. The cosmological constant problem can thus be evaded if we adopt a multi-scale renormalization procedure ($R$ procedure) for quantum condensations that entail { \it multiplications} of divergent Feynman integrals. 

The second category of quantum condensation involves a neutrino-antineutrino pair with right-handed chirality which breaks the BSM symmetries. The effective description of this sort of condensation is the Majorana-Higgs field $\phi_M$. It is a Higgs-like field whose VEV generates mass for the $Z'$ gauge field and the Majorana mass for the right-handed neutrino. The Majorana mass is capable of directly mixing neutrinos from different generations, which is evidenced in the observation of neutrino oscillations\cite{FUK, AHM, EGU}. The Clifford algebra $Cl(0,6)$ allows for a weaker form of charge conjugation which does not invoke particle-antiparticle interchange. Consequently, the Clifford algebraic Majorana mass conserves lepton number, which is different from the traditional Majorana mass term. This might be the underlying reason that no evidence has ever been found for the neutrinoless double beta decay~\cite{NDBD1,NDBD2}.

The third category of quantum condensation involves a standard model fermion-antifermion pair with opposite chirality. Belonging to this category, the standard model Higgs field  $\phi_t$ is an effective description of the top quark condensation, while the other two yet-to-be-detected composite Higgs fields $\phi_{\nu_\tau}$ and $\phi_\tau$ correspond to the tau neutrino and tau lepton condensations, respectively. The VEVs of these three composite Higgs fields break the electroweak symmetries and generate masses for the  $Z^0$/$W^{\pm}$ gauge fields and the Dirac masses for the standard model fermions. The composite Higgs mass is naturally small, since at the electroweak scale there is no elementary Higgs mass term to be modified by any higher order quantum perturbation from external sources. 

The three estimated Higgs VEVs have a hierarchical structure $\upsilon_{t} \approx 246 \; GeV$, $\upsilon_{\nu_{\tau}} \approx 42\;  GeV$ and $\upsilon_{{\tau}} \approx 2.5\;  GeV$. The top-quark Higgs VEV $\upsilon_{t}$ is much larger than the other two. Nonetheless, the tau-neutrino Higgs VEV $\upsilon_{\nu_{\tau}}$ plays a non-negligible role in the electroweak scale saturation, which might be the root cause of the significant deviation of the measured W-boson mass from the standard model prediction~\cite{CDF}. Additionally, given the intrinsic connection between the muon and the tau-neutrino Higgs field $\phi_{\nu_\tau}$, it is worthwhile to investigate the tau-neutrino Higgs field's contribution to the muon anomalous magnetic moment, especially in light of the recent muon $g-2$ measurement which confirms a deviation from the standard model prediction~\cite{G2}. 

The fourth category of quantum condensation involves a standard model fermion-antifermion pair with opposite chirality, similar to the regular composite Higgs field in the third category. However, the Clifford algebra framework  allows for a non-scalar antisymmetric-tensor composite Higgs field $\phi_{AT}$ which could potentially break both the electroweak and Lorentz symmetries. The magnitude of its VEV could be extremely small compared with the electroweak scale, rendering its effects unobservable in laboratories. The ethereal VEV of the antisymmetric-tensor Higgs field might manifest itself as the large-scale anisotropies of the universe~\cite{CMB1, CMB2, CMB3, CMB4, ANISO1,ANISO2}.

The fifth category of quantum condensation involves two standard model fermion-antifermion pairs. There are six composite $\Phi$ fields corresponding to this sort of four-fermion condensations. In contrast to the other four types of composite fields, these scalar $\Phi$ fields are invariant under all the local gauge transformations. Instead, three of the $\Phi$ fields are tied to a $U_\alpha(1)$ {\it global} symmetry, which transforms  all the right-handed fermions by the same phase $e^{\alpha I}$, in a manner similar to the Peccei-Quinn $U(1)_{PQ}$ symmetry. The other three of the $\Phi$ fields are tied to a $U_\beta(1)$ {\it global} symmetry. It transforms the up-type quarks ($u_R$, $c_R$, $t_R$) and down-type leptons ($e_R$, $\mu_R$, $\tau_R$)  by the phase $e^{\beta I}$, whereas it transforms the down-type quarks ($d_R$, $s_R$, $b_R$) and up-type leptons ($\nu_{eR}$, $\nu_{{\mu}R}$, $\nu_{{\tau}R}$) by the opposite phase  $e^{-\beta I}$.  

Upon acquiring nonzero VEVs, these six composite fields break the $U_\alpha(1)$ and $U_\beta(1)$ global symmetries respectively. Their VEVs play a pivotal role in establishing the relative magnitudes of the effective Yukawa coupling constants, and consequently giving rise to the fermion mass hierarchies. According to the effective Yukawa coupling ansatz, the Dirac masses of the $\nu_{\tau}$, ${{\nu_\mu}}$ and ${{\nu_e}}$ neutrinos are estimated as  $29,500 MeV$, $40 MeV$ and $21 MeV$, respectively. Note that these estimations are meant to be the Dirac masses, as opposed to the significantly smaller seesaw effective masses.  

Due to the explicit symmetry breaking originated from the quantum anomaly and instanton effects, the otherwise massless Nambu-Goldstone bosons of the  $\alpha$-type $\Phi$ fields acquire masses and turn into the pseudo-Nambu-Goldstone bosons in a similar fashion as the axions. Historically the axions have  been proposed as a possible solution to the strong CP and dark matter problems. In this regard, we speculate that the (pseudo-)Nambu-Goldstone bosons of the  $\beta$-type $\Phi$ fields could also be viable dark matter candidates. 

In summary, the proposition in this paper is  at the conservative end of the spectrum of physics modeling. Our thesis is that the space-time manifold is 4-dimensional and the fundamental building blocks of the universe are just the garden-variety standard model fermions plus the right-handed neutrinos, accompanied by a handful of ``good old-fashioned'' gauge fields such as the spin connection Lorentz gauge fields and the Yang-Mills-type gauge fields. That is all there is. There are neither compactified extra space dimensions nor exotic wiggling p-branes. The novel bit we bring to the table is the insight that there is a kaleidoscope of quantum condensations which make the world as complex and enchanting as it is. 

\section*{Acknowledgement}
I am grateful to Matej Pav\v{s}i\v{c} and Grigory Volovik for helpful discussions.

\end{document}